\documentclass[11pt]{article}
\pdfoutput=1
\usepackage{graphicx,epsfig}
\usepackage{color}
\usepackage[colorlinks,citecolor=red,linkcolor=blue]{hyperref}
\usepackage{slashed}
\usepackage[left=2.5cm, right=2.5cm, top=2cm]{geometry}

\usepackage{epsfig,float,amsmath,amsfonts,amssymb,graphicx,bm,bbm,array,dcolumn}
\usepackage{subfigure}

\newcommand{\Lagr}{\mathcal{L}}
\newcommand{\VLtc}{\frac{m_{ct}m_{tt} + m_{cc}m_{ct} + m_{cT}m_{tT}}{m_{cc}^2-m_{tt}^2}}
\newcommand{\VLTc}{\frac{\mu_Tm_{cT} + \mu_c m_{cc} + \mu_t m_{ct}}{m_{cc}^2-\mu_T^2}}
\newcommand{\VLTt}{\frac{\mu_Tm_{tT} + \mu_c m_{ct} + \mu_t m_{tt}}{m_{tt}^2-\mu_T^2}}
\newcommand{\VRtc}{\frac{m_{ct}m_{tt} + m_{cc}m_{ct} + \mu_c\mu_t}{m_{cc}^2-m_{tt}^2}}
\newcommand{\VRTc}{\frac{\mu_T\mu_c + m_{cT} m_{cc} + m_{tT} m_{ct}}{m_{cc}^2-\mu_T^2}}
\newcommand{\VRTt}{\frac{m_{ct}m_{cT} + \mu_t \mu_T +  m_{tt}m_{tT}}{m_{tt}^2-\mu_T^2}}
\newcommand{\VLtcneg}{\frac{m_{ct}m_{tt} + m_{cc}m_{ct} + m_{cT}m_{tT}}{m_{tt}^2-m_{cc}^2}}
\newcommand{\VLTcneg}{\frac{\mu_Tm_{cT} + \mu_c m_{cc} + \mu_t m_{ct}}{\mu_T^2-m_{cc}^2}}
\newcommand{\VLTtneg}{\frac{\mu_Tm_{tT} + \mu_c m_{ct} + \mu_t m_{tt}}{\mu_T^2-m_{tt}^2}}
\newcommand{\VRtcneg}{\frac{m_{ct}m_{tt} + m_{cc}m_{ct} + \mu_c\mu_t}{m_{tt}^2-m_{cc}^2}}
\newcommand{\VRTcneg}{\frac{\mu_T\mu_c + m_{cT} m_{cc} + m_{tT} m_{ct}}{\mu_T^2-m_{cc}^2}}
\newcommand{\VRTtneg}{\frac{m_{ct}m_{cT} + \mu_t \mu_T +  m_{tt}m_{tT}}{\mu_T^2-m_{tt}^2}}

\newcommand{\ol}{\overline}

\def\beq{\begin{equation}}
\def\eeq{\end{equation}}
\def\bea{\begin{eqnarray}}
\def\eea{\end{eqnarray}}
\def\bmat{\begin{pmatrix}}
\def\emat{\end{pmatrix}}

\newcommand{ \slashchar }[1]{\setbox0=\hbox{$#1$}   
   \dimen0=\wd0                                     
   \setbox1=\hbox{/} \dimen1=\wd1                   
   \ifdim\dimen0>\dimen1                            
      \rlap{\hbox to \dimen0{\hfil/\hfil}}          
      #1                                            
   \else                                            
      \rlap{\hbox to \dimen1{\hfil$#1$\hfil}}       
      /                                             
   \fi}                                             %

\def\to{\rightarrow}
\def\Chi{{\cal X}}

\begin{document}

\title{
\normalsize{\bf Survey of vector-like fermion extensions of the Standard Model and their phenomenological implications}}

\author{\normalsize{Sebastian A.R. Ellis$^a$\thanks{sarellis@umich.edu}, Rohini M. Godbole$^b$\thanks{rohini@cts.iisc.ernet.in}, Shrihari Gopalakrishna$^c$\thanks{shri@imsc.res.in}, James D. Wells$^a$\thanks{jwells@umich.edu}}
\\
$^a$~\small{Physics Department, University of Michigan, Ann Arbor, MI, USA} \\
$^b$~\small{Center for High Energy Physics, Indian Institute of Science, Bangalore, India}\\
$^c$~\small{Institute of Mathematical Sciences (IMSc), Chennai, India.}
}

\maketitle

\begin{abstract}
With the renewed interest in vector-like fermion extensions of the Standard Model, we present here a study of multiple vector-like theories and their phenomenological implications. Our focus is mostly on minimal flavor conserving theories that couple the vector-like fermions to the SM gauge fields and mix only weakly with SM fermions so as to avoid flavor problems. We present calculations for precision electroweak and vector-like state decays, which are needed to investigate compatibility with currently known data. We investigate the impact of vector-like fermions on Higgs boson production and decay, including loop contributions, in a wide variety of vector-like extensions and their parameter spaces. 
\end{abstract}

\vfill\eject

\tableofcontents

\vfill\eject

\section{Introduction}

The Standard Model of particle physics is a chiral theory under the electroweak gauge symmetries, $SU(2)_L\times U(1)_Y$.  It is for this reason that a condensing Higgs boson is needed to generate elementary particle masses. Non-chiral pairs of fermions, or mirror pairs $f_q$ and $f_{\bar q}$ are able to achieve mass explicitly through the gauge-invariant bilinear interaction $m_ff^\dagger_qf_{\bar q}$. There is no reason why such pairs of vector-like fermions do not exist. The only question is what is their mass, since it is in this case not tied at all to electroweak symmetry breaking.

In the literature there are numerous examples of  theories that require vector-like fermions in the spectrum. As with all cases, the mass scale is uncertain. In supersymmetric theories we know that there must be vector-like fermions in the spectrum, namely the Higgsinos. The Higgsinos form a vector-like pair of fermions with the same quantum numbers as the left-handed leptons and its conjugate. 
Naturalness requires the Higgsinos to be near the weak scale, although the precise mechanism that achieves this
is the subject of rich theoretical analysis -- one that we do not traverse here. 
A general thought does rise with this observation, and that is why not vector-like complements of all the other SM fermions?

Indeed, more fundamental theories, such as string theories and D-brane theories, often do give rise generically to vector-like states.  For example, Dijkstra et al.~\cite{Dijkstra:2004cc} search the landscape of orientifolds of Gepner models for Standard Model-like vacua of three generations and find a plethora of models with vector-like complements of the Standard Model states. In D-brane constructions it is generic to get these extra vector-like states since the computation for chiral fermion representations involves identifying topological intersection numbers, whereas the vector-like states multiplicity is not confined to that, and there can be many more.  The chiral content is the ``left over" chiral fermions that are necessarily lighter because they cannot be paired up to receive mass without the Higgs boson. 

Since all string theories can be related through dualities, it is not surprising, although not a priori necessary, that there would be generic presence of vector-like states in other constructions. Indeed, there are additional cases. For example, the ubiquity of vector-like states is manifest also in orbifold constructions of heterotic string compactifications~\cite{Lebedev:2006kn}. In this case, more realistic models tend to produce vector-like complements of the down-type fermions, ${\bf 5}+\bar {\bf 5}$ in SU(5) language.

Both string examples discussed may also give rise to vector-like states of other representations, and even ones of fractional electric charge (see e.g.,~\cite{Raby:2007hm}). 
Vector-like families are not motivated only by string theory considerations, but also lower energy constructions such as 
top-quark seesaw models~\cite{Dobrescu:1997nm,Chivukula:1998wd,He:1999vp}, 
warped extra dimensions (see for example Ref.~\cite{Gopalakrishna:2013hua} and references therein), 
composite Higgs~\cite{Contino:2006qr,Anastasiou:2009rv,Vignaroli:2012sf,DeSimone:2012fs,Delaunay:2013iia,Gillioz:2013pba}, 
little Higgs theories~\cite{Han:2003wu,Carena:2006jx,Matsumoto:2008fq,Berger:2012ec}, and 
low-scale supersymmetry~\cite{Kang:2007ib,Martin:2009bg,Graham:2009gy,Martin:2010dc,Moroi:2011aa,Martin:2012dg,Fischler:2013tva,GMSB-iwamoto}. 
Ref.~\cite{Martin:2009bg} shows that the Higgs mass can be raised in supersymmetry by the addition of vector-like matter.   
Ref.~\cite{Graham:2009gy} considers the addition of vector-like matter to improve the little hierarchy problem in the MSSM. 
Ref.~\cite{Moroi:2011aa,Martin:2012dg,Fischler:2013tva,GMSB-iwamoto} 
consider the implications of gauge mediated supersymmetry breaking with vector-like matter.
Refs.~\cite{delAguila:1982fs,delAguila:2000rc,delAguila:2008pw,Kearney:2012zi} also deal with 
various aspects of vector-like fermions. 
There is also interest in vector-like states from a purely agnostic phenomenological inquiry to potentially better fit Higgs boson data~\cite{Bonne:2012im,ArkaniHamed:2012kq}. Again, these ideas give rise to vector-like SM complements and also new representations. The implications and phenomenology of these latter states is qualitatively different than vector-like complements of the SM states, and will be discussed elsewhere.

%
There have been many works that have appeared since the discovery of the Higgs attempting to explain the preliminary discrepancies. 
An example is the ``simplified models" approach of Ref.~\cite{Carmi:2012yp}.
Ref.~\cite{Bonne:2012im} attempted to explain the discrepancies in the Higgs data (with limited statistics) prevailing then with vector-like fermions 
having non-standard hypercharge assignments.
Ref.~\cite{ArkaniHamed:2012kq}, with a similar goal, introduces only vector-like leptons with SM hypercharge assignment.  
Refs.~\cite{Joglekar:2012vc,Fairbairn:2013xaa} considers a vector-like lepton generation (including new SU(2) singlets) 
and the possibility of the electromagnetic charge neutral vector-like lepton being a dark matter candidate. 
They analyze precision electroweak bounds, modifications to Higgs observables and vacuum stability bounds for this extension. 
Ref.~\cite{Fairbairn:2013xaa} also investigates baryogenesis and vacuum stability in this context. 

More recently, Ref.~\cite{Dawson:2012di} considers vector-like quarks with SM EM charges. 
Their analysis of the SU(2) doublet case differs from ours in the following respects: 
(a) we add new $SU(2)$ singlet vector-like quarks motivated by replicating the SM structure, while they do not. 
The constraints from shifts to $Zb\bar b$ couplings on their model is very tight, while this will not apply in our case since we do not allow any 
significant Yukawa coupling between the new SU(2) vector-like doublet and SM singlets;
(b) they only consider quarks while we include leptons also;
(c) we include the recent LHC Higgs data while their study was done before the Higgs discovery. 
Refs.~\cite{Fajfer:2013wca,Aguilar-Saavedra:2013wba} also have SM singlets only, not new vector-like singlets like we consider. 
Ref.~\cite{Dawson:2012mk} considers the modification to double Higgs production due to vector-like fermions in such a model. 

Ref.~\cite{Moreau:2012da} performs a model-independent analysis of the recent Higgs data, obtains
preferred regions of the effective couplings, and interprets this in the context of vector-like fermions. 
Since the SU(2) representations of the fermions are left unspecified, electroweak precision constraints
have not been applied. In our work we consider a few concrete SU(2) representations and do apply 
electroweak precision constraints. Refs.~\cite{Chpoi:2013wga} also perform an effective operator analysis of the Higgs data. 
Refs.~\cite{Martin:2009bg,Graham:2009gy,Martin:2012dg,Dawson:2012di} discuss collider searches after taking into
account precision electroweak constraints. 
Ref.~\cite{Azatov:2012rj} takes into account precision electroweak constraints and considers the
$t^\prime$ pair production followed by the $t^\prime \to t h$, $h\to \gamma\gamma, ZZ$ decay modes, 
while Ref.~\cite{Harigaya:2012ir} considers $t^\prime \to t h$ in the multi-$b$-jets channel. 
Ref.~\cite{topPart-Caccia,Aguilar-Saavedra:2013qpa} evaluates precision electroweak and flavor constraints on top-partner vector-like quarks, 
and studies direct LHC signatures.  

In the following sections of this paper, we shall not discuss further the underlying motivations, but study carefully the theory construction and phenomenological implications of vector-like SM complement states.  The masses and couplings will be free parameters for us, except that we will generally confine ourselves to the case of vector-like states mixing weakly with SM states. This is not absolutely required, but we do it so as not to complicate our work with detailed flavor physics. The high-energy manifestations and searchers are our primary consideration, including the impact of vector-like states on the Higgs branching fractions, particularly into two photons.
Since the mass of the vector-like fermions are not generated through the Yukawa couplings, the loop contributions 
involving the Higgs decouple faster than for chiral fermions. 
Hence the constraints from the current Higgs data, precision electroweak observables and direct searches
are less severe for vector-like fermions than for chiral fermions. 

To begin we develop a formalism for vector-like fermion models. In this work we identify properties of vector-like fermions that are consistent with the recently measured Higgs production cross-section and its decay branching ratios at the LHC.  
In particular, we analyze vector-like extensions of the Standard Model (VSM), 
with vector-like quarks and leptons. 
We investigate the precision electroweak constraints from their presence and then the direct collider constraints from LHC searches. We present many numerical results, which will have implications for future vector-like fermion searches. 
And we conclude with a brief discussion on the meaning of the results in the context of some theories of physics beyond the SM, and also discuss what the future may hold in our search for vector-like states.

%

\section{Vector-like-fermion Models}
\label{VLFmodels.SEC}
We add to the SM a vector-like pair of $SU(2)$ doublets and some number of vector-like pair of $SU(2)$ singlets.
We assume all these fields to be charged under $U(1)_Y$. 
Consider the vector-like doublets as two Weyl spinors $\chi_\alpha$ and $\chi^c_\alpha$, 
that transform as conjugates with respect to $SU(2)_L$ and $U(1)_Y$, namely $\chi = (2, Y_\chi)$, $\chi^c = (\bar 2,-Y_\chi)$. 
Although we are considering only one such doublet pair, one can add any number of such pairs and our statements below can be extended to include this case.  
The theory can equivalently be written in terms of a Dirac fermion 
$$\Chi \equiv \bmat \chi_\alpha \\ {\chi^c}^{\dot{\alpha}} \emat \ , $$
where we follow the usual Lorentz index conventions $\alpha$, $\dot{\alpha}$. 
Thus the Dirac spinor $\Chi$ transforms the same way as $\chi$ under $SU(2)_L$ and $U(1)_Y$. 
A vector-like mass term can always be added 
\beq
{\cal L} = - M_\chi \chi \chi^c + h.c. = - M_\chi \bar\Chi \Chi \ .  
\label{Mchi.EQ}
\eeq

Let us consider a vector-like pair of Weyl spinors $\chi = (2,Y_\chi), \chi^c = (\bar 2, -Y_\chi)$. 
Expanding the $SU(2)$ structure we can write
$$\chi = \bmat \chi_1 \\ \chi_2 \emat \ , \quad {\rm or\ equivalently,} \quad \Chi = \bmat \Chi_1 \\ \Chi_2  \emat \ , $$
where $\Chi$, $\Chi_1$ and $\Chi_2$ are Dirac fermions. 

The $W^3$ and $B$ interactions to new vector-like fermions have the structure
\beq
{\cal L} \supset \sum_i
\left[ g W^3_\mu (T^3_{ii}) + g^\prime B_\mu Y^i \right] 
\left[ \bar{\Chi}^i_L \gamma^\mu \Chi^i_L + \bar{\Chi}^i_R \gamma^\mu \Chi^i_R \right] \ ,
\label{2p2barW3B.EQ}
\eeq
with the $W^3$ also coupling to ${\Chi}_R$.  
The vector-like nature is exhibited by the $L,R$ chiralities having the same $T^3$ and $Y$ couplings. 
Furthermore, $SU(2)_L$ invariance requires $Y_L^i = Y_R^i = Y$ for the $SU(2)_L$ component fields.
The $W^1$ interactions to new fermions are given by
\beq
{\cal L} \supset \frac{g}{2} W_\mu^1 (\bar{\Chi}_{1L} \gamma^\mu \Chi_{2L} + \bar{\Chi}_{1R} \gamma^\mu \Chi_{2R}) + h.c. \ .
\label{2p2barW1.EQ}
\eeq

At this level, the $\Chi_1$ and $\Chi_2$ are degenerate owing to the $SU(2)_L$ symmetry. 
By introducing Yukawa couplings to the SM Higgs, this degeneracy can be broken.
In order to write down Yukawa couplings with the SM Higgs and the $\Chi$,
we can introduce two $SU(2)_L$ singlet vector-like fermions $\xi = (1,Y_\chi+1/2) $ and $\Upsilon= (1,Y_\chi-1/2)$, 
written as Dirac fermions.

In this case we can write the Yukawa couplings
\beq
{\cal L}_{\rm Yuk} \supset -\lambda_\xi \bar\Chi\cdot H^* \xi  -\lambda_\Upsilon \bar\Chi H \Upsilon + h.c. \ ,
\label{2p2barYuk.EQ}
\eeq 
where the ``dot'' represents the antisymmetric product. 
EWSB will then mix these new fermions and will split the $\Chi_1$ and $\Chi_2$ masses.
The sign of $\lambda$ is not physical, since in the change $\lambda_{\xi, \Upsilon} \to -\lambda_{\xi, \Upsilon}$, 
the sign can be absorbed away by a redefinition $\xi \to -\xi$ and $\Upsilon \to -\Upsilon$ without affecting anything else. 

The only gauge interactions that the vector-like singlets have is the hypercharge $B_\mu$ interactions
given by 
\beq
{\cal L} \supset \sum_i g^\prime B_\mu \left[ \bar{\xi} \gamma^\mu Y_\xi \, \xi + \bar{\Upsilon} \gamma^\mu Y_\Upsilon \, \Upsilon \right] \ .
\label{BmuUpsXiIntr.EQ}
\eeq

Next we comment on possible mixing terms for the SM hypercharge assignments $Y_\chi = 1/6$ or $Y_\chi = -1/2$.
With only the $\Chi$ added without the $\Upsilon$ and $\xi$, for the SM $Y_\chi$ assignment $Y_\chi = 1/6$, 
we can write the additional terms 
\beq
{\cal L} \supset -M_{q\chi} \bar q \Chi -\lambda_u^\prime \bar\Chi\cdot H^* u_R  -\lambda_d^\prime \bar\Chi H d_R + h.c. \ ,
\eeq
where $q$ is the SM quark doublet, and $u_R, d_R$ are SM SU(2) singlets. 
Alternately, if $Y_\chi = -1/2$, then we can write the additional terms
\beq
{\cal L} \supset -M_{\ell\chi} \bar \ell \Chi  -\lambda_\nu^\prime \bar\Chi\cdot H^* \nu_R  -\lambda_e^\prime \bar\Chi H e_R + h.c. \ ,
\eeq
where $\ell$ is the SM lepton doublet, and $\nu_R, e_R$ are SM SU(2) singlets. 
After EWSB the Yukawa couplings will split the $\Chi_1$ and $\Chi_2$ masses,
and will also mix the new vector-like fermions with either the $u_L$ and $d_L$ in the first case,
or with $\nu_L$ and $e_L$ in the latter case. 
In addition to the $\Chi$, with the $\xi$, $\Upsilon$ also added, 
for $Y_\chi = 1/6$, we can also write the terms 
\beq
{\cal L} \supset -M_{u\xi} \bar u_R \xi - M_{d\Upsilon} \bar d_R \Upsilon -\lambda_{q \xi}^\prime \,  \bar q\cdot H^* \xi  -\lambda_{q \Upsilon}^\prime \, \bar q H \Upsilon + h.c. \ . 
\eeq
Alternately, if $Y_\chi = -1/2$, then we can also write the terms
\beq
{\cal L} \supset -M_{\nu \xi} \bar \nu_R \xi -M_{e\Upsilon} \bar e_R \Upsilon -\lambda_{\ell\xi}^\prime \,  \bar \ell\cdot H^* \xi  -\lambda_{\ell\Upsilon}^\prime  \, \bar \ell H \Upsilon + h.c. \ . 
\label{mixLlep.EQ}
\eeq 
Above, the $M$ and $\lambda^\prime$ are $3\times 3$ Hermitian matrices but the generation indices have been suppressed.
If lepton-number is not a good symmetry, and if $\xi$ is a gauge singlet, 
one can in addition write a Majorana mass term\footnote{We thank Pedro Schwaller
for reminding us of this possibility.} 
\beq
{\cal L} \supset - \bar M_\xi \, \overline{\xi^c} \xi - (\bar M_{\nu\xi} \, \overline{\nu_R^c} \xi  + h.c.) \ ,
\eeq
where $\xi^c$ is the charge conjugated field of the 4-component spinor $\xi$, and similarly for $\nu_R^c$. 

With the mixing $M$ and $\lambda^\prime$ terms added, there is no longer freedom to rotate away the
sign of the $\lambda, \lambda^\prime$, and therefore the sign becomes physical. 
Since in this work, to be safe from flavor constraints, we take the $M_{SM-VL} \ll M_{VL}$, and $\lambda^\prime \ll 1$, 
these physical effects will be suppressed, and we therefore do not investigate it further.  
\footnote{
Refs.~\cite{Dermisek:2013gta,Queiroz:2014zfa} consider models with mixing terms present in the context of the muon $(g-2)$.
Ref.~\cite{Alok-FCNC} obtains the FCNC constraints due to mixing with a down-type singlet vector-like quark.
} 
We will comment further about these possibilities later.

\medskip\noindent\underline{$\Chi^c$ equivalence}: 
Since we are dealing with a vector-like theory, it should not matter 
whether we write the theory in terms of $\Chi$ with hypercharge $Y_\chi$
or in terms of $\Chi^c \equiv -i\gamma^2 \Chi^*$ with hypercharge $-Y_\chi$. 
To show this, let us rewrite the theory in terms of $\Chi^c$. 
To relate conjugated and unconjugated terms, we use the relations
$\bar\Psi^c \Chi^c = \bar\Chi \Psi$, $\bar\Psi^c \gamma^\mu \Chi^c = - \bar\Chi \gamma^\mu \Psi$. 
Using these identities we can write the Yukawa term either as $\lambda_\Upsilon \bar\Chi H \Upsilon$ or as $\lambda_\Upsilon \bar\Upsilon^c H^T \Chi^c$. 
The mass terms, their diagonalization, mass eigenvalues are all identical whether written in terms of $\Chi$ or $\Chi^c$. 
The gauge interactions on the other hand are opposite in sign, which we understand to mean that the conjugated fields
have opposite ``charge''. 

In the theoretical structure just described, we consider in the next few subsections various models. 

\subsection{The $1\bar{1}$ Model}
\label{Mod11.SEC}
For SM-like hypercharge assignments, these vector-like SU(2) singlets can mix with the SM SU(2) singlets 
and will alter Higgs, electroweak, and flavor observables. 
If such mixing is sizable, the constraints from flavor observables can be rather strong, due to which we 
will not consider this possibility here. 
Ref.~\cite{Falkowski:2013jya} discusses the Higgs phenomenology of vector-like leptons with mixing to SM singlets. 

For non-SM-like hypercharge assignments with vector-like pairs of SU(2) singlet fermions added, 
no new (renormalizable) interactions involving the Higgs field can be written down. 
Thus, there is no modification to Higgs observables, to electroweak precision, or to flavor observables.
This will therefore be uninteresting in the present context. 

\subsection{The $2\bar{2}$ Model}
\label{Mod22.SEC}

For SM $Y_\chi$ assignments, as mentioned earlier, Yukawa couplings can be written down 
between the $\chi$ and the SM right-handed singlets.
We do not analyze in much detail the situation when sizable Yukawa couplings with SM SU(2) singlets are allowed 
as constraints on this (for example, coming from shifts in $Zb\bar b$ coupling, FCNC, etc.) are quite severe.   
This is discussed for instance in Ref.~\cite{Dawson:2012di}, where it is argued that after imposing constraints, 
the deviation from the SM case of the Higgs observables are at most a few percent.

For non-SM-like hypercharge assignments, with only a vector-like pair of $SU(2)$-doublets added-in with no singlets, 
no new (renormalizable) interactions
can be written down involving the Higgs field. Such an extension will therefore not lead to any new contributions
to Higgs boson phenomenology that are significant. So this will be uninteresting in the present context, 
and we will move to discussing in the following subsections, models with one or more singlets added.

\subsection{The $2\bar{2}(1\bar{1})_1$ Model}
\label{Mod2211.SEC}

To understand the nature of this model, we consider in turn the model with only one singlet field present at a time in addition to the $\Chi$, 
namely, first with only the $\Upsilon$ present and then with only the $\xi$.

\subsubsection{$\Upsilon$ Model} 
\label{Mod2211Upsilon.SEC}
In addition to the $SU(2)_L$ multiplet $\Chi$, lets consider only one vector-like $SU(2)$ singlet pair in the theory, namely the $\Upsilon$ (without the $\xi$) 
with only the $\lambda_\Upsilon$ term present in Eq.~(\ref{2p2barYuk.EQ}).
This, as we will see, is sufficient to split the $\Chi_1$ and $\Chi_2$ masses.
Since $Y_H = 1/2$ as usual, for $U(1)_Y$ invariance we need 
$$Y_\Upsilon = Y_\chi -1/2 \ .$$  

\begin{figure}
\begin{center}
\includegraphics[width=0.49\textwidth] {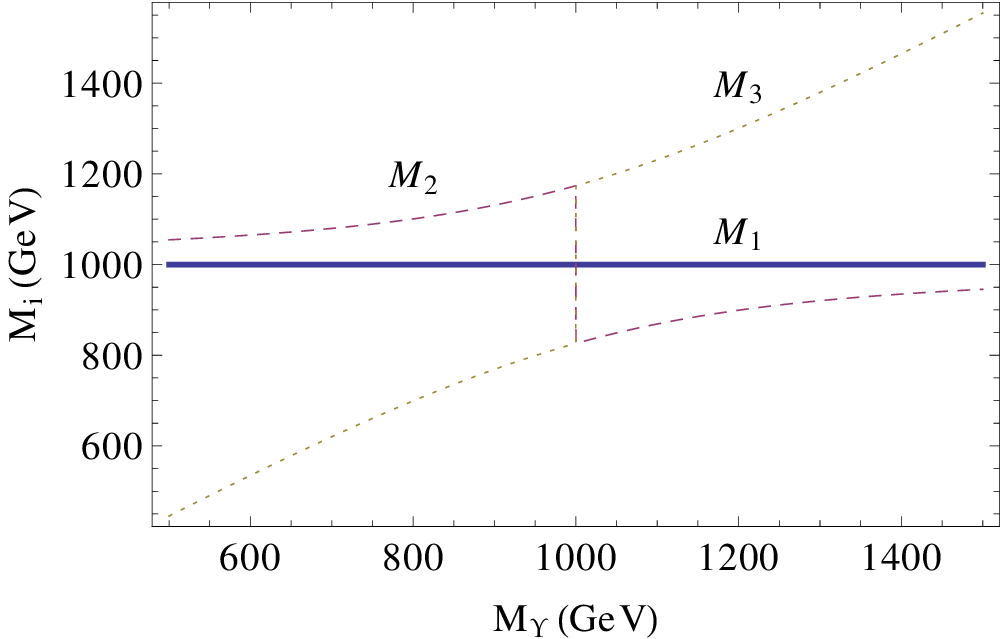}
\includegraphics[width=0.49\textwidth] {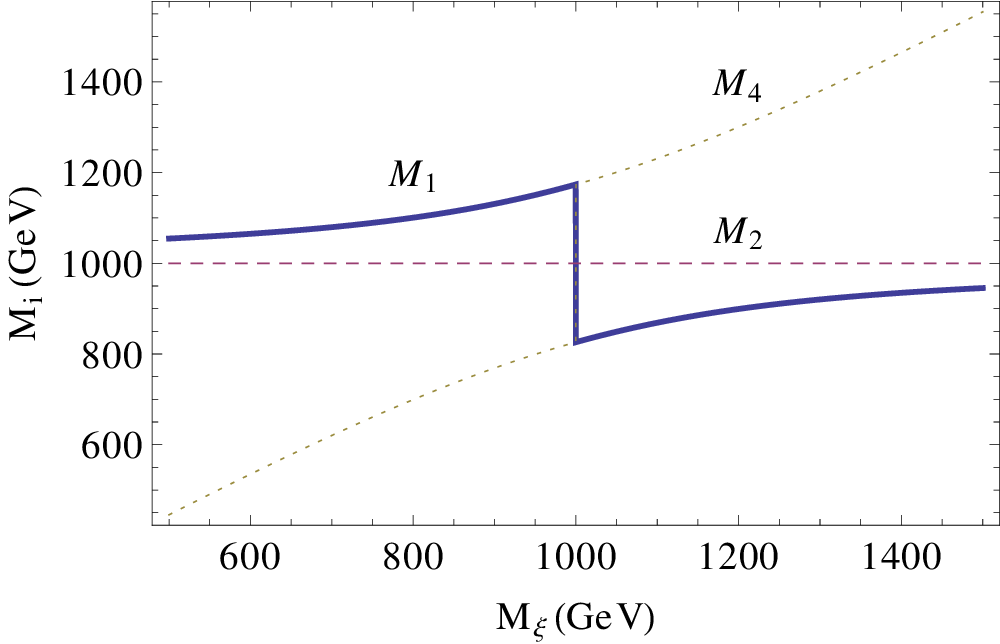}
\caption{The mass eigenvalues $m_i$ as a function of the SU(2) singlet mass taking the SU(2) doublet mass $M_\chi = 1000$~GeV, and $\lambda_{\Upsilon,\xi} = 1$, 
in the $\Upsilon$ model (left) and $\xi$ model (right). 
\label{miXux.FIG}}
\end{center}
\end{figure}

Consistent with all the SM gauge symmetries we can write a vector-like mass term as shown in Eq.~(\ref{Mchi.EQ}), which is
\beq
{\cal L} \supset -M_\chi \bar{\Chi} \Chi = - M_\chi \left( \bar{\Chi}_1 \Chi_1 + \bar{\Chi}_2 \Chi_2 \right) \ .
\eeq 
After EWSB by $\left< H \right> = (0\ v)^T/\sqrt{2}$, we have the mass matrix
\beq
{\cal L}_{\rm mass} \supset - M_{\chi} \bar{\Chi}_1 \Chi_1 - \bmat \bar\Chi_2 & \bar\Upsilon \emat \bmat M_\chi & m \\ m & M_\Upsilon \emat \bmat \Chi_2 \\ \Upsilon \emat \ ,
\label{Lmass2211.EQ}
\eeq 
where $m\equiv \lambda_\Upsilon v/\sqrt{2}$, which we will assume to be real for simplicity. 
The above mass matrix is diagonalized by an $O(2)$ rotation
\beq
\bmat \Chi_2 \\ \Upsilon \emat = \bmat c_V & -s_V \\ s_V & c_V \emat \bmat \Chi_2^\prime \\ \Chi_3^\prime \emat \ ,  
\eeq
where $\{ \Chi_i^\prime \}$ denote the mass eigenstates, and, $c_V \equiv \cos{\theta_V}$, $s_V \equiv \sin{\theta_V}$, with the mixing angle given by
\beq
\tan{2\theta_V} = \frac{2 m}{\left(M_\chi - M_\Upsilon \right)} \ .
\label{tanthV2211.EQ}
\eeq
The mass eigenvalues $M_i$ are
\beq
M_1 = M_\chi \ ; \quad M_{2,3} = \frac{1}{2} \left[ M_\chi + M_\Upsilon \pm \sqrt{(M_\chi - M_\Upsilon)^2 + 4 m^2} \right] \ ,
\label{Mi2211.EQ}
\eeq
which can also be written as
\beq
M_1 = M_\chi \ ; \quad M_2 = M_\chi c_V^2 + M_\Upsilon s_V^2 + 2 m s_V c_V \ ; \quad M_3 = M_\chi s_V^2 + M_\Upsilon c_V^2 - 2 m s_V c_V \ . 
\label{Mi2211-alt.EQ}
\eeq
To have non-negative $M_2$ one requires $M_\chi M_\Upsilon - m^2 \geq 0$. 
In Fig.~\ref{miXux.FIG} we show the mass eigenvalues $M_i$ as a function of $M_\Upsilon$ taking $M_\chi = 1000$~GeV and $\lambda_\Upsilon = 1$.
Henceforth we drop the primes on the mass eigenstate fields for notational ease. 
Taking the EM charges of the mass eigenstates $\Chi_i$ as $Q_i$, we have, $Q_1 = Y_\chi + 1/2$, $Q_2 = Q_3 = Y_\chi - 1/2$. 
We show in Table~\ref{yChiQChi.TAB} the $Q_i$ for various choices of $Y_\chi$ in the $\Upsilon$ model, 
and $Q_4$ should be ignored for this model.  

The $W^1_\mu$, $W^3_\mu$ and $B_\mu$ interaction terms in Eqs.~(\ref{2p2barW3B.EQ}),~(\ref{2p2barW1.EQ})~and~(\ref{BmuUpsXiIntr.EQ}) 
in the fermion mass basis become
\bea
{\cal L} \supset &+&\frac{g}{2} W_\mu^3 \left[ \bar\Chi_1 \gamma^\mu \Chi_1 - \left( c_V^2 \bar\Chi_2 \gamma^\mu \Chi_2 - c_V s_V \bar\Chi_2 \gamma^\mu \Chi_3 - c_V s_V \bar\Chi_3 \gamma^\mu \Chi_2 + s_V^2 \bar\Chi_3 \gamma^\mu \Chi_3 \right)  \right] \nonumber \\
&+& \frac{g}{2} W_\mu^1 \left[ c_V \bar\Chi_2 \gamma^\mu \Chi_1 - s_V \bar\Chi_3 \gamma^\mu \Chi_1 + c_V \bar\Chi_1 \gamma^\mu \Chi_2 - s_V \bar\Chi_1 \gamma^\mu \Chi_3  \right] \nonumber \\
&+& g^\prime B_\mu \left[ Y_\Chi \bar\Chi_1 \gamma^\mu \Chi_1 + (Y_\Chi c_V^2 + Y_\Upsilon s_V^2) \bar\Chi_2 \gamma^\mu \Chi_2 \right. \nonumber \\ 
& &\qquad \left. + (-Y_\Chi + Y_\Upsilon) s_V c_V (\bar\Chi_2 \gamma^\mu \Chi_3 + h.c.) + (Y_\Chi s_V^2 + Y_\Upsilon c_V^2) \bar\Chi_3 \gamma^\mu \Chi_3 \right] \ .
\eea 
Writing this in the $A,Z$ basis we have
\bea
{\cal L} \supset e \left[ \bar\Chi_1  Q_1 \gamma^\mu \Chi_1 +  \bar\Chi_2 Q_2 \gamma^\mu \Chi_2 +  \bar\Chi_3 Q_3 \gamma^\mu \Chi_3 \right] A_\mu  \nonumber \\ 
+  g_Z \left[ \bar\Chi_1 \left( \frac12 - s_W^2 Q_1 \right) \gamma^\mu \Chi_1 
+  \bar\Chi_2 \left( -\frac{c_V^2}{2} - s_W^2 Q_2 \right) \gamma^\mu \Chi_2 \right. \nonumber \\ \left. 
+  \bar\Chi_3 \left( -\frac{s_V^2}{2} - s_W^2 Q_3 \right) \gamma^\mu \Chi_3 
+ \left( \bar\Chi_2  \frac{s_V c_V}{2} \Chi_3  + {\rm h.c.} \right) \right] Z_\mu \ , 
\eea
where $g_Z = g/c_W$, $Q_1 \equiv 1/2 + Y_\Chi$ and $Q_2 = Q_3 = -1/2 + Y_\Chi = Y_\Upsilon$. 

The Higgs interactions are got by replacing $v\to v(1+h/v)$ in Eq.~(\ref{Lmass2211.EQ}).
In the mass basis this gives
\beq
{\cal L} \supset -\frac{\lambda_\Upsilon}{\sqrt{2}} h \left[ s_V c_V \bar\Chi_2 \Chi_2 - s_V c_V \bar\Chi_3 \Chi_3 
+ (c_V^2 - s_V^2) \bar\Chi_3 \Chi_2  \right] + {\rm h.c.} \ . 
\label{hChiChi2211.EQ}
\eeq

We commented below Eqs.~(\ref{2p2barYuk.EQ})~and~(\ref{mixLlep.EQ}) that the sign of the Yukawa couplings can be absorbed away by redefining the fields 
if the Yukawa terms mixing the $\Chi$ and SM fermions are taken negligibly small. 
They are therefore not physical in the limit of the mixing Yukawa couplings not present. 
For later use we note in this model that under $\lambda_\Upsilon \to -\lambda_\Upsilon$, we have 
$m\to -m$, $\theta_V \to -\theta_V$, $s_V \to -s_V$. 
Therefore, off-diagonal (in fermion fields) couplings of the $h$ and $Z_\mu$ change sign (i.e. the $h\Chi_3\Chi_2$ and $Z\Chi_3\Chi_2$ couplings), 
while all the diagonal couplings do not change sign.

\subsubsection{$\xi$ Model} 
\label{Mod2211xi.SEC}
In addition to the $SU(2)_L$ multiplet $\Chi$, lets consider only one vector-like $SU(2)$ singlet pair in the theory, namely the $\xi$ (without the $\Upsilon$) 
with only the $\lambda_\xi$ term present in Eq.~(\ref{2p2barYuk.EQ}).
Since $Y_H = 1/2$ as usual, for $U(1)_Y$ invariance we need 
$$Y_\xi = Y_\chi + 1/2 \ .$$ 
After EWSB by $\left< H \right> = (0\ v)^T/\sqrt{2}$, we have the mass matrix
\beq
{\cal L}_{\rm mass} \supset - M_{\chi} \bar{\Chi}_2 \Chi_2 - \bmat \bar\Chi_1 & \bar\xi \emat \bmat M_\chi & \tilde m \\ \tilde m & M_\xi \emat \bmat \Chi_1 \\ \xi \emat \ ,
\label{Lmass2211xi.EQ}
\eeq 
where $\tilde m\equiv \lambda_\xi v/\sqrt{2}$, which we will assume to be real for simplicity. 
The above mass matrix is diagonalized by an $O(2)$ rotation
\beq
\bmat \Chi_1 \\ \xi \emat = \bmat c_V^\prime & -s_V^\prime \\ s_V^\prime & c_V^\prime \emat \bmat \Chi_1^\prime \\ \Chi_4^\prime \emat \ ,  
\eeq
where $\{ \Chi_i^\prime \}$ denote the mass eigenstates, and, $c_V^\prime \equiv \cos{\theta_V^\prime}$, $s_V^\prime \equiv \sin{\theta_V^\prime}$, 
with the mixing angle given by
\beq
\tan{2\theta_V^\prime} = \frac{2 \tilde m}{\left(M_\chi - M_\xi \right)} \ .
\label{tanthV2211xi.EQ}
\eeq
The mass eigenvalues $M_i$ are
\beq
M_2 = M_\chi \ ; \quad M_{1,4} = \frac{1}{2} \left[ M_\chi + M_\xi \pm \sqrt{(M_\chi - M_\xi)^2 + 4 \tilde m^2} \right] \ ,
\label{Mi2211xi.EQ}
\eeq
which can also be written as
\beq
M_2 = M_\chi \ ; 
\quad M_1 = M_\chi c_V^{\prime 2} + M_\xi s_V^{\prime 2} + 2 \tilde m s_V^\prime c_V^\prime \ ; 
\quad M_4 = M_\chi s_V^{\prime 2} + M_\xi c_V^{\prime 2} - 2 \tilde m s_V^\prime c_V^\prime \ . 
\label{Mi2211xi-alt.EQ}
\eeq
To have non-negative $M_1$ one requires $M_\chi M_\xi - \tilde m^2 \geq 0$. 
Henceforth we drop the primes on the mass eigenstate fields for notational ease. 
Taking the EM charges of the mass eigenstates $\Chi_i$ as $Q_i$, we have, $Q_2 = Y_\chi - 1/2$, $Q_1 = Q_4 = Y_\chi + 1/2$. 
We show in Table~\ref{yChiQChi.TAB} the $Q_i$ for various choices of $Y_\chi$ in the $\xi$ model,
and $Q_3$ should be ignored for this model. 

The $W^1_\mu$, $W^3_\mu$ and $B_\mu$ interaction terms in Eqs.~(\ref{2p2barW3B.EQ}),~(\ref{2p2barW1.EQ})~and~(\ref{BmuUpsXiIntr.EQ}) 
in the fermion mass basis become
\bea
{\cal L} \supset &+&\frac{g}{2} W_\mu^3 \left[  
c_V^{\prime 2} \bar\Chi_1 \gamma^\mu \Chi_1 - c_V^\prime s_V^\prime \bar\Chi_1 \gamma^\mu \Chi_4 - c_V^\prime s_V^\prime \bar\Chi_4 \gamma^\mu \Chi_1 + s_V^{\prime 2} \bar\Chi_4 \gamma^\mu \Chi_4 - \bar\Chi_2 \gamma^\mu \Chi_2    \right] \nonumber \\
&+& \frac{g}{2} W_\mu^1 \left[ c_V^\prime \bar\Chi_2 \gamma^\mu \Chi_1 - s_V^\prime \bar\Chi_2 \gamma^\mu \Chi_4 + h.c.  \right] \nonumber \\
&+& g^\prime B_\mu \left[  (Y_\Chi c_V^{\prime 2} + Y_\xi s_V^{\prime 2}) \bar\Chi_1 \gamma^\mu \Chi_1 
+ (-Y_\Chi + Y_\xi) s_V^\prime c_V^\prime (\bar\Chi_1 \gamma^\mu \Chi_4 + h.c.) 
\right. \nonumber \\ & &\qquad \left. 
+ (Y_\Chi s_V^{\prime 2} + Y_\xi c_V^{\prime 2}) \bar\Chi_4 \gamma^\mu \Chi_4 
+ Y_\Chi \bar\Chi_2 \gamma^\mu \Chi_2
\right] \ .
\eea 
The Higgs interactions are got by replacing $v\to v(1+h/v)$ in Eq.~(\ref{Lmass2211.EQ}).
In the mass basis this gives
\beq
{\cal L} \supset -\frac{\lambda_\xi}{\sqrt{2}} h \left[ s_V^\prime c_V^\prime \bar\Chi_1 \Chi_1 - s_V^\prime c_V^\prime \bar\Chi_4 \Chi_4 
+ (c_V^{\prime 2} - s_V^{\prime 2}) \bar\Chi_4 \Chi_1  \right] + {\rm h.c.} \ . 
\label{hChiChi2211xi.EQ}
\eeq

\subsubsection{Alternate Yukawa coupling}
\label{UpsilonAltYuk.SEC}
Instead of the Yukawa coupling of the $\Upsilon$ shown in Eq.~(\ref{2p2barYuk.EQ}) we can alternately write 
\beq
{\cal L}_{\rm Yuk\, Alt} \supset \hat\lambda_\Upsilon \bar{\tilde\Chi}^c H \Upsilon + h.c. \ ,
\label{2p2barYukAlt.EQ}
\eeq 
where we use the fact that $i\sigma^2 \Chi^*$ transforms the same way as $\Chi$ under $SU(2)$
and have defined $\tilde\Chi^c$ to mean this conjugation in gauge space and also conjugation in spinor space.
In this case we have for $U(1)_Y$ invariance, $\tilde Y_\Upsilon = -Y_\chi - 1/2$. 
While we can write a similar alternate coupling for the $\xi$ also, we do not explicitly write it here.
With the alternate Yukawa coupling Eq.~(\ref{2p2barYukAlt.EQ}) in the $\Upsilon$ model, 
the mass terms are 
\beq
{\cal L}_{\rm mass}^{\rm \Upsilon\, Alt} \supset - M_{\chi} \bar{\Chi}^c_2 \Chi_2^c - \bmat \bar\Chi_1^c & \bar\Upsilon \emat \bmat M_\chi & \hat m \\ \hat m & M_\Upsilon \emat \bmat \Chi_1^c \\ \Upsilon \emat \ ,
\label{Lmass2211Alt.EQ}
\eeq 
where $\hat m \equiv \hat\lambda_\Upsilon v/\sqrt{2}$. 
This mass matrix is identical to that in the theory with the $\xi$ model shown in Eq.~(\ref{Lmass2211xi.EQ}),
but now written for $(\Chi_1^c, \Chi_2^c)$ instead of $(\Chi_1, \Chi_2)$. So the mass eigenvalues are the same as in the $\xi$ model. 
Furthermore, due to the conjugation, all the gauge couplings are with opposite sign to what is in the model with only the $\xi$.

\subsection{The $2\bar{2}(1\bar{1})_2$ Model}
\label{Mod221111.SEC}
Here we consider the addition of both the $SU(2)$ singlet fields $\Upsilon$ and $\xi$, in addition to the $SU(2)_L$ doublet field $\Chi$, 
and turning on both Yukawa couplings in Eq.~(\ref{2p2barYuk.EQ}), namely $\lambda_\Upsilon$ and $\lambda_\xi$.  
For $U(1)_Y$ invariance, we recall that $\Upsilon$ has hypercharge $Y_\chi - 1/2$ and $\xi$ has hypercharge $Y_\chi + 1/2$. 

After EWSB, Eq.~(\ref{2p2barYuk.EQ}) will generate mass mixing terms
\beq
{\cal L}_{\rm mass} \supset 
- \bmat \bar\Chi_1 & \bar\xi \emat \bmat M_\chi & \tilde{m} \\ \tilde{m} & M_\xi \emat \bmat \Chi_1 \\ \xi \emat
- \bmat \bar\Chi_2 & \bar\Upsilon \emat \bmat M_\chi & m \\ m & M_\Upsilon \emat \bmat \Chi_2 \\ \Upsilon \emat \ ,
\label{Lmass221111.EQ}
\eeq 
where $m\equiv \lambda_\Upsilon v/\sqrt{2}$ and $\tilde{m} \equiv \lambda_\xi v/\sqrt{2}$, both of which we will assume to be real for simplicity. 
The above mass matrix is diagonalized by two $O(2)$ rotations
\beq
\bmat \Chi_1 \\ \xi \emat = \bmat c_V^\prime & -s_V^\prime \\ s_V^\prime & c_V^\prime \emat \bmat \Chi_1^\prime \\ \Chi_4^\prime \emat \ ; \quad
\bmat \Chi_2 \\ \Upsilon \emat = \bmat c_V & -s_V \\ s_V & c_V \emat \bmat \Chi_2^\prime \\ \Chi_3^\prime \emat \ ,  
\eeq
where $\{ \Chi_i^\prime \}$ denote the mass eigenstates, and, $c_V \equiv \cos{\theta_V}$, $s_V \equiv \sin{\theta_V}$, with the mixing angle given by
\beq
\tan{2\theta_V^\prime} = \frac{2 \tilde{m}}{\left(M_\chi - M_\xi \right)} \ ; \quad 
\tan{2\theta_V} = \frac{2 m}{\left(M_\chi - M_\Upsilon \right)} \ .
\label{tanthV221111.EQ}
\eeq
The mass eigenvalues $M_i$ are
\bea
M_{1,4} &=& \frac{1}{2} \left[ M_\chi + M_\xi \pm \sqrt{(M_\chi - M_\xi)^2 + 4 \tilde{m}^2} \right] \ ; \quad \nonumber \\
M_{2,3} &=& \frac{1}{2} \left[ M_\chi + M_\Upsilon \pm \sqrt{(M_\chi - M_\Upsilon)^2 + 4 m^2} \right] \ ,
\label{Mi221111.EQ}
\eea
which can also be written as
\bea
M_1 &=& M_\chi c_V^{\prime 2} + M_\xi s_V^{\prime 2} + 2 \tilde{m} s_V^\prime c_V^\prime \ ; \quad M_4 = M_\chi s_V^{\prime 2} + M_\xi c_V^{\prime 2} - 2 \tilde{m} s_V^\prime c_V^\prime \ ; 
\nonumber \\
M_2 &=& M_\chi c_V^2 + M_\Upsilon s_V^2 + 2 m s_V c_V \ ; \quad M_3 = M_\chi s_V^2 + M_\Upsilon c_V^2 - 2 m s_V c_V \ . 
\label{Mi221111-alt.EQ}
\eea
To have non-negative $M_1$ and $M_2$ one requires $M_\chi M_\Upsilon - m^2 \geq 0$ and $M_\chi M_\xi - \tilde{m}^2 \geq 0$ respectively. 
Henceforth we drop the primes on the mass eigenstate fields for notational ease. 
Taking the EM charges of the mass eigenstates $\Chi_i$ as $Q_i$, we have, $Q_1 = Q_4 = Y_\chi + 1/2$, $Q_2 = Q_3 = Y_\chi - 1/2$. 
We show in Table~\ref{yChiQChi.TAB} the $Q_i$ for various choices of $Y_\chi$. 
\begin{table}
\caption{$Q_{i}$ for various choices of $Y_{\chi} = \pm Y_{SM}$. 
$Q_4$ should be ignored in the $2\bar{2}(1\bar{1})_1$ $\Upsilon$ model, 
$Q_3$ should be ignored in $2\bar{2}(1\bar{1})_1$ $\xi$ model, 
and all four states are present in the $2\bar{2}(1\bar{1})_2$ model.
\label{yChiQChi.TAB}}
\begin{centering}
\begin{tabular}{|c||c|c|c|c|}
\hline 
$Y_{\chi}$ & -1/2 & -1/6 & 1/6 & 1/2\tabularnewline
\hline 
\hline 
$Q_{1}, Q_{4}$ & 0 & 1/3 & 2/3 & 1\tabularnewline
\hline 
$Q_{2}$, $Q_{3}$ & -1 & -2/3 & -1/3 & 0\tabularnewline
\hline 
\end{tabular}
\par\end{centering}
\end{table}

The $W^1_\mu$, $W^3_\mu$ and $B_\mu$ interaction terms in Eqs.~(\ref{2p2barW3B.EQ}),~(\ref{2p2barW1.EQ})~and~(\ref{BmuUpsXiIntr.EQ}) 
in the fermion mass basis become
\bea
{\cal L} \supset 
&+&\frac{g}{2} W_\mu^3 \left[ 
 \left( c_V^{\prime 2} \bar\Chi_1 \gamma^\mu \Chi_1 - c_V^\prime s_V^\prime \bar\Chi_1 \gamma^\mu \Chi_4 - c_V^\prime s_V^\prime \bar\Chi_4 \gamma^\mu \Chi_1 + s_V^{\prime 2} \bar\Chi_4 \gamma^\mu \Chi_4 \right) \right. \nonumber \\ 
& &  \left. - \left( c_V^2 \bar\Chi_2 \gamma^\mu \Chi_2 - c_V s_V \bar\Chi_2 \gamma^\mu \Chi_3 - c_V s_V \bar\Chi_3 \gamma^\mu \Chi_2 + s_V^2 \bar\Chi_3 \gamma^\mu \Chi_3 \right)  \right] \\
&+& \frac{g}{2} W_\mu^1 \left[ 
c_V c_V^\prime \bar\Chi_2 \gamma^\mu \Chi_1 - c_V s_V^\prime \bar\Chi_2 \gamma^\mu \Chi_4 - s_V c_V^\prime \bar\Chi_3 \gamma^\mu \Chi_1  +  s_V s_V^\prime \bar\Chi_3 \gamma^\mu \Chi_4 + h.c. \right] \nonumber \\
+ g^\prime B_\mu \left[ 
\right. && \hspace*{-1cm} \left. (Y_\Chi c_V^{\prime 2} + Y_\xi s_V^{\prime 2}) \bar\Chi_1 \gamma^\mu \Chi_1 
+ (-Y_\Chi + Y_\xi) s_V^\prime c_V^\prime (\bar\Chi_1 \gamma^\mu \Chi_4 + h.c.) + (Y_\Chi s_V^{\prime 2} + Y_\xi c_V^{\prime 2}) \bar\Chi_4 \gamma^\mu \Chi_4  \right.  \nonumber \\
& & \hspace*{-2cm} \left. + (Y_\Chi c_V^2 + Y_\Upsilon s_V^2) \bar\Chi_2 \gamma^\mu \Chi_2 
+ (-Y_\Chi + Y_\Upsilon) s_V c_V (\bar\Chi_2 \gamma^\mu \Chi_3 + h.c.) + (Y_\Chi s_V^2 + Y_\Upsilon c_V^2) \bar\Chi_3 \gamma^\mu \Chi_3 \right] \ . \nonumber
\eea 

The Higgs interactions are got by replacing $v\to v(1+h/v)$ in Eq.~(\ref{Lmass2211.EQ}), which in the 
fermion mass basis is
\bea
{\cal L} \supset \left\{ \hspace*{-1cm} \phantom{\frac{d}{d}}
\right. && \left. -\frac{\lambda_\xi}{\sqrt{2}} h \left[ s_V^\prime c_V^\prime \bar\Chi_1 \Chi_1 - s_V^\prime c_V^\prime \bar\Chi_4 \Chi_4 
+ (c_V^{\prime 2} - s_V^{\prime 2}) \bar\Chi_4 \Chi_1  \right] \nonumber \right. \\ \left.
\right. && \left. -\frac{\lambda_\Upsilon}{\sqrt{2}} h \left[ s_V c_V \bar\Chi_2 \Chi_2 - s_V c_V \bar\Chi_3 \Chi_3 
+ (c_V^2 - s_V^2) \bar\Chi_3 \Chi_2  \right] \right\}
+ {\rm h.c.} \ . 
\label{hChiChi221111.EQ}
\eea

\subsection{Some vector-like Extensions of the SM}
We add to the SM some number of vector-like colored (quark) and uncolored (lepton) doublets and singlets corresponding to the 
$2\bar{2}(1\bar{1})_1$ or $2\bar{2}(1\bar{1})_2$ framework outlined in the previous subsection. 
For notational ease, we refer to the quark SU(2) doublet $\Chi_Q$ simply as $Q$, quark SU(2) singlets $\xi_U$ as $U$ and $\Upsilon_D$ as $D$, 
and, the lepton SU(2) doublet $\Chi_L$ as $L$, SU(2) singlets $\Upsilon_E$ as $E$ and $\xi_N$ as $N$.    
From the context it should be clear that we are referring to the new vector-like fermions and not the SM ones. 
These vector-like quarks and leptons have EM charges same as the corresponding SM quarks and leptons only for the specific 
hypercharge $Y_\chi$ assignments of $Y_Q=1/6$ and $Y_L=-1/2$ respectively.  
We give below the nomenclature of the models with the addition of various vector-like SU(2) doublet and singlet quarks and leptons to the SM:
\begin{itemize}
\item Minimal vector-like Quark: 
The $2\bar 2(1\bar 1)_1$ model in the $SU(3)_c$ fundamental representation, with an $SU(2)$ doublet quark $\Chi_Q$ and {\em one} $SU(2)$ singlet quark;  
with only the $\Upsilon_D$ will be the MVQD$_1$, while with only the $\xi_U$ will be the MVQU$_1$. 
$n$ copies of these will be called MVQD$_n$ and MVQU$_n$.
\item vector-like quark (VQ): 
The $2\bar 2(1\bar 1)_2$ model in the $SU(3)_c$ fundamental representation, with an $SU(2)$ doublet quark $\Chi_Q$ and two $SU(2)$ singlet quarks 
$\Upsilon_D$ and $\xi_U$. 
$n$ copies of these will be called VQ$_n$.
\item Minimal vector-like lepton: 
The $2\bar 2(1\bar 1)_1$ model which is an $SU(3)_c$ singlet, with an $SU(2)$ doublet lepton $\Chi_L$ and {\em one} $SU(2)$ singlet lepton;
with only the $\Upsilon_E$ will be the MVLE$_1$, while with only the $\xi_N$ will be the MVLN$_1$. 
$n$ copies of these will be called MVLE$_n$ and MVLN$_n$.
\item vector-like lepton (VL): 
The $2\bar 2(1\bar 1)_2$ model which is an $SU(3)_c$ singlet, with an $SU(2)$ doublet lepton $\Chi_L$ and two $SU(2)$ singlet leptons $\Upsilon_E$ and $\xi_N$. 
$n$ copies of these will be called VL$_n$.
\item The minimal vector-like extension of the SM (MVSM): 
$MVQ\{D,U\} + MVL\{E,N\}$ forms the 1-generation MVSM$_{\{D,U,E,N\}}$. 
$n$ copies of these will form the n-generation MVSM. 
\item The vector-like extension of the SM (VSM): 
$VQ + VL$ forms the 1-generation VSM (VSM$_1$). $n$ copies of these will form the n-generation VSM, called here VSM$_n$. 
\end{itemize}

For some $Y_Q$ and $Y_L$ assignments and in some region of parameter-space it is possible that the lightest vector-like particle is uncolored and EM neutral. 
Such a state, if stable, can be a possible dark matter candidate. 
For example, in the $MVLE$ model with $Y_L = 1/2$ there are two EM neutral uncolored states, and if stable, the lighter of these can be a dark matter candidate.
Alternately, if the lightest vector-like particle is colored and stable, is it possible that a color-singlet bound-state of this (either with another 
vector-like particle or with an SM quark) could be the dark matter?  
A detailed investigation of this possibility is beyond the scope of this work, but such a strongly interacting dark matter (SIMP) candidate appears to be 
heavily disfavored as discussed for example in 
Refs.~\cite{Hemmick:1989ns,Yamagata:1993jq,Starkman:1990nj,Fargion:2005xz,macketal.BIB,Mavromatos:2011kd}.

The addition of vector-like fermions changes the running of the Higgs quartic coupling and the vacuum can become unstable 
at some energy scale $\Lambda$.
We will treat the theory we have written down as an effective theory valid below the scale $\Lambda$.
Refs.~\cite{ArkaniHamed:2012kq,Joglekar:2012vc} derive the vacuum stability constraint with vector-like leptons only,
in a model similar to our $VL_n$. 
Ref.~\cite{Joglekar:2012vc} finds, for example for $\lambda = 1$, that $\Lambda \approx 2.5$~TeV 
for the lightest charged lepton mass of 100~GeV. 
For vector-like lepton masses of $200, 450$~GeV (for the two mass eigenstates), Ref.~\cite{ArkaniHamed:2012kq} 
finds $\Lambda \sim 10$~TeV. 
Since their main motivation is obtaining a large $\mu_{\gamma\gamma} \approx (1.5, 2)$, 
they need large Yukawa couplings for large vector-like masses,
and therefore one would expect stronger constraints in their case than in ours. 
Ref.~\cite{Reece:2012gi} analyzes the vacuum stability issue when new physics contribution changes the sign of the
$ggh$ and $\gamma\gamma h$ effective vertices (while keeping the magnitude the same as in the SM). 
Ref.~\cite{Xiao:2014kba} works out the vacuum stability constraints in the case when the vector-like fermion mass is tied to the 
scale of EWSB. 
Ref.~\cite{Davoudiasl:2012tu} considers the possibility of a strongly first-order electroweak phase transition 
driven by heavy vector-like fermions. 
Working out the precise value of $\Lambda$ in our case will be taken up in future work.  

We focus here on SM-like $SU(3)$, $SU(2)_L$ and $U(1)_Y$ representations, and a more general analysis with 
non-SM-like representations will be postponed to future work.
Our focus in this work will be the modifications of precision electroweak observables, and, LHC Higgs production and decay, 
due to the presence of additional vector-like fermions in SM-like representations. 

\section{Precision Electroweak Observables}
\label{precEW.SEC}
One of the great successes of the SM is the near perfect agreement between the indirect constraints on the 
Higgs mass from precision electroweak observables~\cite{Baak:2011ze,Baak:2012kk,Ciuchini:2013pca} 
and the mass of the observed state at the LHC. 
The precision electroweak constraints have been investigated in the context of a chiral fourth generation 
elsewhere (see for example Ref.~\cite{Dighe:2012dz} and references therein). 
In this section, we present the oblique corrections~\cite{PeskinTakeuchiSTU,AltBarbObq,Maksymyk:1993zm} 
written in terms of the $S,T,U$ parameters~\cite{PeskinTakeuchiSTU} due to vector-like fermions.  
The $S,T,U$ parameters are given by
\bea 
S &=& -16 \pi \Pi_{3Y}^\prime (0) \ , \label{Sdef.EQ} \\
T &=& \frac{4\pi}{s_W^2 m_W^2} \left[ \Pi_{11}(0) - \Pi_{33}(0) \right] \ , \label{Tdef.EQ} \\
U &=& 16 \pi \left[ \Pi_{11}^\prime (0) - \Pi_{33}^\prime (0)  \right]   \label{Udef.EQ} \ .
\eea
Vector-like quark contributions to $S$, $T$ and $U$ have been presented earlier in Ref.~\cite{Lavoura:1992np}. 
Here we derive these including contributions from vector-like quarks and leptons, 
and in the following sections make contact with properties of the recently observed Higgs boson. 

\subsection{$S$, $T$, $U$ Constraints}
The allowed regions in the $S$ and $T$ plane is shown for instance in Refs.~\cite{Beringer:1900zz,Kribs:2007nz}.
We take the reference Higgs mass and top mass values as $m_h=125.5~$GeV~\cite{Aad:2013wqa,CMS-HiggsCombi} 
and $m_t=173.2~$GeV~\cite{Lancaster:2011wr,LHCmtCombi}.  
We include the $\Delta S$, $\Delta T$ and $\Delta U$ contributions due to vector-like fermions and 
ascertain whether they lie within the 68~\% or 95~\% allowed regions.  

The $U$ parameter best-fit value is $0.08 \pm 0.11$~\cite{Beringer:1900zz}. 
Since contributions to $U$ are typically quite small, it does not impose any nontrivial constraints on models. 
It is therefore acceptable to set $U$ to zero and obtain constraints from $S$ and $T$; we adopt this method here. 

\begin{figure}
\begin{center}
\includegraphics[width=0.30\textwidth] {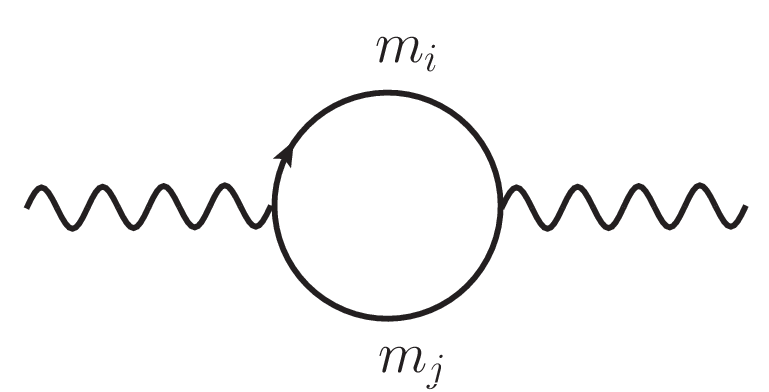} 
\caption{The vector-like fermion contribution to the gauge 2-point function $\Pi^{(m_i m_j)}$. 
\label{VV-mimj.FIG}}
\end{center}
\end{figure}

New vector-like fermions $\Chi^i$ will generate additional contributions to $S$, $T$ and $U$  
due to the new $W^1$, $W^3$ and $B$ interactions going into the gauge 2-point functions $\Pi$ shown in Fig.~\ref{VV-mimj.FIG}.
The detailed computation of the $\Pi$ with vector-like fermions is given in Appendix~\ref{VV2pt-VL.SEC}.
There we define $\Pi_{LL}$ with a chirality projector $P_L$ inserted in the first vertex and $P_L$ in the second vertex, 
and similarly for $\Pi_{RR}$, $\Pi_{RL}$ and $\Pi_{LR}$, 
and, $\Pi \equiv \Pi_{LL} + \Pi_{LR}$, 
and also, $\Pi^{\{m_i,m_j\}} = \Pi^{(m_i m_j)} + \Pi^{(m_j m_i)}$, $\Pi^{(m_i)} = \Pi^{(m_i m_i)}$.
We similarly define $\Pi^\prime$.   
In terms of these $\Pi$s computed in Appendix~\ref{VV2pt-VL.SEC}, we present next the $S$, $T$, and $U$ due to vector-like fermions 
in the $2\bar{2}(1\bar{1})_1$ and $2\bar{2}(1\bar{1})_2$ models of Sections~\ref{Mod2211.SEC}~and~\ref{Mod221111.SEC} respectively. 

\subsection{$S$, $T$, $U$ in the $2\bar{2}(1\bar{1})_1$ model}

We work out in turn the $S$, $T$, $U$, first for the model of Sec.~\ref{Mod2211Upsilon.SEC} with only the $\Upsilon$ present,
and then for the model of Sec.~\ref{Mod2211xi.SEC} with only the $\xi$ present.

\subsubsection{$\Upsilon$ Model}
With only the $\Upsilon$ present,
in the $2\bar{2}(1\bar{1})_1$ vector-like model described in Sec.~\ref{Mod2211Upsilon.SEC},
from the interactions in Eqs.~(\ref{2p2barW3B.EQ}),~(\ref{2p2barW1.EQ})~and~(\ref{BmuUpsXiIntr.EQ})
we compute the gauge boson 2-point functions of Fig.~\ref{VV-mimj.FIG} and then $S$, $T$ and $U$.
We compute $\Pi^\prime_{3Y}(0)$ and then the $S$ parameter from Eq.~(\ref{Sdef.EQ}) as 
\bea
S = -32 \pi d_3 \left[ 
T^3_{11} Y_\Chi {\Pi^\prime}^{(m_1)}(0) + 
T^3_{22} \left( c_V^2 (Y_\Chi c_V^2 + Y_\Upsilon s_V^2) {\Pi^\prime}^{(m_2)}(0) \nonumber \right. \right. \\ \left. \left.
+ s_V^2 (Y_\Chi s_V^2 + Y_\Upsilon c_V^2) {\Pi^\prime}^{(m_3)}(0) - c_V^2 s_V^2 (-Y_\Chi + Y_\Upsilon) {\Pi^\prime}^{\{m_2, m_3\}}(0) \right) \right] \ ,
\label{S2211Intr.EQ}
\eea
where $d_3$ is the dimension of the $SU(3)$ representation of the vector-like fermion (for example, for a fundamental of $SU(3)$, $d_3 = N_c = 3$). 
The divergent $2/\epsilon$ pieces (along with the $-\gamma + \log{4\pi}$ pieces) cancel leaving a finite $S$.
We compute next the $T$-parameter, for which we compute $\Pi_{11}(0)$ and $\Pi_{33}(0)$ as
\bea
\Pi_{11} &=& 2 d_3 (T^1_{12})^2 \left( c_V^2 \Pi^{\{m_1, m_2 \}} + s_V^2 \Pi^{\{m_1, m_3 \}} \right) \ , \label{2211Pi11.EQ} \\
\Pi_{33} &=& 2 d_3 \left\{ (T^3_{11})^2 \Pi^{(m_1)} 
+  (T^3_{22})^2 \left[ c_V^4 \Pi^{(m_2)} + s_V^2 c_V^2 \Pi^{\{m_2,m_3\}} + s_V^4 \Pi^{(m_3)} \right] \right\} \ .
\label{2211Pi33.EQ}
\eea 
$T$ is then given by Eq.~(\ref{Tdef.EQ}). 
Again, the divergent $2/\epsilon$ pieces (along with the $-\gamma + \log{4\pi}$ pieces) cancel leaving a finite $T$.
$U$ is computed using Eq.~(\ref{Udef.EQ}) for which the $\Pi^\prime_{11}(0)$ and $\Pi^\prime_{33}(0)$ are got from 
equations identical to Eqs.~(\ref{2211Pi11.EQ})~and~(\ref{2211Pi33.EQ}) except for replacing $\Pi \to \Pi^\prime$. 

From the $\Chi^c$ equivalence noted in Sec.~\ref{VLFmodels.SEC}, we infer that the vector-like fermion contributions
to $S$ and $T$ parameters is the same whether written in terms of the $\Chi$ or $\Chi^c$.
This is because the 2-point functions shown in Fig.~\ref{VV-mimj.FIG} have two gauge interaction vertices 
although each with opposite sign compared between $\Chi$ and $\Chi^c$ formulations,
and the loop functions have the same value since the mass eigenvalues are exactly the same in the two formulations. 

This is not to say, however, that $Y_\chi$ and $-Y_\chi$ should necessarily give the same $S$, 
and we do indeed find that $S$ can change sign with the sign of $Y_\chi$. 
This is because isospin offers a reference, i.e., in $\Chi$, a given isospin component has hypercharge $Y_\chi$,
while in $\Chi^c$ it has hypercharge $-Y_\chi$. 
Furthermore, the Yukawa interaction terms in Eq.~(\ref{2p2barYuk.EQ}) couples the vector-like sector to a 
chiral SM sector, in particular to the Higgs with a {\rm specific} hypercharge assignment, namely $Y_H = +1/2$.
For the two cases with $Y_\chi$ and $-Y_\chi$, 
the $U(1)_Y$ invariance of this Yukawa coupling implies different $\Upsilon$ (or $\xi$) hypercharges and consequently different electromagnetic charges,  
and observables can be sensitive to this difference, as for instance the $S$-parameter.
 
It is interesting that vector-like fermions can easily give a negative $S$-parameter. 
This is unlike in the case of chiral fermions where one finds that the $S$-parameter is usually positive leading to a tight constraint on
dynamical EWSB BSM sectors~\cite{PeskinTakeuchiSTU}. 
Some examples of theories that give a negative $S$-parameter are also discussed in Refs.~\cite{Dugan:1991ck,Bamert:1994yq}.

\subsubsection{$\xi$ Model}
Let us turn to the $\xi$ model described in Sec.~\ref{Mod2211xi.SEC}. 
For obtaining the gauge 2-point functions, we notice that we can take the results for the model of Sec.~\ref{Mod2211Upsilon.SEC} for the $\Upsilon$ model
and interchange $T^3_{11} \leftrightarrow T^3_{22}$ in the $W^3_\mu$ interactions while keeping $W^1_\mu$ and $B_\mu$ interactions the same. 
This implies that the $S$-parameter for this model will be the same as that obtained in the model of Sec.~\ref{Mod2211Upsilon.SEC} but for 
{\em opposite} sign of $Y_\chi$, i.e., $S_{\rm \xi\, Model}(Y_\chi) = S_{\rm \Upsilon\, Model}(-Y_\chi)$. 
The $T$-parameter is the same in both the models of Sec.~\ref{Mod2211Upsilon.SEC} and Sec.~\ref{Mod2211xi.SEC}.
We have checked that our numerical results are consistent with these expectations.

\subsubsection{$\Upsilon$ Model with alternate Yukawa coupling}
For the $\Upsilon$ model with the alternate Yukawa coupling of Eq.~(\ref{2p2barYukAlt.EQ}), 
we noted in Sec.~\ref{UpsilonAltYuk.SEC} that the mass matrices are identical to that of the $\xi$ model.
Also, the gauge couplings are of opposite sign, but since the vector-like fermion contributions to the gauge 2-point functions have two such couplings, 
this will not affect the $S$ and $T$ parameters.
The last change is the different hypercharge assignment $\tilde Y_\Upsilon = -Y_\chi - 1/2$.     
The $S$ and $T$ parameters for the $\Upsilon$ coupled as in ${\cal L}_{\rm Yuk\, Alt}$ will in general be different when compared to the $\Upsilon$ model 
or $\xi$ model. For $Y_\chi = -1/2$, however, we have $\tilde Y_\Upsilon = Y_\chi$, and we therefore have 
$$S_{\rm \Upsilon\, Model}^{\rm Yuk\, Alt}(Y_\chi = -1/2) = S_{\rm \xi\, Model}^{\rm Yuk}(Y_\chi = -1/2) \ .$$ 

\subsection{$S$, $T$, $U$ in the $2\bar{2}(1\bar{1})_2$ model}
In the $2\bar{2}(1\bar{1})_2$ vector-like model described in Sec.~\ref{Mod221111.SEC},
from the interactions in Eqs.~(\ref{2p2barW3B.EQ}),~(\ref{2p2barW1.EQ})~and~(\ref{BmuUpsXiIntr.EQ})
we can compute the gauge boson 2-point functions of Fig.~\ref{VV-mimj.FIG}. 
We find
\bea
\Pi_{11} \hspace{-0.25cm} &=& \hspace{-0.25cm} 
2 d_3 (T^1_{12})^2 \left[
c_V^2 c_V^{\prime 2} \Pi^{\{M_1,M_2\}} + s_V^2 c_V^{\prime 2} \Pi^{\{M_1,M_3\}} + c_V^2 s_V^{\prime 2} \Pi^{\{M_2,M_4\}} + s_V^2 s_V^{\prime 2} \Pi^{\{M_3,M_4\}} \right] \ , \nonumber \\
\Pi_{33} \hspace{-0.25cm} &=& \hspace{-0.25cm} 2 d_3 \left\{
(T^3_{11})^2 \left[c_V^{\prime 4} \Pi^{(M_1)} + s_V^{\prime 2} c_V^{\prime 2} \Pi^{\{M_1,M_4\}} + s_V^{\prime 4} \Pi^{(M_4)} \right] \nonumber \right. \\ 
&& \left. + (T^3_{22})^2 \left[c_V^4 \Pi^{(M_2)} + s_V^2 c_V^2 \Pi^{\{M_2,M_3\}} + s_V^4 \Pi^{(M_3)}   \right] \right\} \ , \nonumber \\
\Pi^\prime_{3Y} \hspace{-0.25cm} &=& \hspace{-0.25cm} 2 d_3 \left\{
(T^3_{11}) \left[ 
c_V^{\prime 2} (Y_\chi c_V^{\prime 2} + Y_\xi s_V^{\prime 2}) {\Pi^\prime}^{(M_1)}  - c_V^{\prime 2} s_V^{\prime 2} (-Y_\chi + Y_\xi) {\Pi^\prime}^{\{M_1,M_4\}} 
+ s_V^{\prime 2} (Y_\chi s_V^{\prime 2} + Y_\xi c_V^{\prime 2}) {\Pi^\prime}^{(M_4)}
   \right] \nonumber \right. \\
&& \left. \hspace{-0.4cm} + (T^3_{22}) \left[ c_V^{2} (Y_\chi c_V^2 + Y_\Upsilon s_V^2) {\Pi^\prime}^{(M_2)}  - c_V^{2} s_V^2 (-Y_\chi + Y_\Upsilon) {\Pi^\prime}^{\{M_2,M_3\}} 
+ s_V^{2} (Y_\chi s_V^2 + Y_\Upsilon c_V^2) {\Pi^\prime}^{(M_3)}   \right]  \right\}  \nonumber \\
\eea
where the $\Pi$ and $\Pi^\prime$ are as given in Eqs.~(\ref{PiVLMassDep.EQ})~and~(\ref{PiPriVL.EQ}) respectively. 
The $S$, $T$ and $U$ are then given by Eqs.~(\ref{Sdef.EQ}),~(\ref{Tdef.EQ})~and~(\ref{Udef.EQ}) respectively.

\subsection{Shift in the $Zb\bar{b}$ coupling}
For $Y_\chi = 1/6, -1/6$ there are vector-like states with EM charge $-1/3$, and if they also have color, these states can mix with the SM $b$-quark after EWSB.
This mixing will imply a shift in the $Zb\bar{b}$ coupling which is constrained by LEP. 
The measurements $R_b = 0.21629 \pm 0.00066$ and $\Gamma_{hadrons} = 1.7444 \pm 0.0020$~GeV at LEP~\cite{Beringer:1900zz},
imply constraints on the $Zb\bar{b}$ coupling. 
A Yukawa coupling between the SM $b$ and the vector-like $b'$ induces a mixing, which if it's only in the $L$ sector, i.e. between $b_L \leftrightarrow b'_L$ only, 
\footnote{For a concrete example of a theory where this is the case, see Refs.~\cite{Gopalakrishna:2013hua,Gopalakrishna:2011ef}.} 
the $Z b_L \bar{b}_L$ coupling shifts to $-1/2 (c_L^2) + (1/3) s_W^2$ where $c_L \equiv \cos{\theta_L}$ and the mixing angle is given by
$\tan(2\theta_L) = (\lambda/\sqrt{2}) v/M_{b'}$.  
This mixing is constrained by the above data and will imply the limit 
\beq
M_{b'}/\lambda \geq 3509~{\rm GeV\ at\ } 1\, \sigma\ ;\ \  {\rm and\ } M_{b'}/\lambda \geq 2481~{\rm GeV\ at\ }2\, \sigma \ .
\label{ZbbLimit.EQ}
\eeq  
This limit can be evaded by ensuring that the Yukawa couplings between the SM $b$ and $b'$ is not too large, 
as dictated by Eq.~(\ref{ZbbLimit.EQ}). 
We will assume that this is the case in the rest of our analysis. 
This limit will be evaded completely in the model where a custodial symmetry protects the $Zb\bar b$ coupling (see Ref.~\cite{Agashe:2006at}). 
Also, this limit will not be relevant for $Y_\chi$ assignments that do not result in a charge $-1/3$ vector-like state.

\section{Direct Collider Constraints}
\label{ColliderConstraints}

\subsection{Mass and gauge eigenstates}

In order to consider the constraints from colliders, we look at the decays of the vector-like matter. In particular, we are interested in decays via the Higgs and the electroweak gauge bosons. 
We write the following Lagrangian for the mass sector of the model. We denote the VSM quarks as $T_R$ and $B_R$, where these are the singlet states under SU(2) that are referred to in previous sections as ${\xi}_U$ and $\Upsilon_D$. The states denoted as $T_{\bar{R}},~B_{\bar{R}}$ are the left-handed projections of the vector-like singlets. We assume that we are dealing with only one new generation of VSM quarks and leptons, i.e. the VSM$_1$ in order to get an idea of the constraints. For the sake of simplicity, we do not consider the effects of mixing with the doublet under SU(2), previously denoted as ${\Chi}_Q$, as it will not substantially alter the results. Aside from the assumption that off-diagonal couplings are small relative to the diagonal entries, we do not make any a priori assumptions about their sizes relative to each other, so the mixing between T and the 2nd and 3rd generations can in principle be of similar sizes. Motivated by simplicity and data, we do not consider mixing with the first generation, although this could be implemented, given that the off-diagonal couplings are small. We can write the mass terms as 
\beq
\label{MassLagrangian.EQ}
\begin{split}
\Lagr_{mass} &= m_{cc}\bar{c}_L c_R + m_{ct}\bar{c}_L t_R + m_{cT}\bar{c}_L T_R + m_{tc}\bar{t}_L c_R + m_{tt}\bar{t}_L t_R + m_{tT}\bar{t}_L T_R
\\
&+ \mu_c \bar{T}_{\bar{R}} c_R + \mu_t \bar{T}_{\bar{R}} t_R + \mu_T \bar{T}_{\bar{R}} T_R
\\
&+ (c \leftrightarrow s,~~t\leftrightarrow b, ~~T \leftrightarrow B) \ , 
\end{split}
\eeq
where $m_{ij}$ is used to denote a mass obtained due to the Higgs VEV in the Yukawa interactions of 
Sec.~\ref{VLFmodels.SEC}, and $\mu_i$, which appeared in Sections~\ref{VLFmodels.SEC}~and~\ref{precEW.SEC} 
as $M_\xi$, is a vector-like mass. 
Because this produces a non-diagonal mass matrix, in order to perform decay calculations we must diagonalise to the mass-eigenvalue basis. We write the up-type mass matrix as follows:
\beq
M_u=M_u^0 + \delta M_u \equiv
\bmat
m_{cc} & 0 & 0 \\
0 & m_{tt} & 0 \\
0 & 0 & \mu_T 
\emat
+
\bmat
0 & m_{ct} & m_{cT} \\
m_{tc} & 0 & m_{tT} \\
\mu_c & \mu_t & 0
\emat \ ,
\eeq
which will allow us to use a perturbative approach in $m_{offdiag}/m_{diag}$ to find the diagonalising matrices. The resulting left-handed diagonalization matrix,$V^u_L$, is given by
\beq
V^u_L=
\bmat
1 & \frac{m_{ct}m_{tt} + m_{cc}m_{ct} + m_{cT}m_{tT}}{m_{tt}^2-m_{cc}^2} & \frac{\mu_Tm_{cT} + \mu_cm_{cc} + \mu_tm_{ct}}{\mu_T^2-m_{cc}^2} \\
\frac{m_{ct}m_{tt} + m_{cc}m_{ct} + m_{cT}m_{tT}}{m_{cc}^2-m_{tt}^2} & 1 & \frac{\mu_Tm_{tT} + \mu_cm_{ct} + \mu_tm_{tt}}{\mu_T^2-m_{tt}^2} \\
\frac{\mu_Tm_{cT} + \mu_cm_{cc} + \mu_tm_{ct}}{m_{cc}^2-\mu_T^2} & \frac{\mu_Tm_{tT} + \mu_cm_{ct} + \mu_tm_{tt}}{m_{tt}^2-\mu_T^2} & 1
\emat \ .
\eeq
Using the relation ${V_L^u}^{\dagger} M_uM_u^\dagger V^u_L$ in order to determine $V^u_L$.
Similarly, $V^u_R$ is found to be
\beq
V^u_R=
\bmat
1 & \frac{m_{ct}m_{tt} + m_{cc}m_{ct} + \mu_c \mu_t}{m_{tt}^2-m_{cc}^2} & \frac{\mu_T \mu_c + m_{cT}m_{cc} + m_{tT}m_{ct}}{\mu_T^2-m_{cc}^2} \\
\frac{m_{ct}m_{tt} + m_{cc}m_{ct} + \mu_c \mu_t}{m_{cc}^2-m_{tt}^2} & 1 & \frac{m_{ct}m_{cT} + \mu_t\mu_T + m_{tt}m_{tT}}{\mu_T^2-m_{tt}^2} \\
\frac{\mu_T \mu_c + m_{cT}m_{cc} + m_{tT}m_{ct}}{m_{cc}^2-\mu_T^2} & \frac{m_{ct}m_{cT} + \mu_t\mu_T + m_{tt}m_{tT}}{m_{tt}^2-\mu_T^2}  & 1
\emat \ .
\eeq
Given that ${V_L^u}^{\dagger} M_u V^u_R$ diagonalises $M_u$ to $M^{diag}_u$, we can look at the mass terms
\beq
\begin{split}
\Lagr_{mass}&=
\bmat
\bar{c}'_L & \bar{t}'_L & \bar{T}'_{\bar{R}}
\emat
M_{diag}
\bmat
c'_R \\
t'_R \\
T'_R
\emat
\\
&=\bmat
\bar{c}'_L & \bar{t}'_L & \bar{T}'_{\bar{R}}
\emat
{V_L^u}^{\dagger} M_u V^u_R
\bmat
c'_R \\
t'_R \\
T'_R
\emat
\end{split} 
\label{Masses.EQ}
\eeq
such that the mass eigenstates are given by
\beq
\begin{split}
&c'_L = c_L  +\left[\VLtcneg\right] t_L  + \left[\VLTcneg\right] T_{\bar{R}}
\\
&t'_L = \left[ \VLtc \right] c_L + t_L + \left[ \VLTtneg\right] T_{\bar{R}}
\\
&T'_{\bar{R}} = \left[ \VLTc \right] c_L + \left[ \VLTt \right] t_L + T_{\bar{R}}
\end{split} 
\eeq
and
\beq
\begin{split}
&c'_R = c_R  +\left[\VRtcneg\right] t_R  + \left[\VRTcneg\right] T_R
\\
&t'_R = \left[ \VRtc \right] c_R+ t_R + \left[ \VRTtneg\right] T_R
\\
&T'_{R} = \left[ \VRTc \right] c_R + \left[ \VRTt \right] t_R + T_R
\end{split} 
\eeq
The same procedure is followed for the down-type quarks, yielding the same results as above, only with $c \leftrightarrow s,~~t\leftrightarrow b, ~~T \leftrightarrow B$.

\subsubsection*{Higgs sector}
Since the mass matrix we wrote above includes vector-like mass terms that do not arise via interaction with the Higgs, there is a misalignment between the mass matrix and the Yukawa interaction matrix. The Lagrangian for the Yukawa interactions with the Higgs is given by
\beq
\label{YukawaLagrangian.EQ}
\begin{split}
\Lagr_{Higgs} &= -\lambda_{cc}h \bar{c}_L c_R  -\lambda_{ct}h \bar{c}_L  t_R  -\lambda_{cT}h \bar{c}_L  T_R  -\lambda_{tc}h \bar{t}_L  c_R  -\lambda_{tt}h \bar{t}_L  t_R  -\lambda_{tT}h \bar{t}_L  T_R 
\\
&-\lambda_{ss}h \bar{s}_L s_R  -\lambda_{sb}h \bar{s}_L b_R  -\lambda_{sB}h \bar{s}_L B_R  -\lambda_{bs}h \bar{b}_L s_R  -\lambda_{bb}h \bar{b}_L b_R  -\lambda_{bB}h \bar{b}_L B_R 
\end{split} \ ,
\eeq
which can be written in matrix form as
\begin{align}
\Lagr_{Higgs} = - h\bmat
\bar{c}_L & \bar{t}_L & \bar{T}_{\bar{R}}
\emat
\bmat
\lambda_{cc} & \lambda_{ct} & \lambda_{cT} \\
\lambda_{tc} & \lambda_{tt} & \lambda_{tT} \\
0 & 0 & 0
\emat
\bmat
c_R \\
t_R \\
T_R
\emat
\nonumber+ c \leftrightarrow s,~~t\leftrightarrow b, ~~T \leftrightarrow B
\\
= - h \bmat
\bar{c}_L & \bar{t}_L & \bar{T}_{\bar{R}}
\emat
\Lambda_u
\bmat
c_R \\
t_R \\
T_R
\emat
\nonumber+ c \leftrightarrow s,~~t\leftrightarrow b, ~~T \leftrightarrow B \ .
\end{align}
It is clear that applying the same rotation matrices that diagonalised to the mass eigenstate basis will not diagonalise the Yukawa interaction matrix, giving rise to non-zero off-diagonal couplings via the Higgs.

We can write the Lagrangian in terms of the mass eigenstates as 
\begin{align}
\Lagr_{Higgs}
&=-h\bmat
\bar{c}'_L & \bar{t}'_L & \bar{T}'_{\bar{R}}
\emat
\Lambda_u^{diag}
\bmat
c'_R \\
t'_R \\
T'_R
\emat \ ,
\end{align}
with 
\beq
\label{LambdaDiag.EQ}
\Lambda_u^{diag} = {V^u_L}^\dagger \Lambda_u V^u_R = \bmat \alpha_{cc} & \alpha_{ct} & \alpha_{cT} \\ \alpha_{tc} & \alpha_{tt} & \alpha_{tT} \\ \alpha_{Tc} & \alpha_{Tt} & \alpha_{TT} \emat \ , 
\eeq
which is not diagonal. The $\alpha_{ij}$ coefficients are listed in Appendix~\ref{MixCoeff.APP}. They give the size of the mixing through the Higgs between generations in the mass eigenstate basis. To make our expressions more clear, we define quantities $a^{u,d}_{i}$ where $i=1,...,6$ in the Appendix, where $a^{u,d}_{1,2,3}$ correspond to the off-diagonal entries in the up (down)-type left diagonalization matrix, and $a^{u,d}_{4,5,6}$ correspond to the right diagonalization matrix.

\subsubsection*{Electroweak Sector}
Due to the addition of the new family of quarks, we must modify the CKM matrix accordingly. The matrix is defined as 
\beq
V_{CKM, (4\times4)} = (V^u_{L,(4\times4)})^\dagger V^d_{L,(4\times4)}
\eeq
with the full $4\times4$ matrices approximated as
\beq
V^u_{L,(4\times4)} = 
\bmat
1 & 0 & 0 & 0\\
0 & 1 & a^u_1 & a^u_2\\
0 & -a^u_1 & 1 & a^u_3\\
0 & -a^u_2 & -a^u_3 & 1
\emat
,~~V^d_{L,(4\times4)} =
\bmat
1 & 0 & 0 & 0\\
0 & 1 & a^d_1 & a^d_2\\
0 & -a^d_1 & 1 & a^d_3\\
0 & -a^d_2 & -a^d_3 & 1
\emat \ ,
\eeq
where we use the coefficients $a^{u,d}_{i}$ defined in Appendix~\ref{MixCoeff.APP}, and explained above, such that the new CKM matrix is approximately given by
\beq
V_{CKM, (4\times4)} =
\bmat
1 & 0 & 0 & 0\\
0 & 1 + a^u_1 a^d_1 + a^u_2 a^d_2 & -a^u_1 + a^d_1+ a^u_2 a^d_3 & -a^u_2 + a^d_2- a^u_1 a^d_3\\
0 & a^u_1 - a^d_1+ a^u_3 a^d_2 & 1 + a^u_1 a^d_1 + a^u_3 a^d_3 & -a^u_3 + a^d_3 + a^u_1 a^d_2\\
0 & a^u_2 - a^d_2- a^u_3 a^d_2 & a^u_d - a^d_3+ a^u_2 a^d_1 & 1 + a^u_2 a^d_2+ a^u_3 a^d_3
\emat \ .
\eeq 
Because we did not include the first generation earlier, we have approximated $V_{us},~V_{ub},~V_{cd},V_{td} =0$, but doing a full $4\times4$ analysis would generate these properly as required. 

Flavour changing neutral currents (FCNCs) may arise through couplings to the Z boson at tree level. The quark neutral current is given by 
\beq
\begin{split}
j^\mu_{NC} = \frac{g}{\cos\theta_W}\sum_i &\bar{q}_L^i \gamma^\mu \left[ t_3^i - \sin^2\theta_W Q_i\right] q_L^i
\\
&+ \bar{q}_R^i \gamma^\mu \left[ -\sin^2\theta_W Q_i \right] q_R^i \ .
\end{split} 
\eeq
The right handed current does not change with the addition of the $T_R$ state, as it carries the same charge as the other up-type quarks. However, the addition of the $T_{\bar{R}}$ singlet to the left handed states will result in flavour-changing neutral currents, as it does not carry the same isospin as the other left handed quarks. This can be seen by writing down the current in matrix form as
\beq
\begin{split}
\Lagr_{NC} = \frac{g}{\cos\theta_W}
\bmat
\bar{c}_L & \bar{t}_L & \bar{T}_{\bar{R}}
\emat
\gamma^\mu Z_\mu \Bigg\{&
\bmat
\frac{1}{2}-\frac{2}{3}\sin^2\theta_W & 0 & 0
\\
0 & \frac{1}{2}-\frac{2}{3}\sin^2\theta_W & 0
\\
0 & 0 & \frac{1}{2}-\frac{2}{3}\sin^2\theta_W
\emat
\\
&+
\bmat
0 & 0 & 0
\\
0 & 0 & 0
\\
0 & 0 & -\frac{1}{2}
\emat
\Bigg\}
\bmat
c_L
\\
t_L
\\
T_{\bar{R}}
\emat \ .
\end{split} 
\eeq
This allows it to be seen that there will be a FCNC induced due to the appearance of the $-\frac{1}{2}$ in the matrix (henceforth denoted as $\delta g_{Z\bar{q}q}$ added to the diagonal matrix, which accounts for the isospin of the $T_{\bar{R}}$ being $0$).
It is then quite clear that the matrix entries for our FCNC will be given by $\bm{\beta} = g_{Z\bar{q}q} + V_L^\dagger \delta g_{Z\bar{q}q} V_L$, and is found to be:
\beq
\bm{\beta} = 
\bmat
\beta_{11} & \beta_{12} & \beta_{13}
\\
\beta_{21} & \beta_{22} & \beta_{23}
\\
\beta_{31} & \beta_{32} & \beta_{33}
\emat \ ,
\eeq
where the coefficients $\beta_{ij}$ are defined in Appendix~\ref{MixCoeff.APP}, and give the sizes of the modified interactions between the Z boson and the SM and vector-like quark. Thus the interaction Lagrangian can be written as
\beq
\Lagr_{NC} = \frac{g}{\cos\theta_W}
\bmat
\bar{c'}_L & \bar{t'}_L & \bar{T'}_{\bar{R}}
\emat
\gamma^\mu Z_\mu \bm{\beta} 
\bmat
c'_L
\\
t'_L
\\
T'_{\bar{R}}
\emat \ .
\eeq
We have shown here the analysis for the up-type sector, and a similar analysis follows for the down-type sector.

\subsection{Computation of decay widths}

Having introduced these small couplings between the 3rd generation SM quarks and the VL quarks, we now want to check that there is prompt decay via W to b, or H to t, while also ensuring minimal effects on the gluon fusion rate and the diphoton production rate. 
The Lagrangian term governing the decay of the T to a W boson and b quark is given by $\Lagr = \frac{g}{\sqrt{2}}W_\mu^+\ol{T}_{\bar{R}} \gamma^\mu V_{Tb} b_L$. 
This generates the decay partial width
\begin{align}
\Gamma(T\to Wb) =\frac{g^2 |V_{Tb}|^2 M_T^3}{8M_W^2}\frac{1}{8\pi} \left( 1-\frac{M_W^2}{M_T^2}\right)^2\left(1 + \frac{2M_W^2}{M_T^2}\right) \ ,
\end{align}
where we can see that this will depend on $|V_{Tb}|^2$.
We denote the fermion mass eigenvalues as $M_t$ and $M_T$. 
For the decay of the T to a Z and a SM top, the Lagrangian term is $\Lagr = \frac{g}{\cos\theta_W}Z_\mu \ol{T}_{\bar{R}} \gamma^\mu \beta_{32}t_L$.
This generates the decay partial width
\begin{align}
\Gamma(T\to Zt) = &\frac{g^2|\beta_{32}|^2 M_T}{4 M_Z^2 \cos^2\theta_W}\frac{1}{8\pi} \left[(M_T^2-(M_t+M_Z)^2)(M_T^2 - (M_t -M_Z)^2\right]^{1/2} \nonumber
\\
&\cdot \left[ \frac{M_Z^2}{M_T^2}\left(1 + \frac{M_t^2-M_Z^2}{M_T^2}\right) + \left( 1 - \frac{2M_t^2}{M_T^2} + \frac{M_t^4 - M_Z^4}{M_T^4}\right)\right] \ .
\end{align}
We can see that this depend on $\beta_{32}^2$.
For a T decaying to a SM top and a Higgs, the Lagrangian term is $\Lagr = -\alpha_{tT} h \bar{t}_L{T}_R$. 
This leads to the decay partial width
\begin{align}
\label{gammaTHt}
\Gamma(T \to Ht) & = \frac{\alpha_{tT}^2}{32 \pi M_T} \left[ (M_T^2 -(M_t + M_H)^2)(M_T^2 -(M_t -M_H)^2)\right]^{1/2} \left( 1 + \frac{M_t^2 -M_H^2}{M_T^2}\right) \ .
\end{align}
So this depends on $\alpha_{tT}^2$.

\subsection{Experimental Limits from VLQuark searches}
\label{VLQlimits.SEC}
In this section we review the current experimental limits from ATLAS and CMS on searches for vector-like quarks, taking the example of a top-like quark. 
The decays of such a vector-like quark have a dependence on the off-diagonal Yukawa couplings of the $T$ to the $t$, as well as the other off-diagonal couplings (see definition of 
$\alpha_{tT}$ in Appendix~\ref{MixCoeff.APP}). 
The decay of $T$ to $tH$ will provide a bound on ($\lambda_{tT}^2\cdot \mu_T$), as the decay width is $\Gamma(T\to Ht) \sim \alpha_{tT}^2\cdot M_T/(32\pi^2)$, 
and to leading order in $M_{t}/M_{T}$, we have $\alpha_{tT} \sim \lambda_{tT}$ and $M_T \sim \mu_T$, such that $\Gamma(T\to Ht) \sim \lambda_{tT}^2\cdot \mu_T/(32\pi^2)$. 
This approximate expression only applies for $M_T \gg M_t$ (see Eq.~(\ref{gammaTHt})), and $\lambda_{ij} \ll 1$ for $i \neq j$.  
The condition $\lambda_{ij} \ll 1$ comes from the expression for $M_T$ as a function of $\lambda_{ij}$ that can be seen from Eq.~(\ref{EigenMasses.EQ}), recalling that $m_{ij}$ are the masses 
obtained due to the Higgs VEV in the Yukawa interactions of Sec.~\ref{VLFmodels.SEC} such that $m_{ij} = \lambda_{ij} v/ \sqrt{2}$. 

Depending on the decay width, the decay can either be prompt or displaced. 
We define a prompt decay to be one where $c\tau < 100 \mu$m, such that it would be smaller than $c\tau$ for the $D^0$ meson \cite{Beringer:1900zz}, 
which requires $(\lambda_{tT}^2\cdot \mu_T) \gtrsim 1.6 \times 10^{-10}$ GeV. 
Based on 19.6 fb$^{-1}$, CMS studied the properties of a possible t-like VL quark with decays into bW, tZ and tH. 
They find limits of between 687 and 782 GeV \cite{CMS:2013}. The search at ATLAS used 14.3 fb$^{-1}$ of data, with possible t-like decays into bW and tH. 
They propose limits of between 640 and 790 GeV \cite{ATLAS:2013}.

If the VL quark is long-lived such that $c\tau \gtrsim 3$m, it will decay outside the detector, 
which requires $(\lambda_{tT}^2\cdot \mu_T) \lesssim 6 \times 10^{-15}$ GeV. 
Heavy charged long-lived states such as these usually have low, but not zero, energy loss in the detector and may charge oscillate.
Limits on long-lived VL quarks can be derived from the interpretation of ATLAS and CMS searches~\cite{Buchkremer:2012dn}. 
Using CMS results \cite{Chatrchyan:2012sp} one can exclude $m_Q < 800$ GeV for $c\tau > 3 m$, assuming behaviour similar to that of long-lived heavy scalar top quarks. 
A caveat is that these limits assume pair-production of the VL quarks, with kinematics independent of the lifetime of the particle \cite{Buchkremer:2012dn}. 
The MoEDAL collaboration discusses \cite{Acharya:2014nyr} in more detail the limits on production cross-section-dependent limits for various decay lifetimes of long-lived VL quarks.

The current FCNC experimental limits also place an upper bound on the off-diagonal Yukawa couplings. 
In particular, we compute the decay width of $t\rightarrow c X$, where $X=Z,H$ and use that as a means of limiting the mixing between the VL generation and the SM quarks. 
The current limit on FCNC top decays to a charm and Higgs are given by $\sqrt{|\alpha_{tc}|^2 + |\alpha_{ct}|^2} < 0.14$ \cite{CMS:2014qxa}. 
The current limit on top decays into up-type quarks and Z is given by $BR(t\rightarrow Zq) < 0.05$ \cite{Chatrchyan:2013nwa}, with eventual limits expected to reach $BR(t\rightarrow Zq) < 10^{-5}$ \cite{CMSNote13-002}. 
We simplify our analysis by setting all the off-diagonal Yukawa couplings equal to each other, $\lambda_{ij} \equiv \delta$. 
We find that $\delta \lesssim 0.24$ is required to avoid the constraint on $\sqrt{|\alpha_{tc}|^2 + |\alpha_{ct}|^2}$ for $\mu_T \gtrsim 600$ GeV. 
This bound comfortably avoids the current bounds set by the decay of the top into Z and an up-type quark.

\subsection{Experimental Limits from VLLepton searches}
\label{VLLlimits.SEC}
We do not compute limits on vector-like leptons here, so we summarise briefly here the results of Ref.~\cite{Falkowski:2013jya}. In the case of singlet vector-like leptons, the constraints are not very tight, with $M_l \gtrsim 100$ GeV if it decays into an electron or muon, and a lower limit less than 100 GeV if it decays into a tau. In the case of doublet vector-like leptons, the constraints are more strict, with $M_l \gtrsim 450$ GeV if it decays into an electron or muon, and $M_l \gtrsim 300$ GeV if it decays into a tau. 
These lower bounds motivate our choice of lightest mass eigenvalues for the vector-like fermions in Table~\ref{MVLcats.TAB}.

\section{Higgs Production and Decay Processes}
In the SM, the $hgg$ and $h\gamma\gamma$ couplings occur at loop level, the leading order being at one-loop. 
If vector-like fermions couple to the Higgs boson, they could potentially contribute in these loops
and affect these amplitudes. 
Colored vector-like fermions that couple to the Higgs boson contribute to the 
$gg \to h$ loop amplitude, 
and electromagnetically charged ones to the $h\to \gamma\gamma$ loop amplitude. 
Here we present these contributions. 
We use leading order expressions since we always deal with ratios of these quantities to 
corresponding SM ones. These ratios are quite insensitive to higher order corrections; 
for example, Ref.~\cite{Furlan:2011uq} finds that due to vector-like quarks in a composite-Higgs model, the ratios are 
changed by at most a few percent. 
Ref.~\cite{Gori:2013mia} computes the fermion contribution to the $ggh$ effective coupling using the 
low-energy Higgs theorem. They show the typical size of modification of this vertex in a few models of new physics.
They argue that QCD corrections should largely cancel in the ratio $\Gamma_{gg}/\Gamma_{gg}^{SM}$ where the numerator 
includes the contribution of heavy new physics.

The $h\to gg$ partial width at leading order (1-loop) in the SM is given for example in Ref.~\cite{Djouadi:2005gi}.
vector-like fermions that carry color also contribute to the $h\to gg$ decay rate while uncolored ones do not.  
We generalize this result to include the contributions due to vector-like fermions in the fundamental representation of $SU(3)$ (i.e. vector-like quarks), 
denoted as $\chi$, by rescaling with the Higgs coupling shown below in parenthesis $(...)$ as
\beq
\Gamma(h\to gg) = \frac{G_\mu \alpha_s^2 m_h^3}{36 \sqrt{2} \pi^3} \left|\frac{3}{4} \sum_{Q} \left(\frac{\kappa_{hQQ}v}{\sqrt{2} M_Q}\right) A_{1/2}^h(\tau_Q) \right|^2 \ ,
\label{Gmh2gg.EQ}
\eeq
where  $\tau_Q = m_h^2/4M_Q^2$, 
$Q$ now includes the SM and the new vector-like quarks, 
$\kappa_{hQQ}$ are the Higgs couplings to the quarks, 
i.e. the usual Yukawa couplings for the SM quarks and the couplings shown in Eqs.~(\ref{hChiChi2211.EQ}),~(\ref{hChiChi2211xi.EQ})~and~(\ref{hChiChi221111.EQ})
for the new vector-like quarks $\Chi$.
The function $A_{1/2}$ can be found in Ref.~\cite{Djouadi:2005gi}. 
The rescaling is as shown above since the SM $hgg$ and $h\gamma\gamma$ fermion contributions 
are proportional to the fermion Yukawa coupling. 
Also, such a rescaling of the SM (chiral) result is applicable in the vector-like case 
since a vector-like Weyl fermion pair counts as a Dirac fermion, similar to the SM situation, and the $\gamma$ and the $g$ interactions are vector-like
with respect to the corresponding unbroken gauge-group and do not see the distinction between chiral and vector-like fermions. 
For our numerical analysis, we take $\alpha_s$ at the scale $M_Z$. 
The $gg\to h$ production can be obtained from the $h\to gg$ decay rate. 
At leading order this is (see for example, Ref.~\cite{Spira:1995rr})
\beq
\sigma(gg\to h) = \frac{8\pi^2}{m_h^3} \Gamma(h\to gg) \ .
\eeq

The $h\gamma\gamma$ decay rate at leading order (1-loop) in the SM is  
given for example in Ref.~\cite{Djouadi:2005gi}.
All states that couple to the Higgs and have nonzero electromagnetic charge contribute to this process. 
We generalize this result to include the contributions due to vector-like fermions (i.e. vector-like quarks and leptons), 
denoted as $\chi$, by rescaling with the Higgs coupling shown below in parenthesis $(...)$ as
\beq 
\Gamma(h\to \gamma\gamma) = \frac{G_\mu \alpha^2 m_h^3}{128 \sqrt{2} \pi^3} 
\left| \sum_{f} N_c Q_f^2 \left(\frac{\kappa_{hff}v}{\sqrt{2} M_f}\right) A_{1/2}^h(\tau_f) + A_1^h(\tau_W) \right|^2 \ ,
\label{Gmh2AA.EQ}
\eeq
where  $\tau_i = m_h^2/4M_i^2$, 
$f$ now includes the SM and the new vector-like fermions, 
$\kappa_{hff}$ are the Higgs couplings to fermions, 
i.e. the usual Yukawa couplings for the SM fermions and the couplings shown in Eqs.~(\ref{hChiChi2211.EQ}),~(\ref{hChiChi2211xi.EQ})~and~(\ref{hChiChi221111.EQ}) 
for the new vector-like fermions $\Chi$.
The function $A_{1/2}$ and $A_1$ can be found in Ref.~\cite{Djouadi:2005gi}.   
All EM charged fermions, both colored and uncolored, contribute here; $N_c=3$ for quarks as usual, and abusing notation, $N_c=1$ for uncolored fermions (leptons).
It is noted there that in the $h\gamma\gamma$ amplitude, since the photon is real, one takes $\alpha_{EM}(q^2=0)$.

\begin{figure}
\begin{center}
\includegraphics[width=0.6\textwidth] {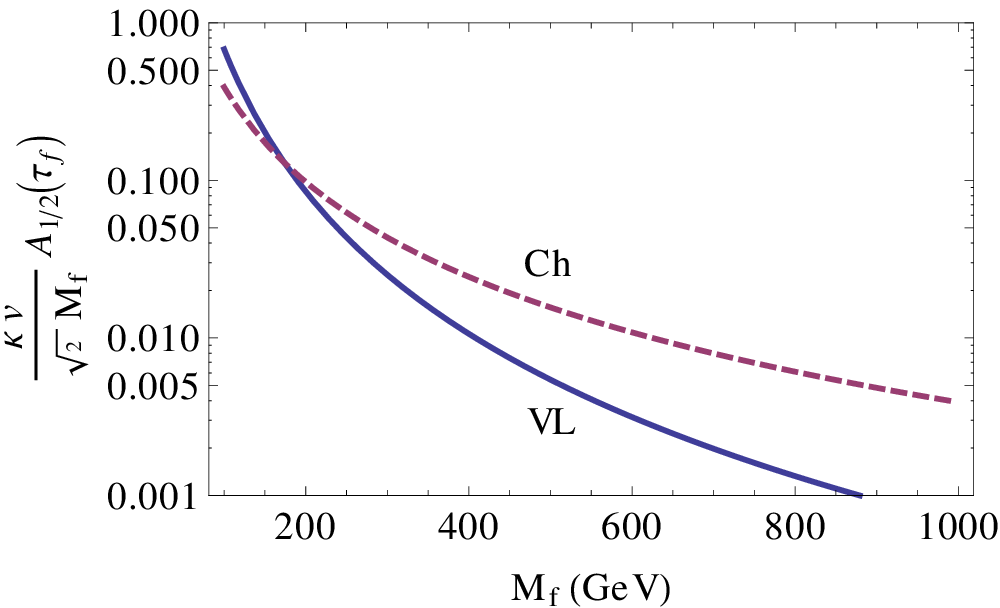}
\caption{
The $hgg$ and $h\gamma\gamma$ triangle loop quantity $\kappa_{hff} v/(\sqrt{2} M_f)~A_{1/2}(\tau_f)$ showing how it decouples as a function of $M_f$, 
contrasting the vector-like case marked ``VL'' taking $\kappa_{hff}=1$ with the chiral case marked ``Ch'' taking $\kappa_{hff} = \sqrt{2} M_f/v$.
\label{KapAhalfMf.FIG}}
\end{center}
\end{figure}

To study the decoupling behavior of the fermion contributions to the $ggh$ and $\gamma\gamma h$ loop amplitudes,
we plot in Fig.~\ref{KapAhalfMf.FIG} the quantity $\kappa_{hff} v/(\sqrt{2} M_f)~A_{1/2}(\tau_f)$ 
that appear in Eqs.~(\ref{Gmh2gg.EQ})~and~(\ref{Gmh2AA.EQ}) contrasting the vector-like case (marked ``VL'') 
with the chiral case (marked ``Ch'').
For the vector-like case, we fix $\kappa_{hff}=1$ and $M_f$ varied, 
while for the chiral case we take $\kappa = \sqrt{2} M_f/v$ since $M_f$ entirely arises from the Yukawa coupling.    
In fact, the existence of a chiral fourth generation increases the $gg\to h$ production rate by a factor of
9 compared to the SM, and after including the contribution to the $h\to \gamma\gamma$ BR, 
a fourth generation appears to be severely disfavored~\cite{Djouadi:2012ae} in a single Higgs doublet model. 

Another Higgs production channel is via vector-boson fusion. Although the signal cross-section is lower than 
Higgs production via gluon fusion, the presence of the forward tagging jets help suppress background, making this mode also promising. 
We will separate the Higgs production in these categories by including $ggh$ and $VBF$ superscripts respectively. 
For brevity, we denote $\Gamma_{h\to XX}$ as $\Gamma_{XX}$, etc. 
 
The ``signal strength'' for the $h\to XX$ decay mode is defined as
\beq
\mu_{XX} \equiv \frac{\left[\sigma(pp\to h) * BR(h \to XX)\right]_{\phantom{SM}}}{\left[\sigma(pp\to h) * BR(h \to XX)\right]_{SM}} \ .
\eeq
Processes that have contributions at the tree-level are modified at loop-level by a relatively small amount, 
and therefore the contributions of heavy vector-like fermions to such processes can be neglected. 
For instance, $\Gamma_{ZZ} \approx \Gamma_{ZZ}^{SM}$, $\sigma^{VBF} \approx \sigma^{VBF}_{SM}$, etc.
Furthermore, with heavy vector-like fermions present, the total Higgs decay width remains dominated by tree-level decays 
as it is in the SM, and to a very good approximation, $\Gamma_{tot} \approx \Gamma_{tot}^{SM}$. 
Thus, in the case of heavy vector-like fermion extensions of the SM, these imply the approximate relations 
\beq
\mu_{\gamma\gamma}^{VBF} \approx \frac{\Gamma_{\gamma\gamma}}{\Gamma^{SM}_{\gamma\gamma}} \ ; \quad
\mu_{ZZ}^{ggh} \approx \frac{\Gamma_{gg}}{\Gamma^{SM}_{gg}} \ ; \quad
\mu_{\gamma\gamma}^{ggh} \approx \frac{\Gamma_{gg}}{\Gamma^{SM}_{gg}} \frac{\Gamma_{\gamma\gamma}}{\Gamma^{SM}_{\gamma\gamma}}\ ; \quad
\frac{\mu_{\gamma\gamma}^{ggh}}{\mu_{ZZ}^{ggh}} \approx \frac{\Gamma_{\gamma\gamma}}{\Gamma^{SM}_{\gamma\gamma}} \approx \mu_{\gamma\gamma}^{VBF} \ .
\label{muRels.EQ}
\eeq

\section{Numerical Results for Electroweak and Higgs Boson Observables}

Here we explore the possible impact of vector-like fermions on precision electroweak observables and LHC Higgs observables. 
We begin with some general comments.
Whenever $\mu_{ZZ}$ deviates from 1, it is entirely due to the vector-like quark contributions to the $ggh$ vertex
as the $hZZ$ vertex is not shifted from its tree-level value. 
Thus when there are no vector-like quarks present $\mu_{ZZ} = 1$, which for instance is the case with only vector-like leptons present. 
In the latter case, $\mu_{\gamma\gamma}$ may be shifted. 
$\mu_{ZZ}$, $\mu_{WW}$ and $\mu_{bb}$ are all equal to each other since all of them measure the $ggh$ vertex.  
In the following subsections we present some illustrative examples and show the deviations in $\mu_{\gamma\gamma}$ and $\mu_{ZZ}$
we find in the various models presented. 

In all the models we consider, the vector-like quark contribution adds constructively with the SM top contribution
and increases the magnitude of the $ggh$ vertex when compared to the SM; 
as a consequence, $\mu_{ZZ,WW,bb} \geq 1$ always.  
The $h\gamma\gamma$ coupling receives a contribution from the $W^\pm$ as well which is larger 
and of opposite sign to the SM quark/lepton contribution. 
The vector-like fermion contribution being of the same sign as the SM quark/lepton contribution therefore 
decreases the magnitude of the $h\gamma\gamma$ coupling when compared to the SM; 
as a consequence, $\mu_{\gamma\gamma}^{VBF} \leq 1$ always. 
$\mu_{\gamma\gamma}^{ggh}$ being the product of the above two (see Eq.~(\ref{muRels.EQ})) can be either bigger than
or lesser than 1. 

We argued below Eqs.~(\ref{2p2barYuk.EQ})~and~(\ref{mixLlep.EQ}) that the sign of the Yukawa couplings are not physical in the
limit of negligibly small Yukawa couplings mixing SM fermions with vector-like fermions, as we take here. 
We demonstrate this here explicitly by taking the $\Upsilon$ model as an example. 
For this, we use the argument for how various quantities change under $\lambda_\Upsilon \to -\lambda_\Upsilon$ we gave below Eq.~(\ref{hChiChi2211.EQ}). 
It is easy to see that in the vector-like fermion triangle-loop diagrams for $h\gamma\gamma$ and $hgg$ couplings,
there are either zero or two couplings that change sign in a particular diagram, thus making the overall sign unchanged. 
Also, the precision electroweak observables, $S,T,U$, given in terms of $\Pi_{11},\Pi_{33},\Pi_{3Y}$ (and their derivatives with respect to $p^2$) 
all depend on $s_V^2$, and thus remain unchanged under $\lambda_\Upsilon \to -\lambda_\Upsilon$.
Although we explicitly demonstrate here for the $\Upsilon$ model, 
this is true for all of the models we have discussed in Sec.~\ref{VLFmodels.SEC}
due to the general arguments given below Eqs.~(\ref{2p2barYuk.EQ})~and~(\ref{mixLlep.EQ}).  
We therefore restrict ourselves to positive Yukawa couplings in the following, without loss of generality.

\subsection{MVLE$_1$ vector-like Leptons Model}
\begin{figure}
\begin{center}
\includegraphics[width=0.6\textwidth]{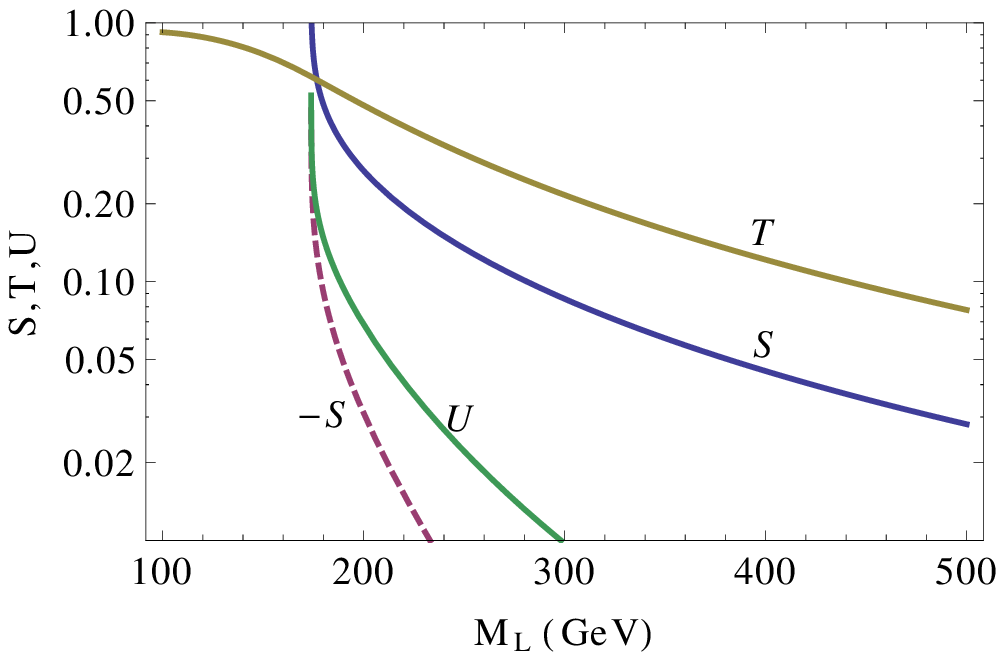}
\caption{
$S$, $T$, $U$ for MVLE$_1$ as a function of $M_L$ with $\lambda_E=1$, $M_E=M_L$, for $Y_L=-1/2$ (solid) 
and $Y_L=1/2$ (dashed, for which $-S$ is shown). 
\label{ST2211ML.FIG}}
\end{center}
\end{figure}

In Fig.~\ref{ST2211ML.FIG} we show $S$, $T$, $U$ as a function of $M_L$ for MVLE$_1$ with $M_E = M_L$ and $\lambda_E = 1$. 
Note that $T$ and $U$ are independent of $Y_L$.
Furthermore, $U \lesssim 0.02$ for $M_{L,E} > 250~$GeV, and will not place any nontrivial constraints; we therefore ignore $U$ in the rest of our analysis.   

For $Y_L=1/2$ ($-1/2$), the $\chi_2,\chi_3$ ($\chi_1$) mass eigenstates are EM neutral (and color singlet) as can be seen from Table~\ref{yChiQChi.TAB}. 
These assignments of $Y_L$ are therefore interesting for dark matter if the lightest $\chi$ is stable
and in regions of parameter space where it is EM neutral.  
In the $Y_L=1/2$ ($-1/2$) $\Upsilon$ ($\xi$) model case, 
the EM charged state $\chi_1$ ($\chi_2$) does not couple to the Higgs as seen from 
Eq.~(\ref{hChiChi2211.EQ}),~(\ref{hChiChi2211xi.EQ}) 
and thus $\mu_{\gamma\gamma}$ will be 1. 
Since there are no new colored fermions, the $gg\to h$ production is unchanged, and $\mu_{ZZ,WW,bb}$ etc. will also be 1. 

\begin{figure}
\begin{center}
\includegraphics[width=0.4\textwidth]{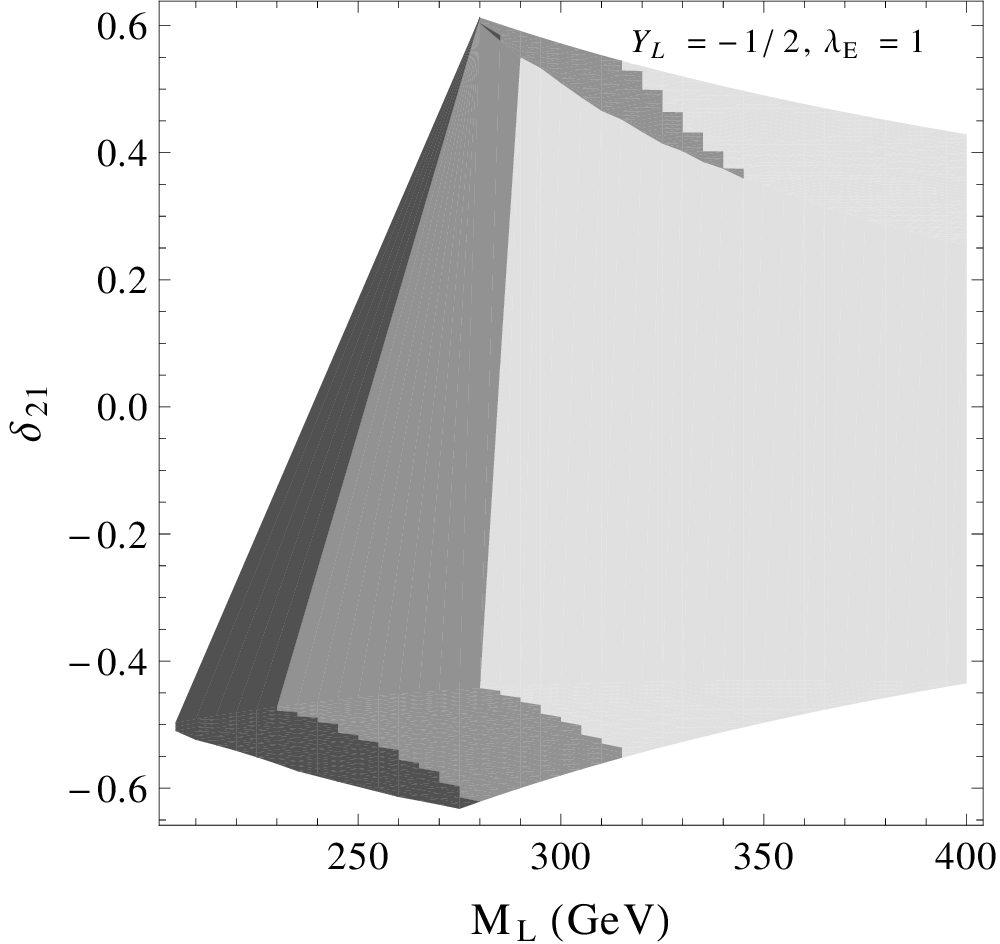}
\vspace*{0.5cm}
\includegraphics[width=0.4\textwidth]{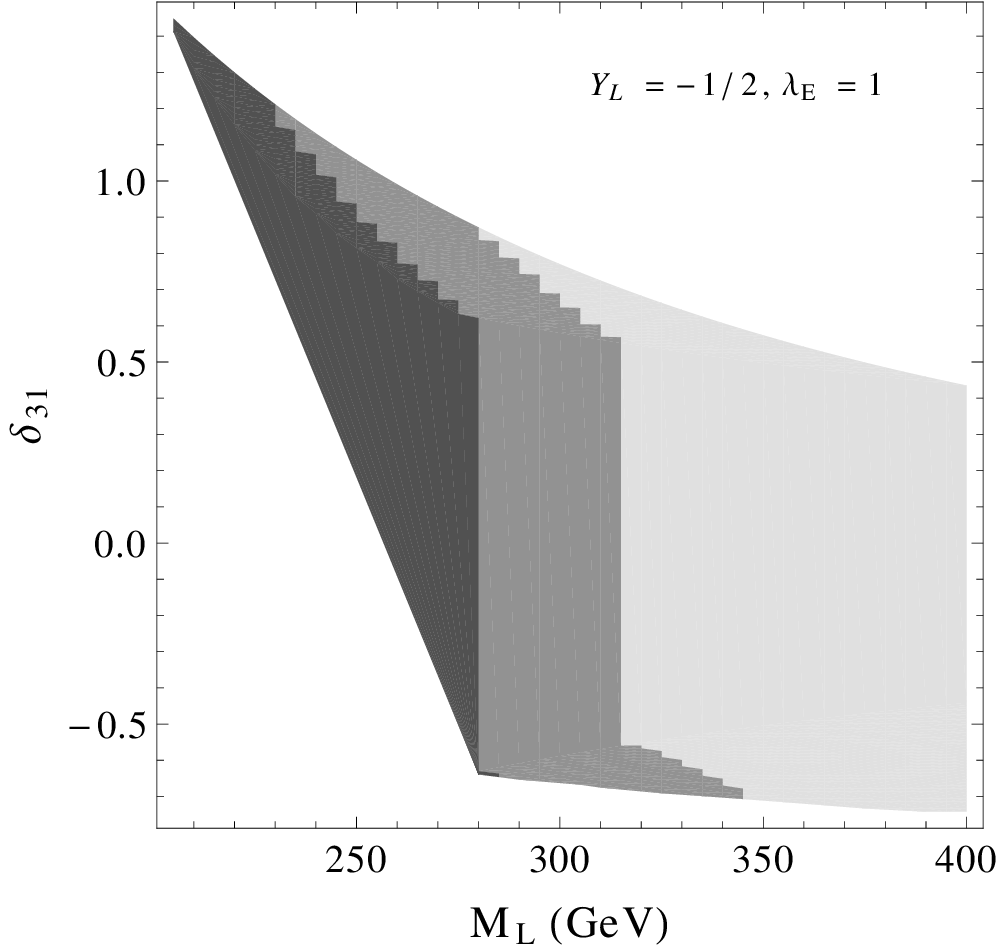}
\includegraphics[width=0.4\textwidth]{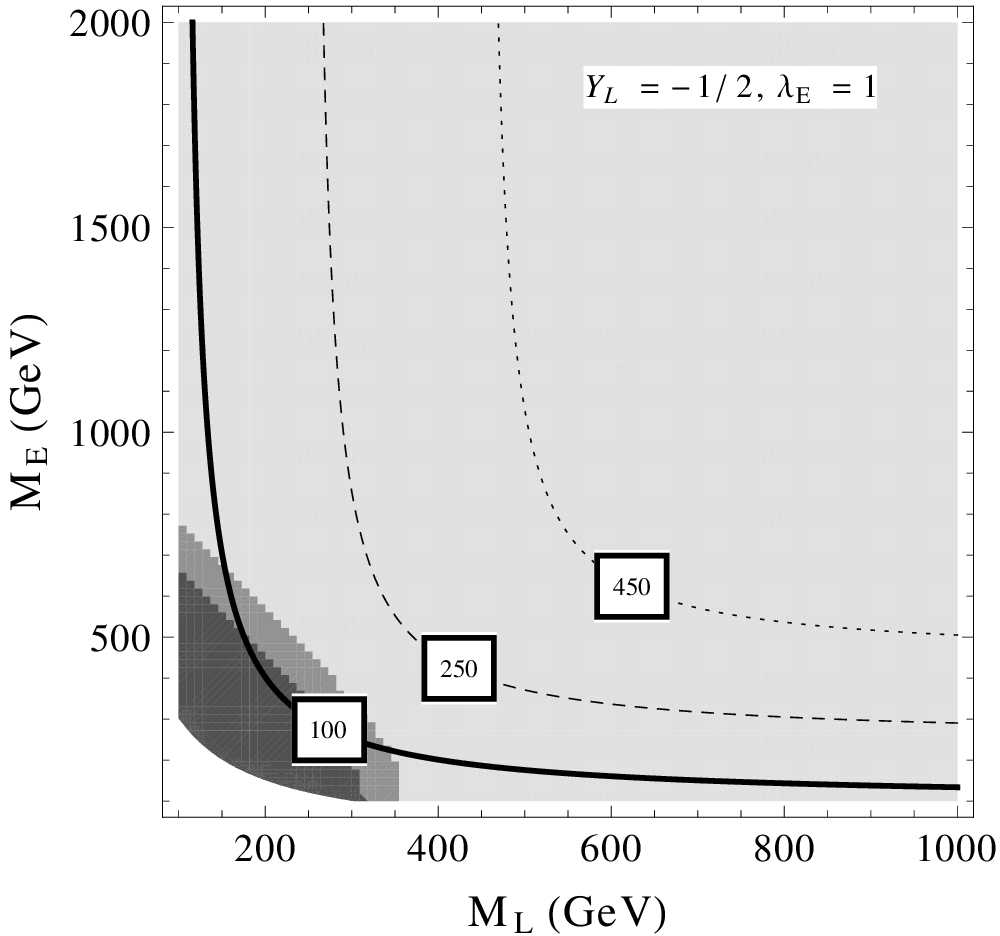}
\vspace*{0.5cm}
\includegraphics[width=0.4\textwidth]{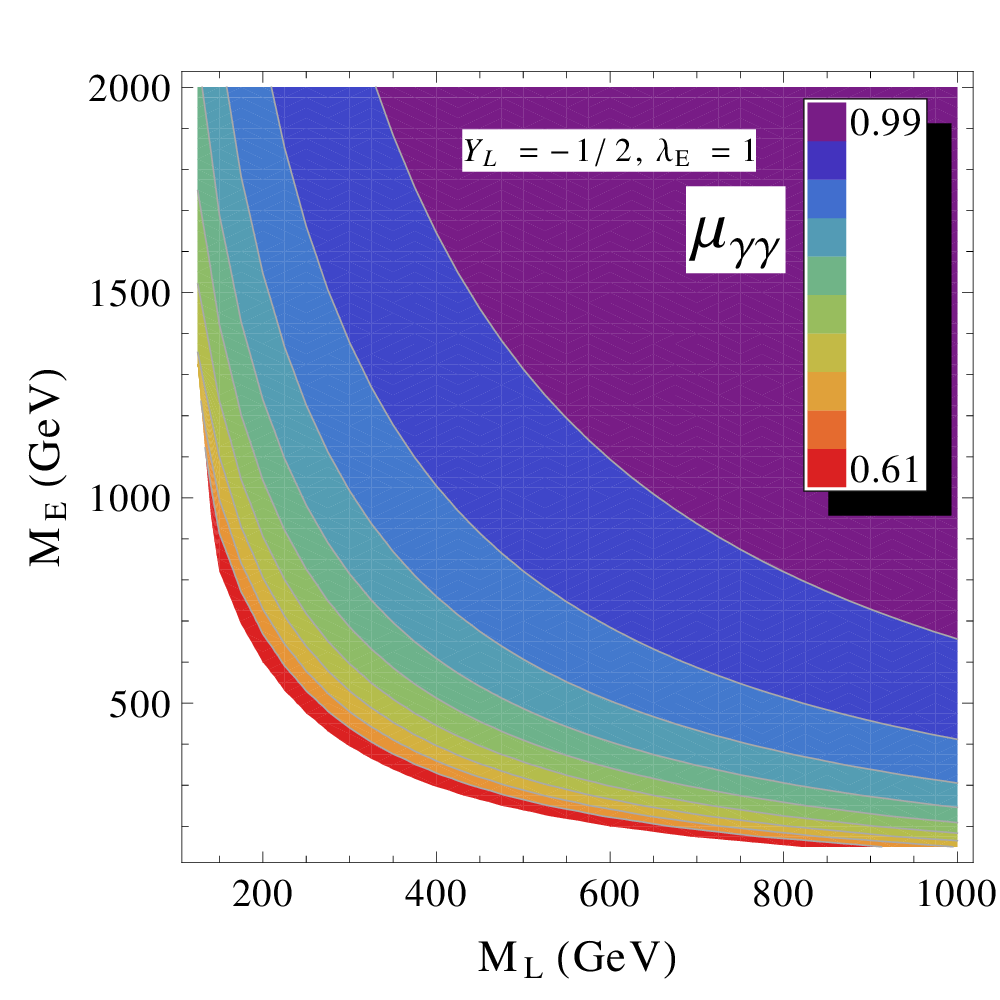}
\includegraphics[width=0.4\textwidth]{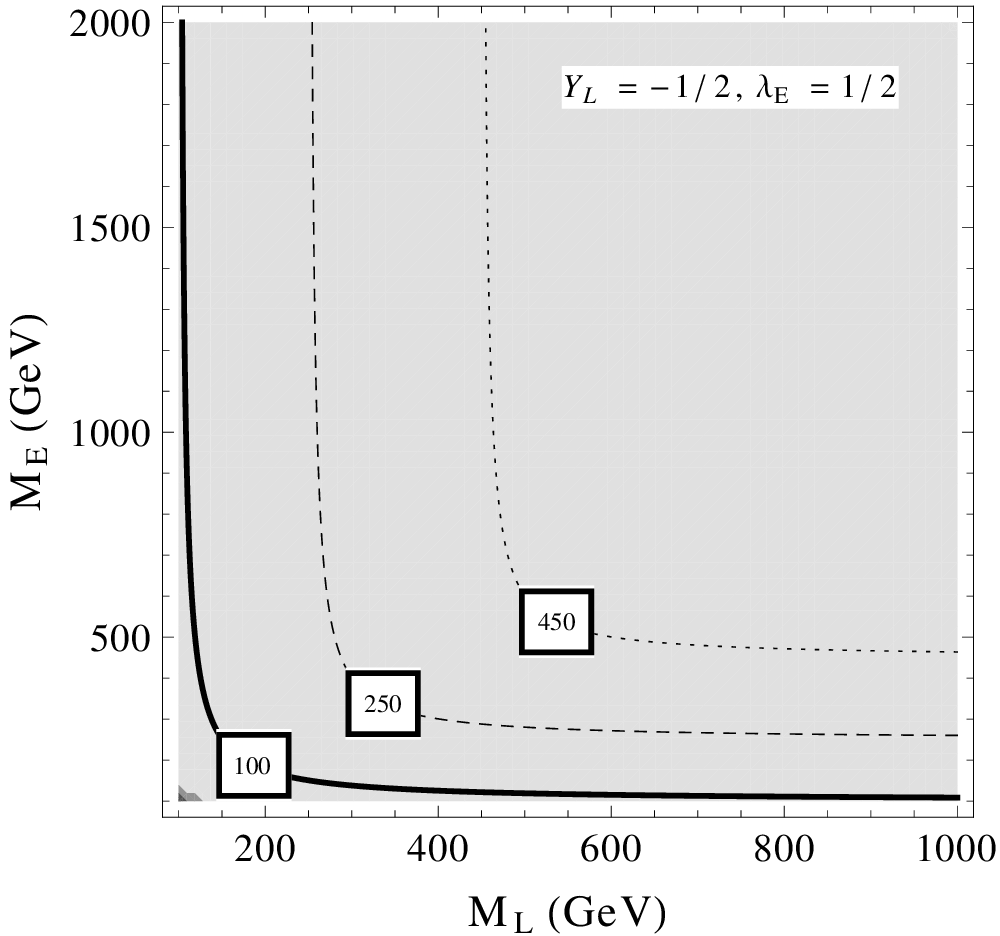}
\includegraphics[width=0.4\textwidth]{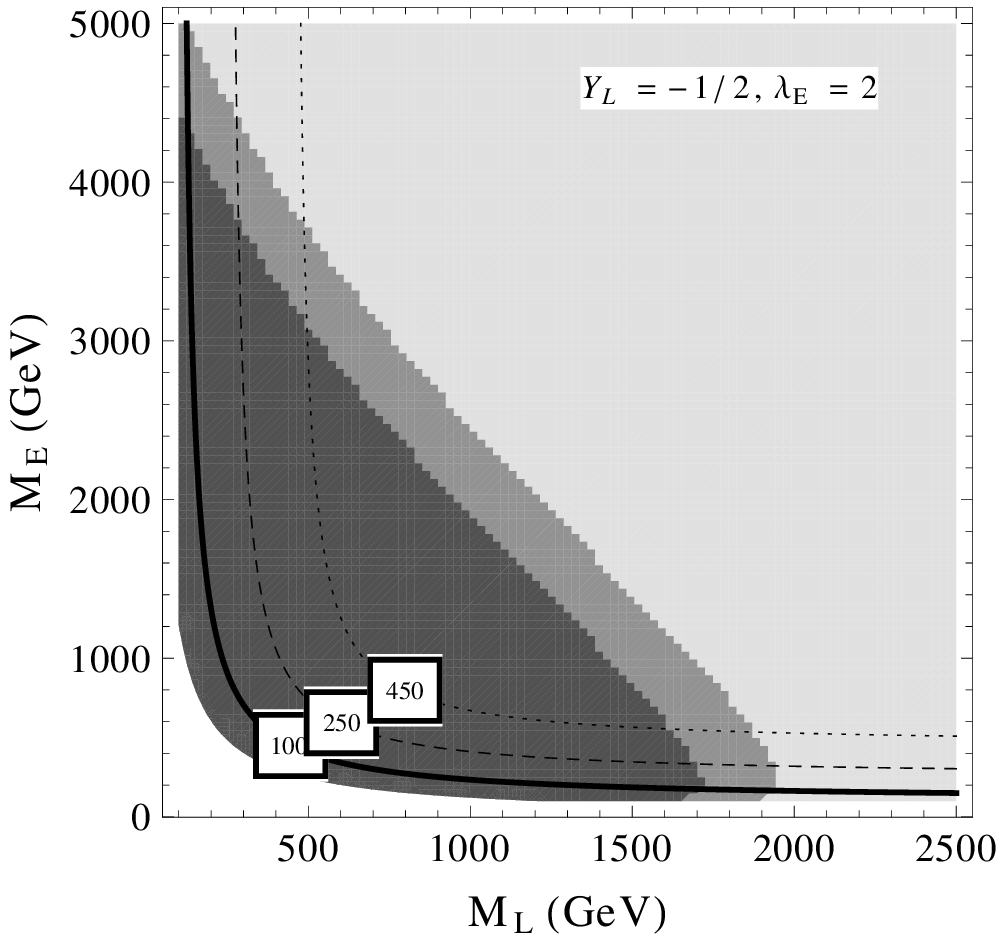}
\caption{
For the MVLE$_1$ model with $Y_L = -1/2$, $\lambda_E = 0.5$ (first four figures) and 
$\lambda_E = 1$ (bottom-left), $\lambda_E = 2$ (bottom-right), 
$S$ and $T$ allowed regions are shown in all the plots, except middle-right which shows the $\mu_{\gamma\gamma}^{ggh}$ 
(in which the values increase from bottom-left corner to top-right). 
$S$ and $T$ are within the 68~\% CL ellipse in the light gray region, between 68~\% and 95~\% ellipse in the medium gray region,
and excluded worse than 95~\% CL in the dark gray region. 
The boxed numbers label (in GeV) contours of the minimum mass eigenvalue, i.e. $min(m_1,m_2,m_3)$. 
\label{MVLE1-ST-mu.FIG}}
\end{center}
\end{figure}

We show in Fig.~\ref{MVLE1-ST-mu.FIG} the constraint from $S$ and $T$ parameters and $\mu_{\gamma\gamma}^{ggh}$.
$S$ and $T$ are within the 68~\% CL ellipse in the light gray region, between 68~\% and 95~\% ellipse in the medium gray region,
and excluded worse than 95~\% CL in the dark gray region. 
We also show the $(100,250,450)$~GeV contours (with boxed numbers) of the the minimum mass eigenvalue, i.e. $min(m_1,m_2,m_3)$.  
It is possible that in some specially constructed models such light masses may still be allowed; 
for a discussion of the limits on vector-like leptons, see for example Ref.~\cite{Falkowski:2013jya}. 
To the left of the black solid line there is a mass eigenvalue $<250$~GeV. 
Therefore the safest region in most models from both $S$ and $T$ parameters and from direct collider constraints 
is the light gray region to the right of all black lines. 
With only vector-like leptons added, the $hgg$ coupling is unaltered and the deviations of interest are all in the 
$h\gamma\gamma$ coupling. 
Thus, $\mu_{\gamma\gamma}^{ggh}$ is altered as shown, whereas $\mu_{ZZ,bb}^{ggh} = 1$ and $\Gamma_{gg} = \Gamma_{gg}^{SM}$ and are not shown.
%

\subsection{MVQD$_1$ vector-like Quarks Model}
\begin{figure}
\begin{center}
\includegraphics[width=0.6\textwidth]{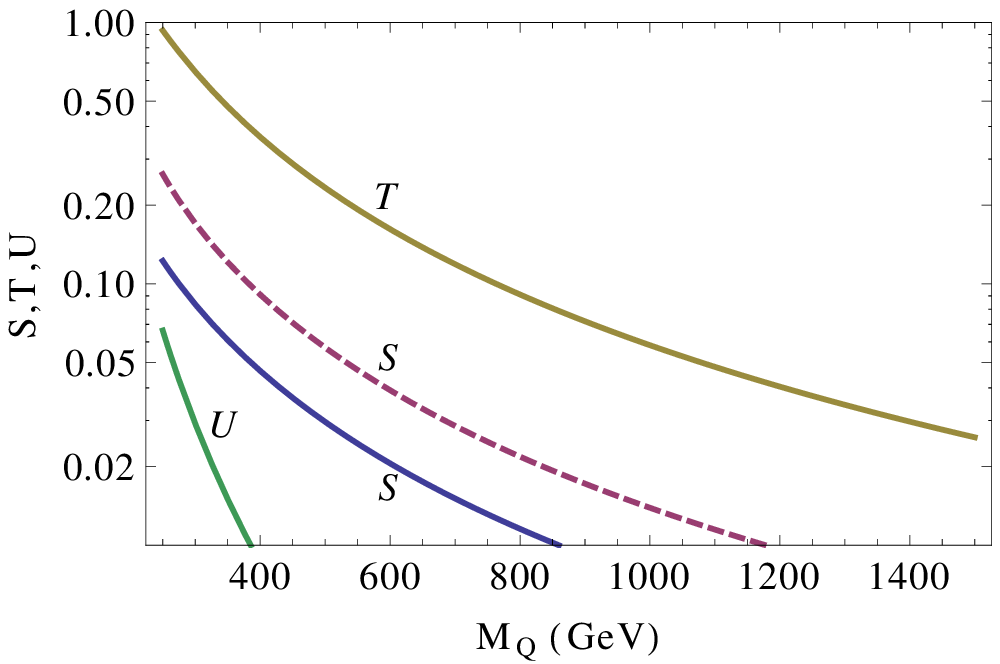}
\caption{
$S$, $T$, $U$ for MVQD$_1$ with $\lambda_D=1$ and $M_D=M_Q$ for $Y_Q = 1/6$ (solid) and $Y_Q = -1/6$ (dashed). 
\label{STU2211MQ.FIG}}
\end{center}
\end{figure}

\begin{figure}[!ht]
\begin{center}
\includegraphics[width=0.39\textwidth]{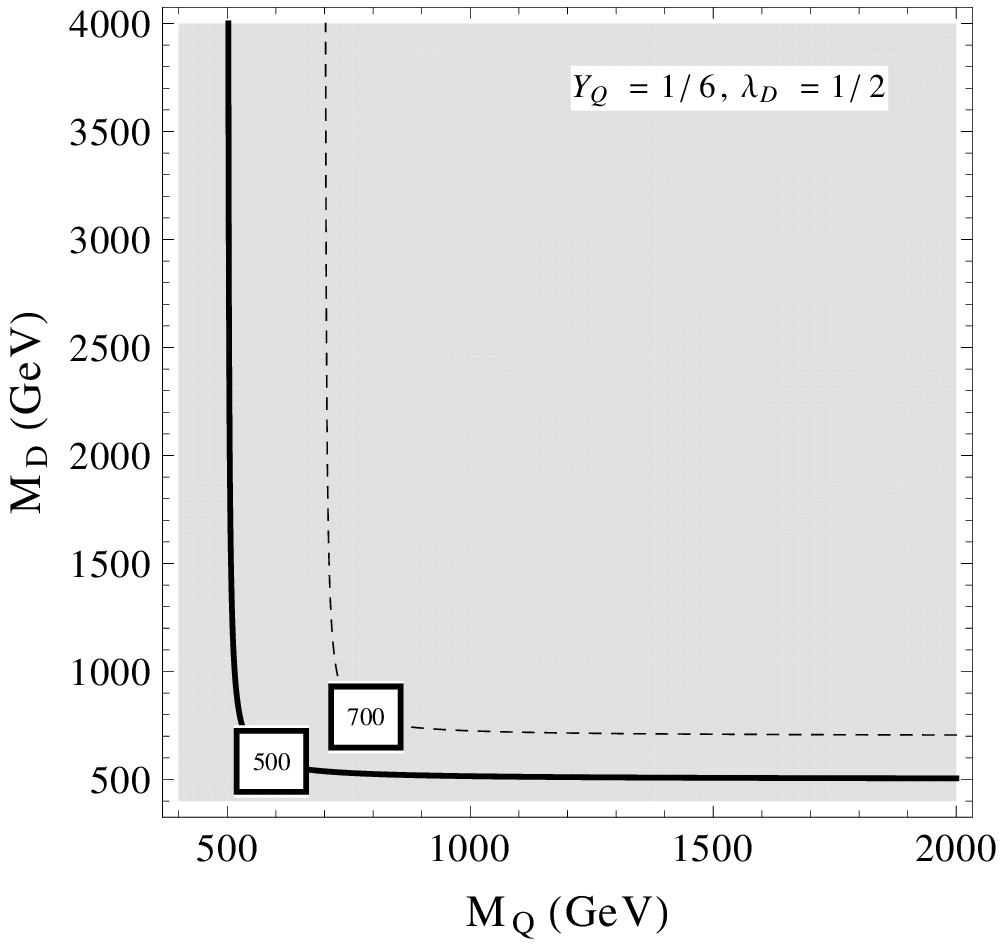}
\vspace*{0.5cm}
\includegraphics[width=0.39\textwidth]{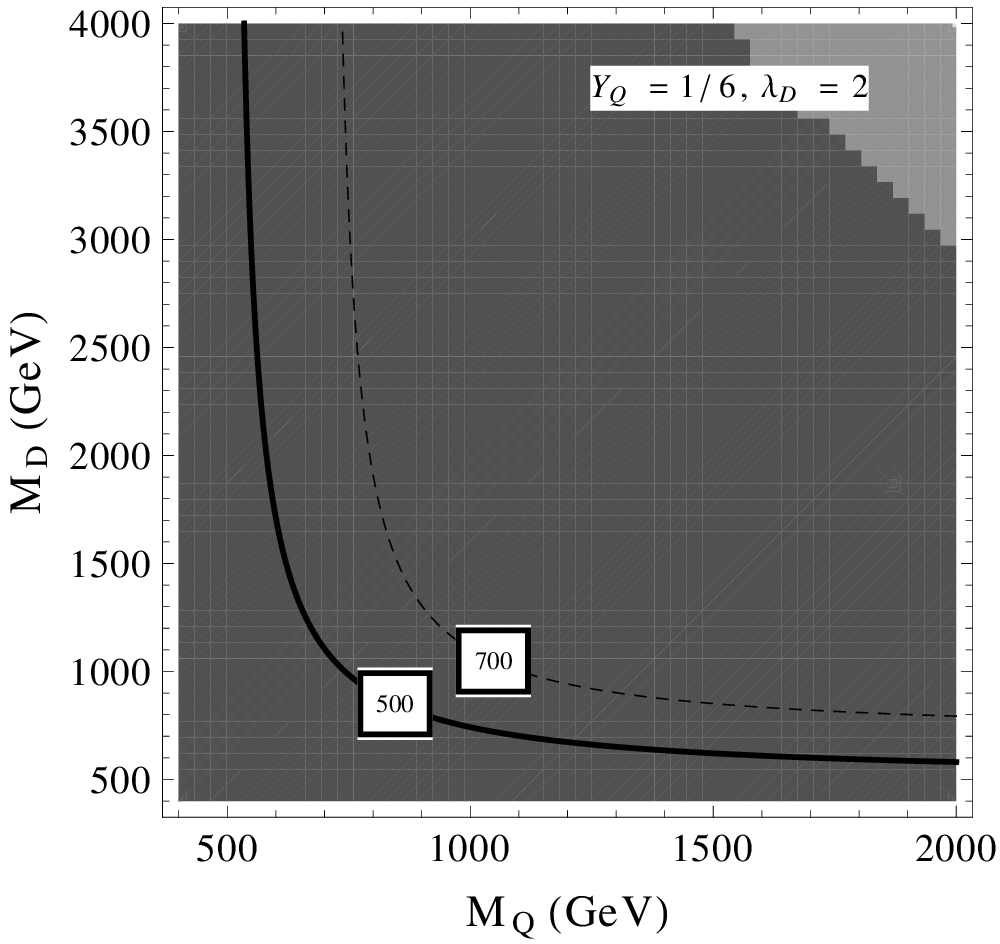}
\includegraphics[width=0.39\textwidth]{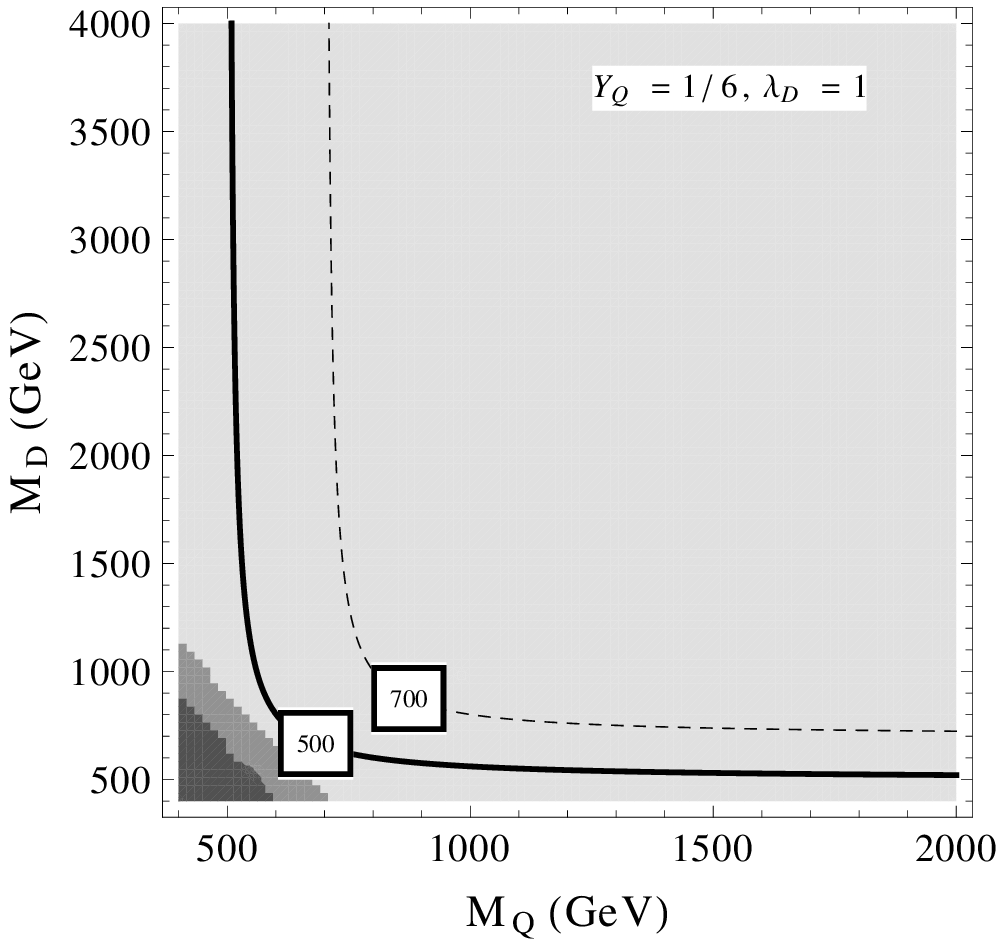}
\includegraphics[width=0.39\textwidth]{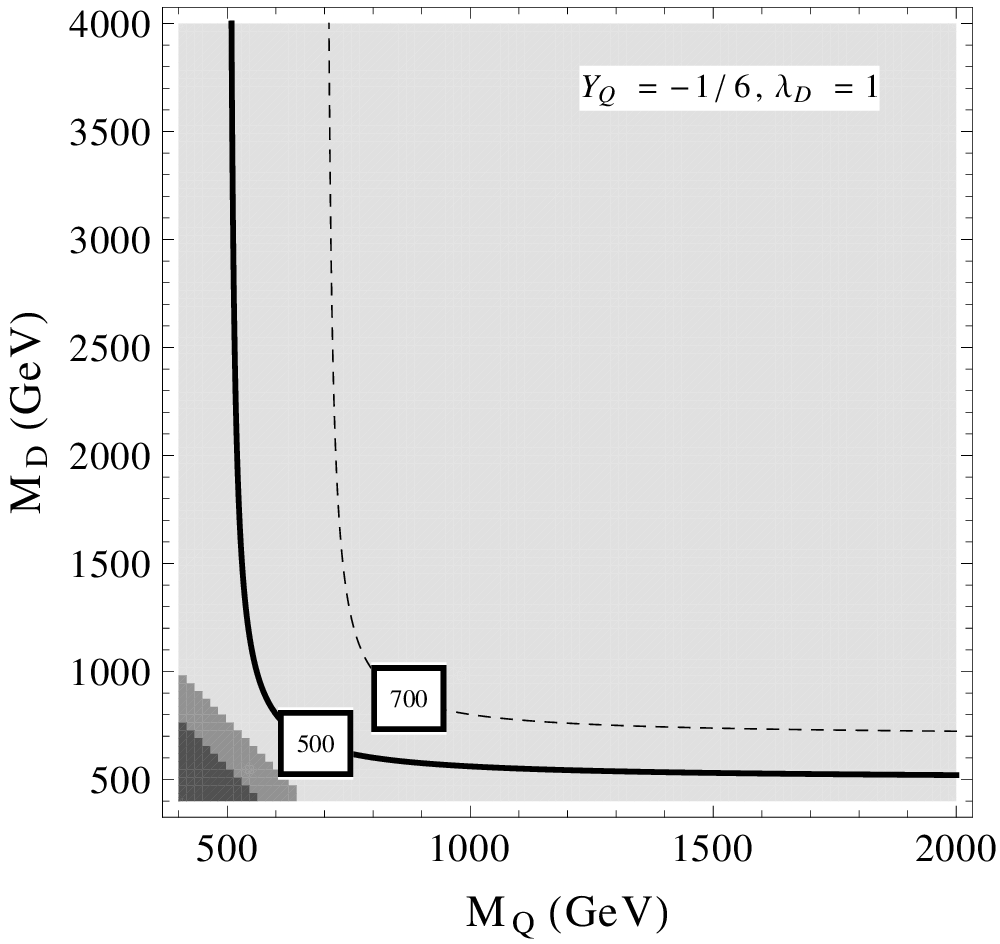}
\caption{
$S$ and $T$ parameters for the MVQD$_1$ model with $Y_Q = 1/6$, $\lambda_D = 0.5$, $1$, $2$, and for $Y_Q = -1/6$, $\lambda_D = 1$. 
$S$ and $T$ are within the 68~\% CL ellipse in the light gray region, between 68~\% and 95~\% ellipse in the medium gray region,
and excluded worse than 95~\% CL in the dark gray region. 
The boxed numbers label (in GeV) contours of the minimum mass eigenvalue, i.e. $min(m_1,m_2,m_3)$. 
\label{MVQD1-ST.FIG}}
\end{center}
\end{figure}

\begin{figure}
\begin{center}
\includegraphics[width=0.39\textwidth]{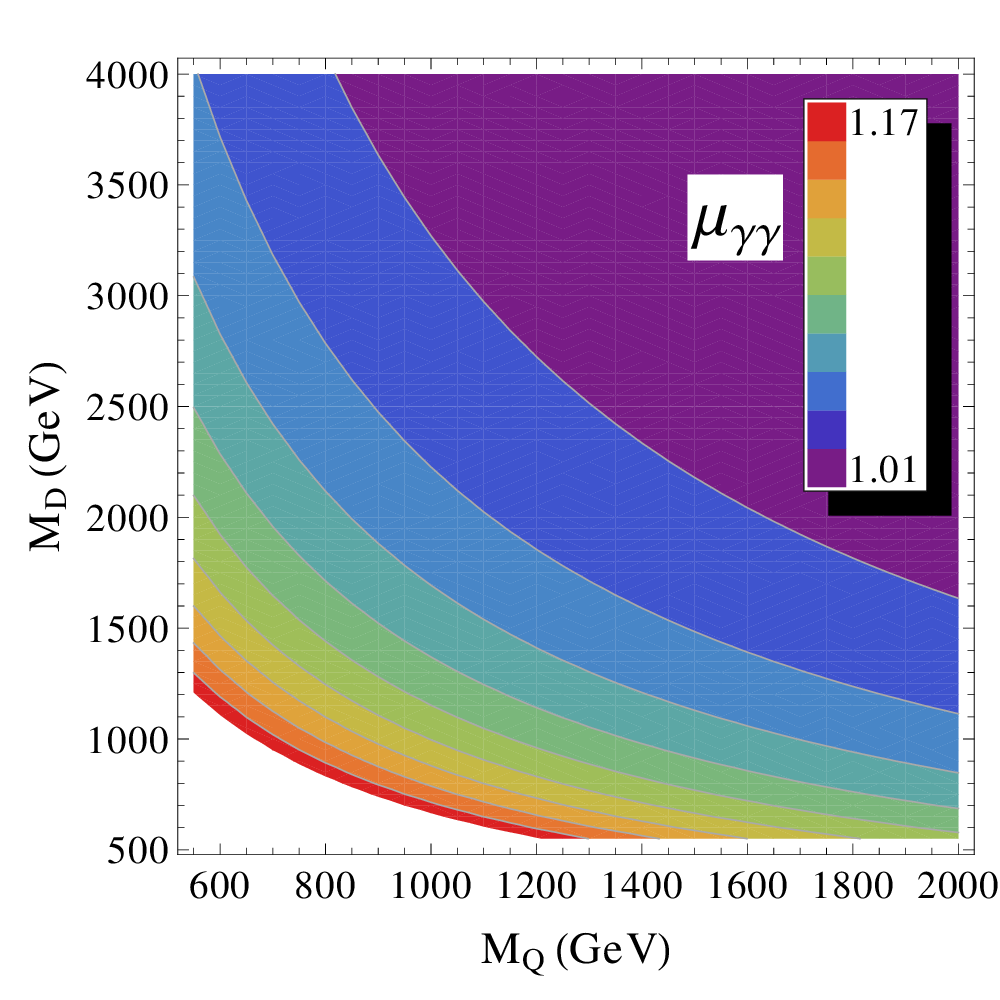}
\vspace*{0.5cm}
\includegraphics[width=0.39\textwidth]{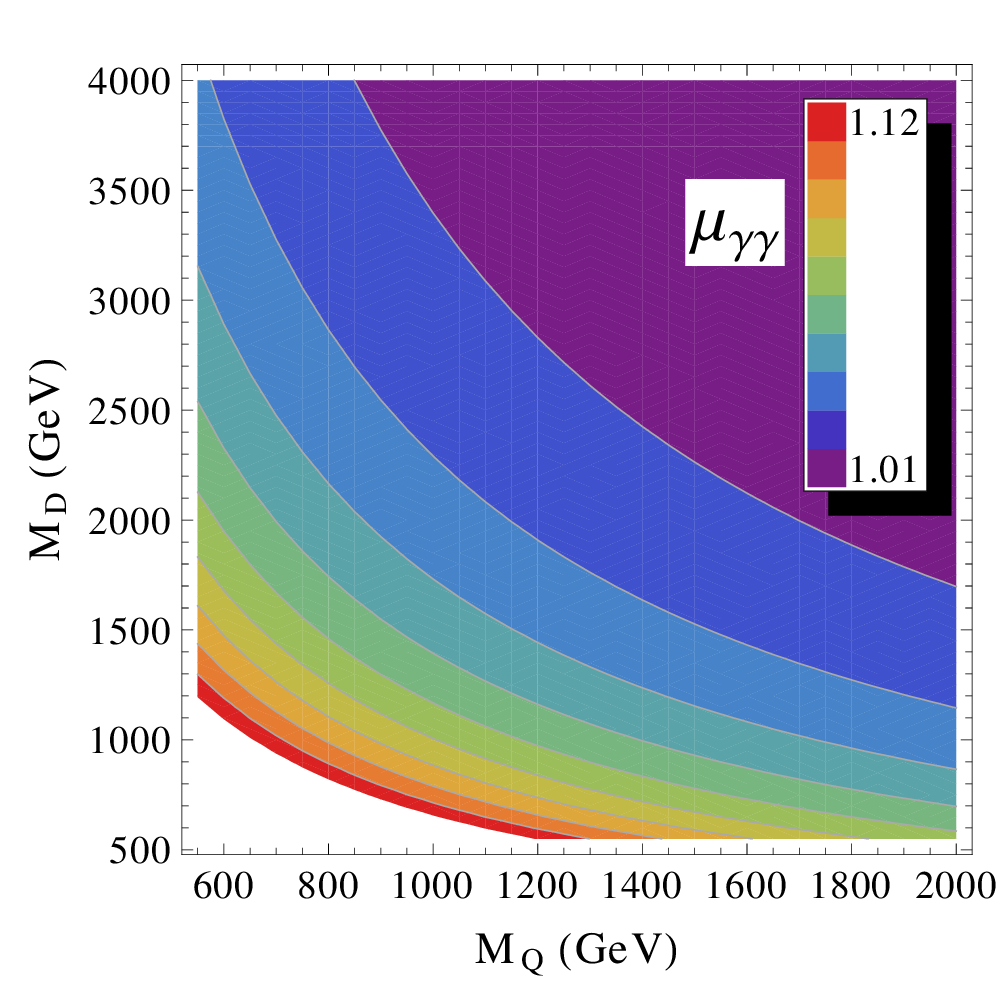}
\includegraphics[width=0.39\textwidth]{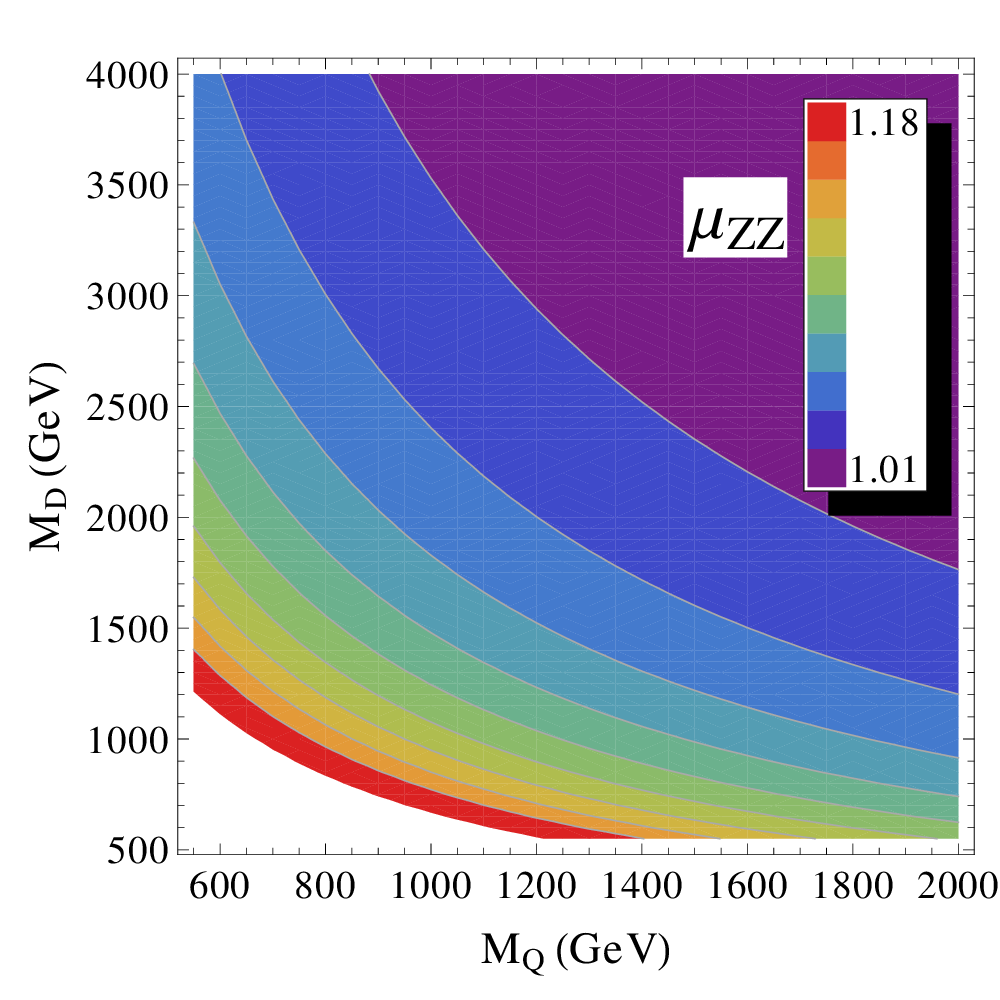}
\vspace*{0.5cm}
\includegraphics[width=0.39\textwidth]{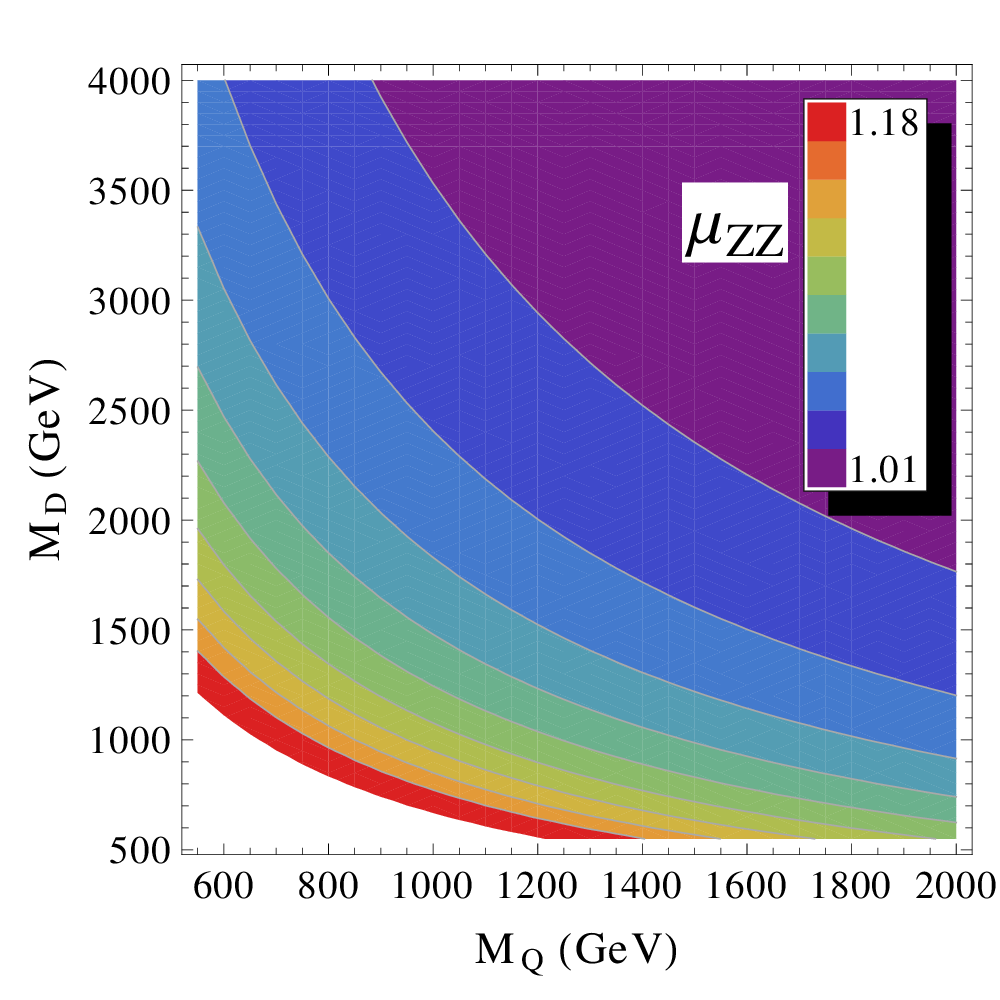}
\includegraphics[width=0.39\textwidth]{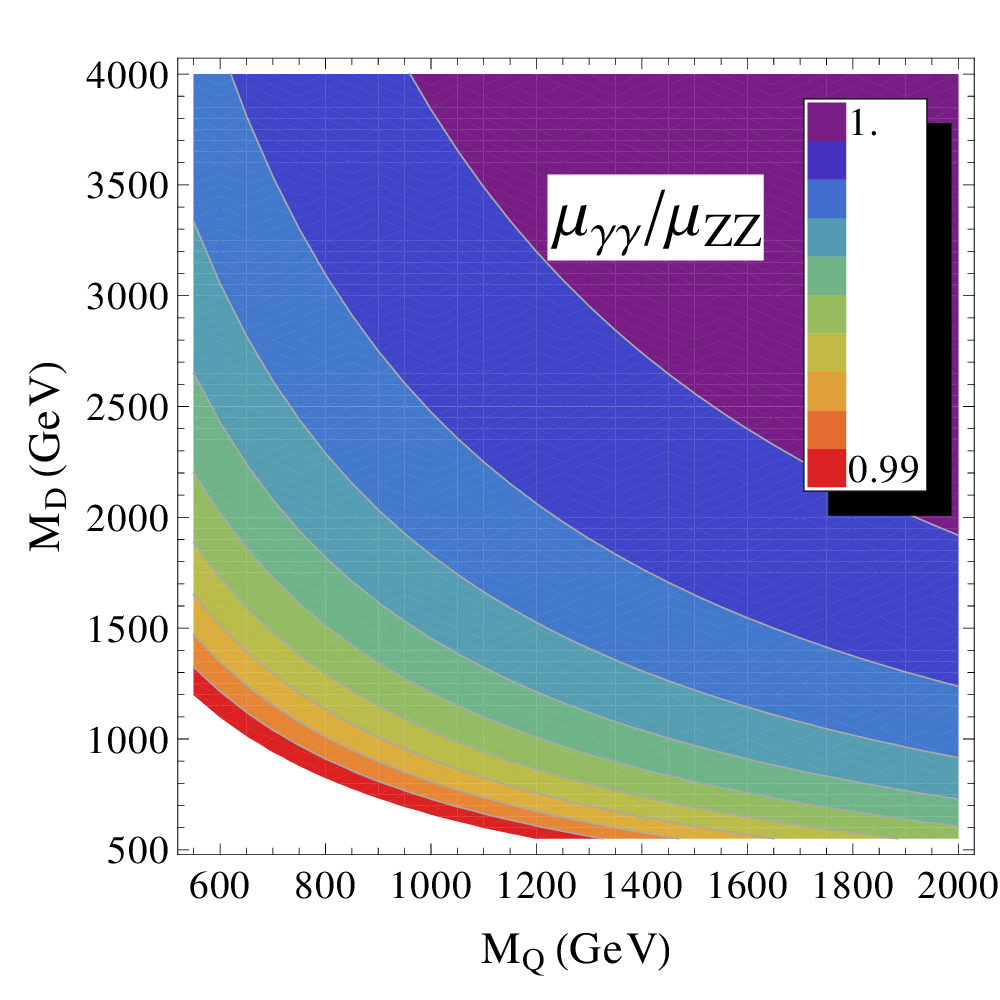}
\includegraphics[width=0.39\textwidth]{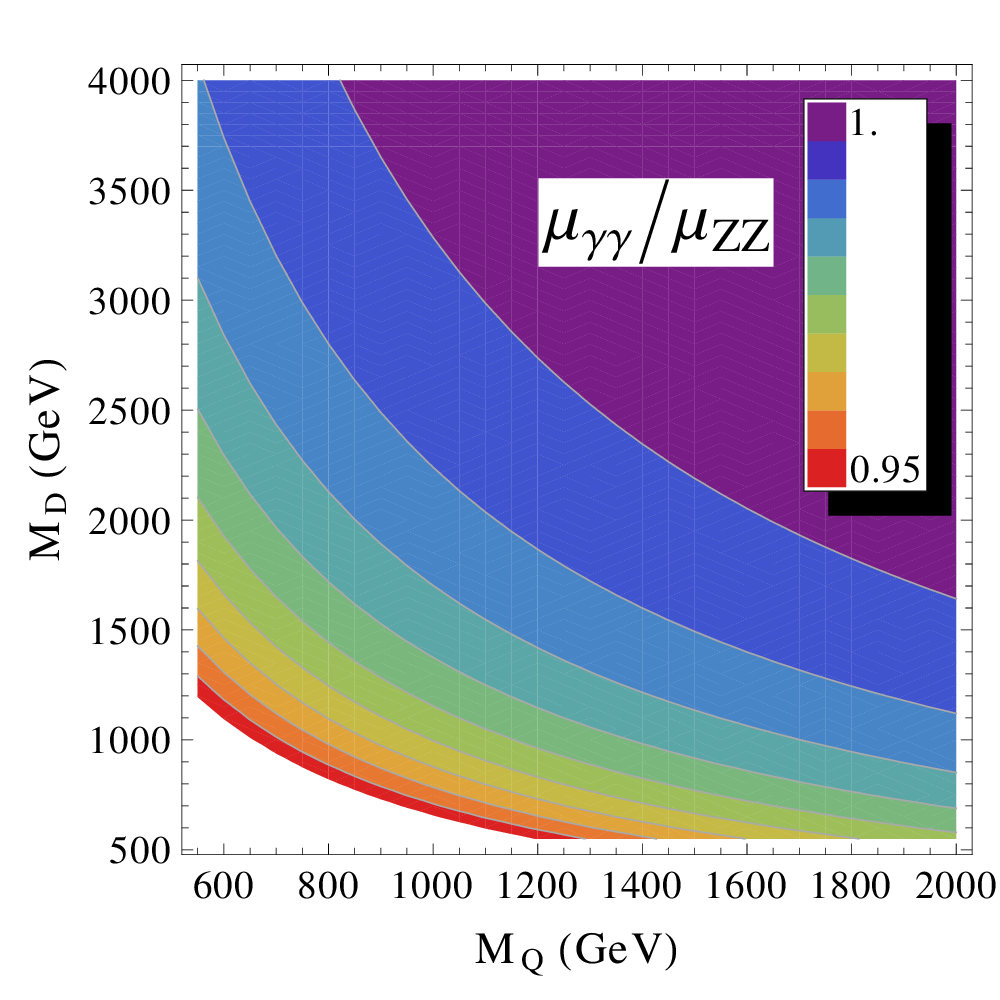}
\caption{For the MVQD$_1$ model, $\mu_{\gamma\gamma}^{ggh}$, $\mu_{ZZ}^{ggh}$ and $\mu_{\gamma\gamma}^{ggh}/\mu_{ZZ}^{ggh}$ for 
$\lambda_D = 1$, $Y_Q = 1/6$ (left panel) and $Y_Q = -1/6$ (right panel).
The colored region has all mass eigenvalues $>500~$GeV.
In every plot, the $\mu$ values progressively approach 1 as we go toward heavier masses (i.e. top-right corner). 
\label{MVQD1-muAAZZ.FIG}}
\end{center}
\end{figure}

In Fig.~\ref{STU2211MQ.FIG} we show $S$, $T$, $U$ as a function of $M_Q$ for MVQD$_1$. 
Note that $T$ is always positive and independent of $Y_Q$. 
$U \lesssim 0.003$ for $M_{Q,D} > 500~$GeV, and will not place any nontrivial constraints; we therefore ignore $U$ in the rest of our analysis.
In Fig.~\ref{MVQD1-ST.FIG} we show the the constraint from $S$ and $T$ parameters as shaded regions
for $Y_Q = 1/6$, $\lambda_D = 0.5$, $1$, $2$, and for $Y_Q = -1/6$, $\lambda_D = 1$.
We also show the $(500,700)$~GeV contours (with boxed numbers) of the the minimum mass eigenvalue, i.e. $min(m_1,m_2,m_3)$. 
In Fig.~\ref{MVQD1-muAAZZ.FIG} we show $\mu_{\gamma\gamma}^{ggh}$, $\mu_{ZZ}^{ggh}$ and 
$\mu_{\gamma\gamma}^{ggh}/\mu_{ZZ}^{ggh}$ for $Y_Q = 1/6$ and $-1/6$ for $\lambda_D = 1$.
Keeping in mind the direct LHC limits on a vector-like $b'$ and $t'$ we ensure that all mass eigenvalues are 
$>500~$GeV and show them as the colored region.    
The deviation in $\mu_{ZZ}^{ggh}$ is entirely due to a change in the $ggh$ vertex since the $hZZ$ vertex is unchanged, 
and this change is the same for either sign of $Y_Q$.
In general, in $gg\to h \to \gamma\gamma$, i.e. for $\mu_{\gamma\gamma}^{ggh}$, both production and decay vertices are shifted; 
the $ggh$ vertex is shifted due to the presence of new colored vector-like fermions, and the $h\gamma\gamma$ vertex is shifted since
these states carry EM charge as well. 
Interestingly, as can be seen from Eq.~(\ref{hChiChi2211.EQ}), 
the $h\gamma\gamma$ coupling shift is small in this model since $\Chi_1$ does not couple to the $h$, 
while the $h$ couplings to $\Chi_2$ and $\Chi_3$ are opposite sign and will cancel up to mass differences. 
In all cases, the $\mu$ asymptote to 1 as the vector-like fermion contributions decouple.  
%

\subsection{VSM$_1$ vector-like Standard Model}
\begin{figure}[!ht]
\begin{center}
\includegraphics[width=0.39\textwidth] {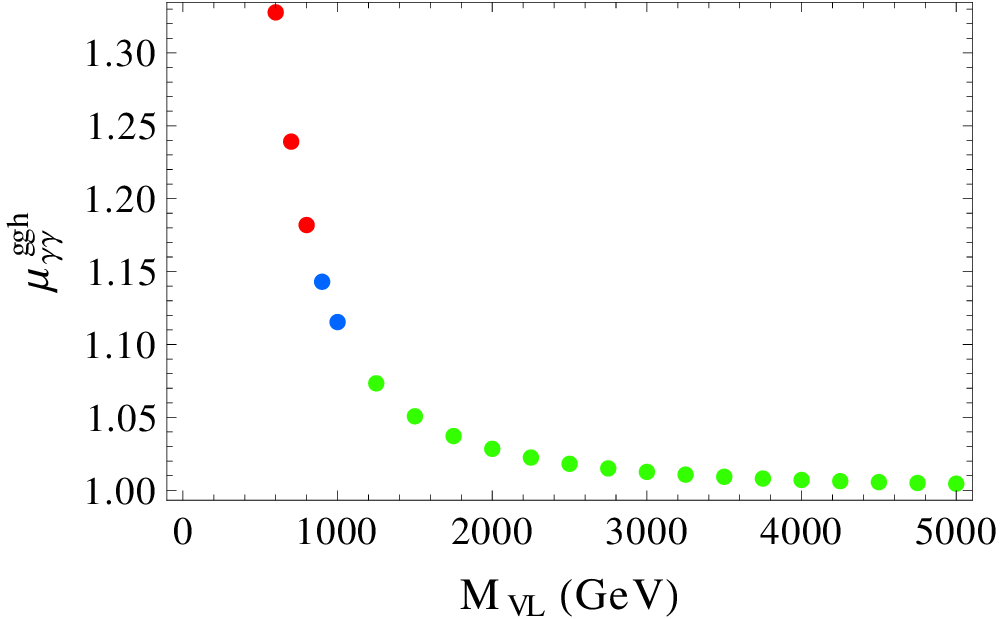}    
\vspace*{0.5cm}
\includegraphics[width=0.39\textwidth] {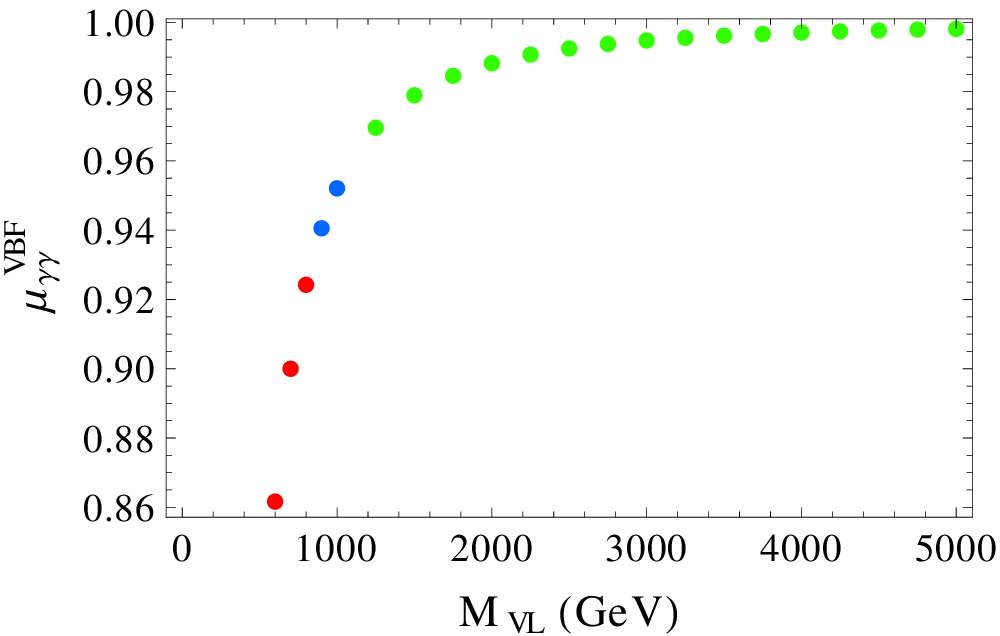}
\includegraphics[width=0.39\textwidth] {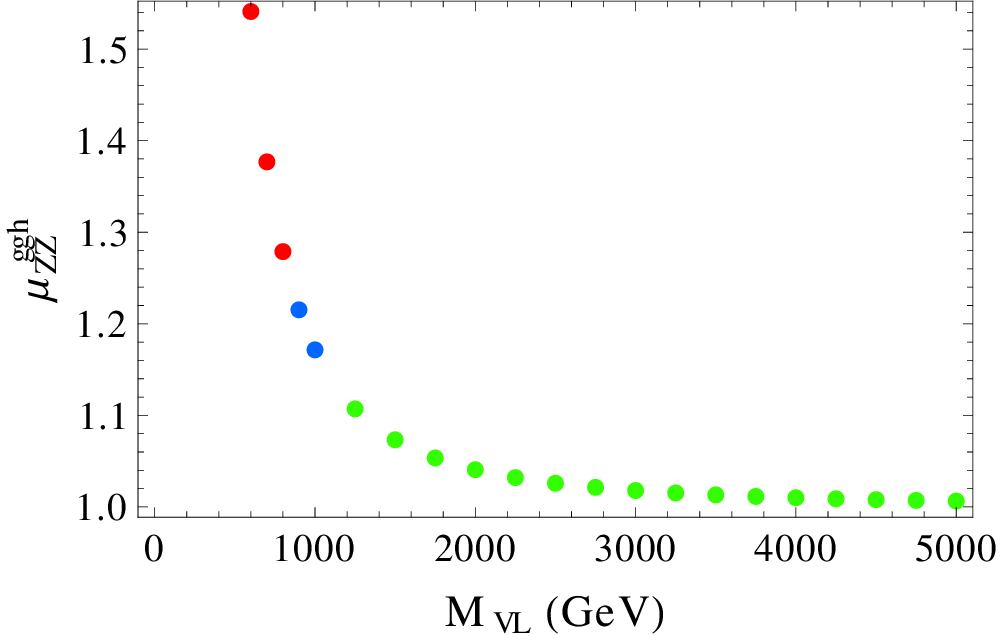}
\vspace*{0.5cm}
\includegraphics[width=0.39\textwidth] {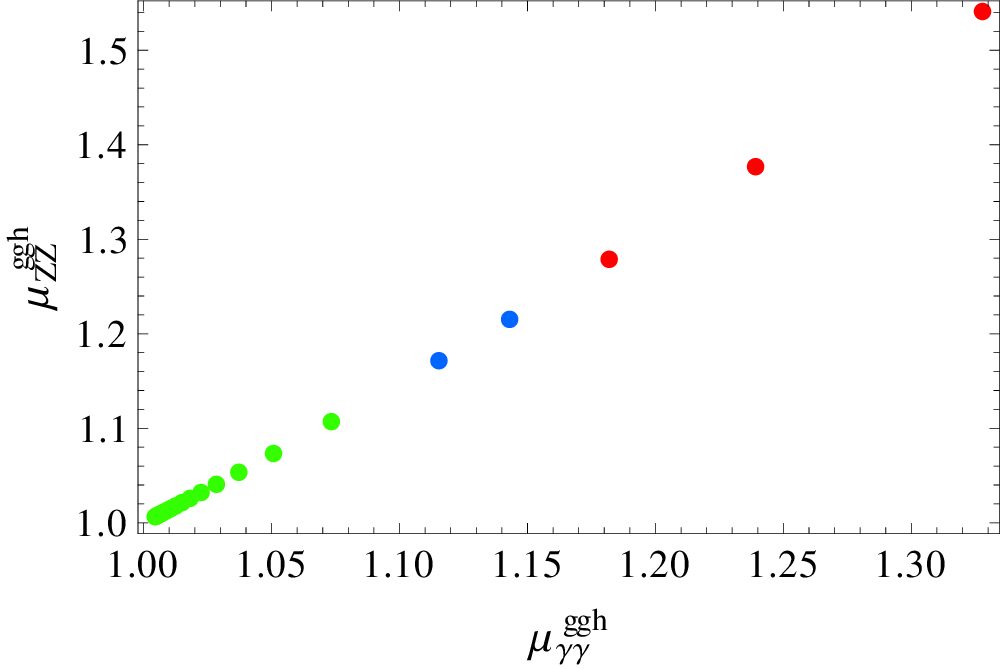}  
\includegraphics[width=0.39\textwidth] {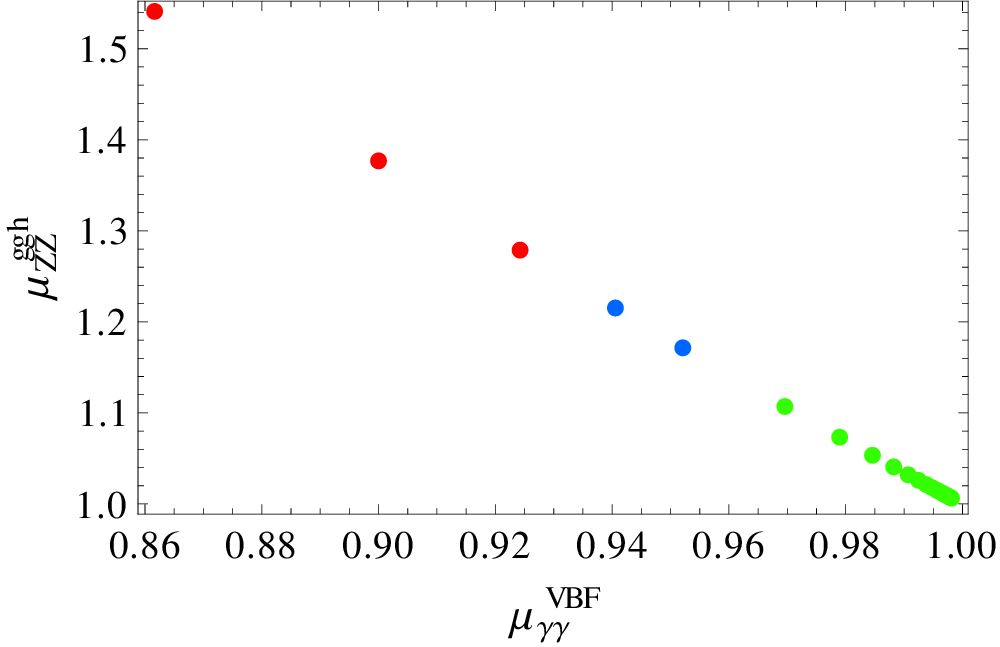}  
\caption{
$\mu$ in the VSM$_1$ model with all the vector-like quark and lepton masses equal to the value shown in the $X$-axis, 
i.e. $M_{ \{ Q,U,D,L,E,N \} } = M_{VL}$,
with $Y_Q=1/6$ and $Y_L=-1/2$, and all the Yukawa couplings $\lambda=1$. 
All these points satisfy the $S$ and $T$ constraints at or better than $2\, \sigma$ level.  
The color (or shade of gray, if viewing in gray-scale) of the dots denote the lightest mass eigenvalue; 
the red (dark gray), blue and green (light gray) dots respectively stand for light, medium and heavy mass categories given in Table~\ref{MVLcats.TAB}.  
\label{mu-VSM1-MV-lam1.FIG}}
\end{center}
\end{figure}

\begin{figure}
\begin{center}
\includegraphics[width=0.39\textwidth] {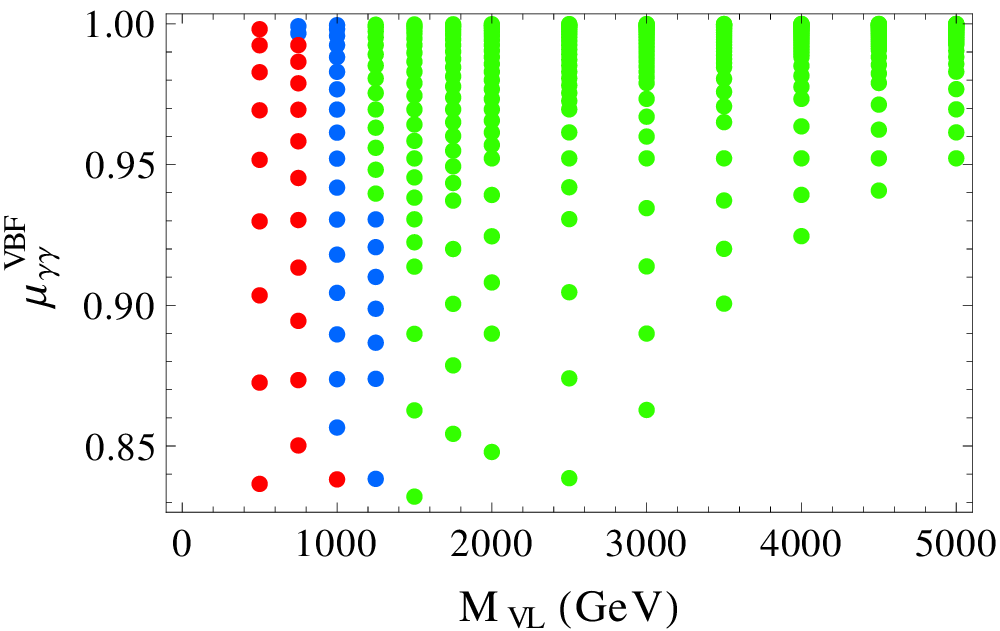}
\vspace*{0.5cm}
\includegraphics[width=0.39\textwidth] {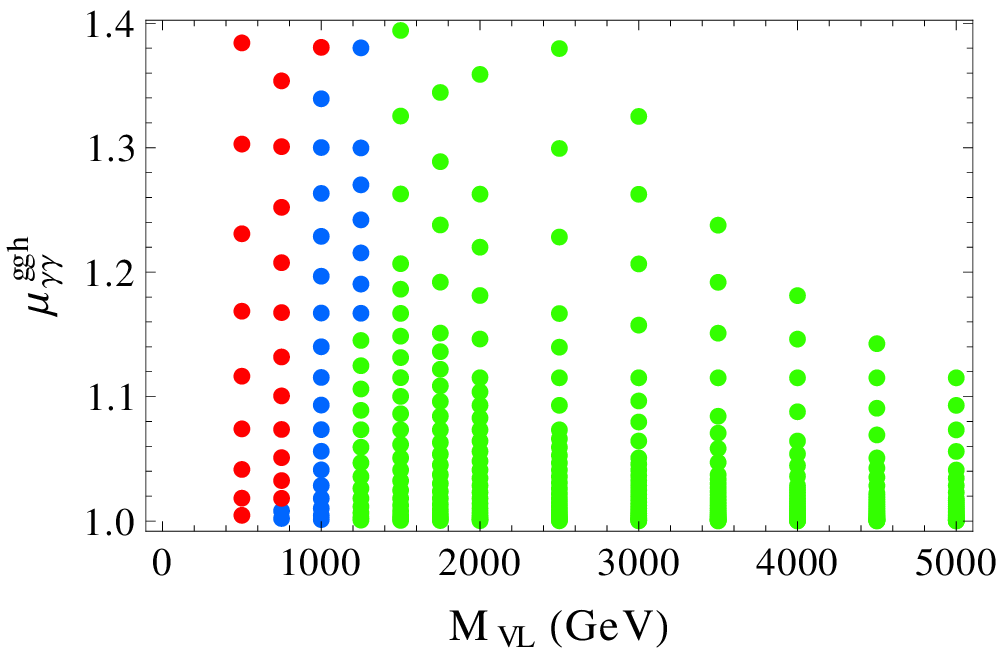}      
\includegraphics[width=0.39\textwidth] {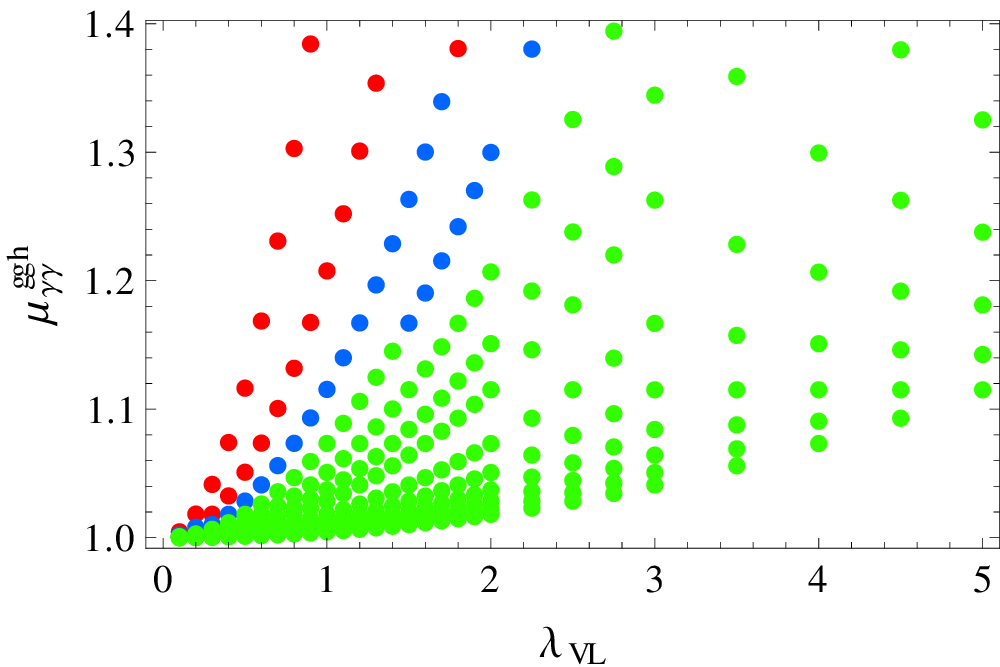}   
\vspace*{0.5cm}
\includegraphics[width=0.39\textwidth] {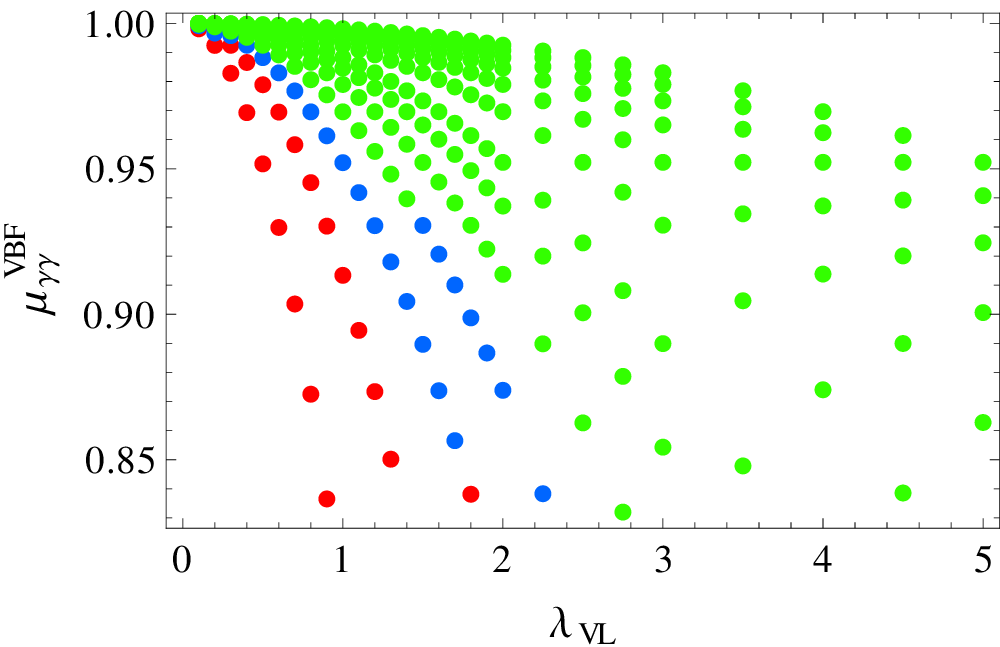}  
\includegraphics[width=0.39\textwidth] {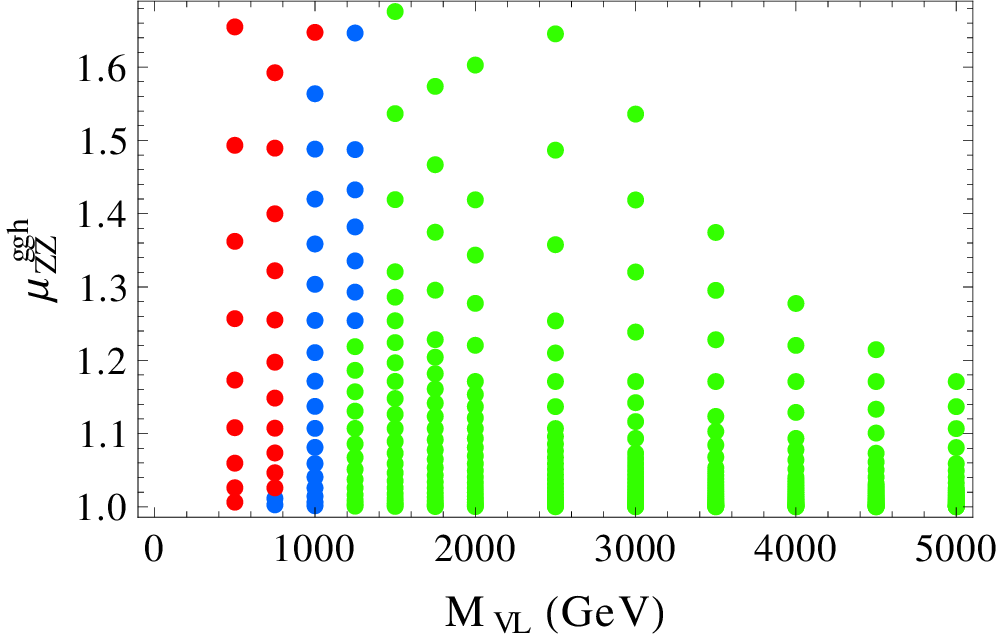}
\vspace*{0.5cm}
\includegraphics[width=0.39\textwidth] {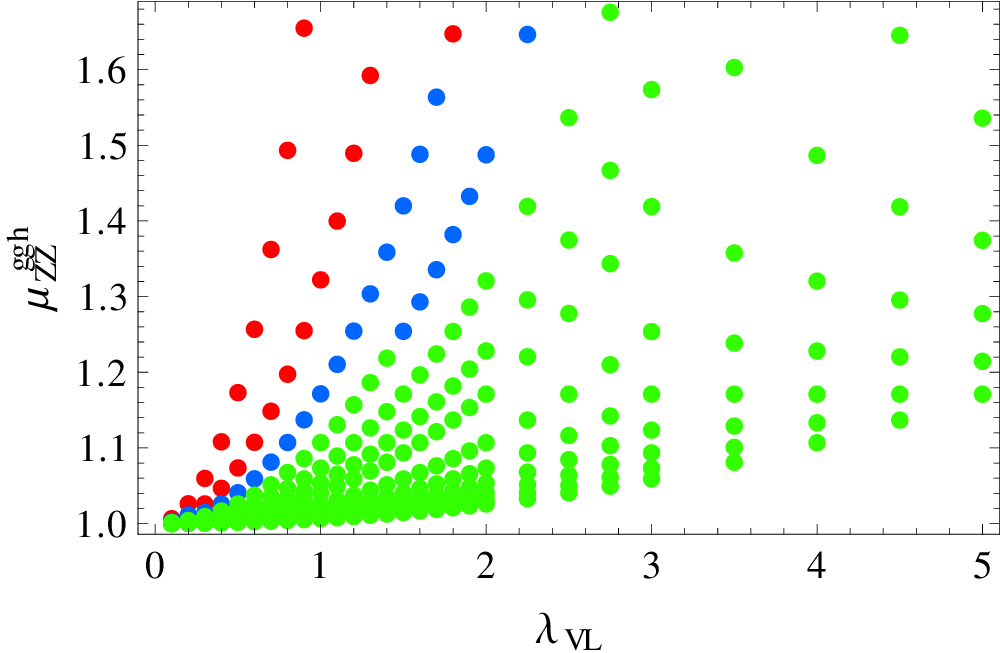}  
\includegraphics[width=0.39\textwidth] {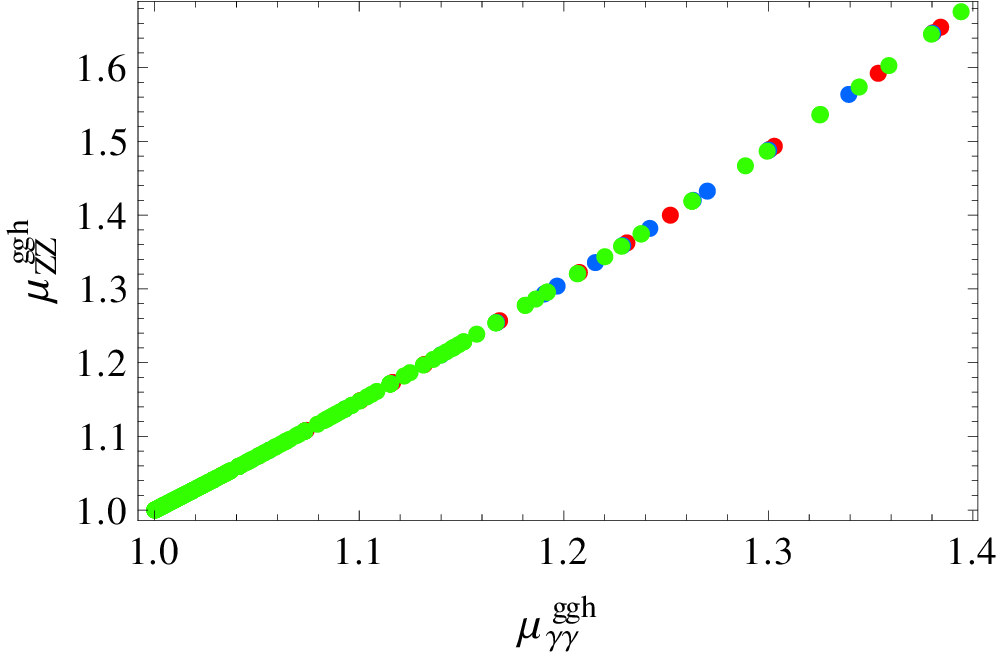}  
\includegraphics[width=0.39\textwidth] {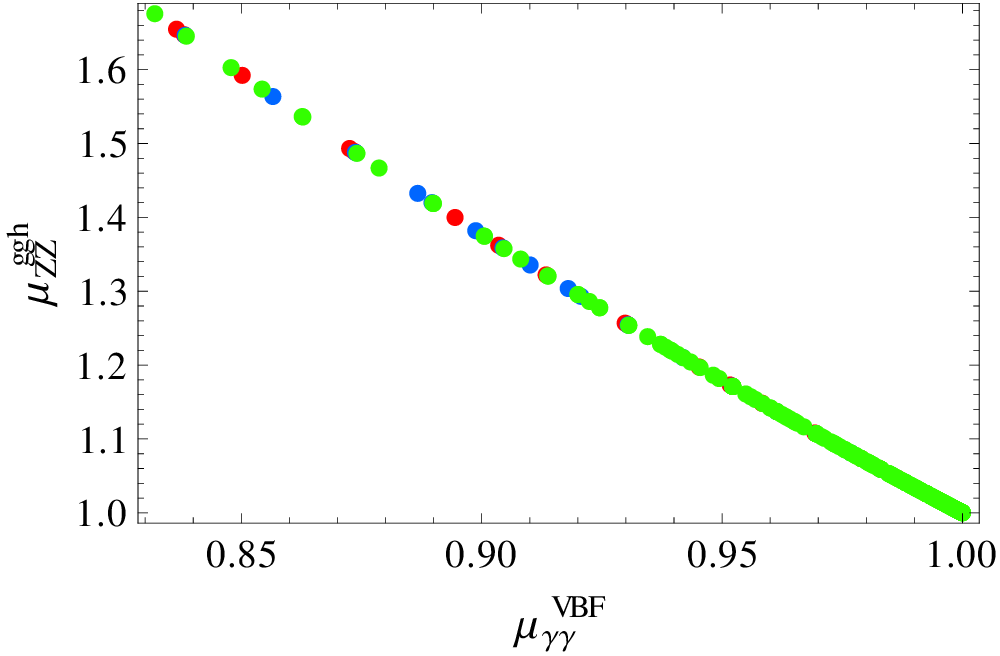}
\caption{
$\mu$ in the VSM$_1$ model with all the vector-like quark and lepton masses set equal ($M_{ \{ Q,U,D,L,E,N \} } = M_{VL}$),
and all the $\lambda$ set equal ($\lambda_{ \{ U,D,E,N \} } = \lambda_{VL}$), for $Y_Q = 1/6$ and $Y_L = -1/2$.
All these points satisfy the $S$ and $T$ constraints at or better than $2\, \sigma$ level. 
The color (or shade of gray, if viewing in gray-scale) of the dots denote the lightest mass eigenvalue; 
the red (dark gray), blue and green (light gray) dots respectively stand for light, medium and heavy mass categories given in 
Table~\ref{MVLcats.TAB}. 
\label{mu-VSM1-MV-lam.FIG}}
\end{center}
\end{figure}

Keeping in mind the direct collider limits, we restrict to the parameter-space with all vector-like 
quark mass eigenvalue $\geq 500$~GeV and lepton mass eigenvalues $\geq 250$~GeV. 
In 
Fig.~\ref{mu-VSM1-MV-lam1.FIG} we show the signal strength in the VSM$_1$ 
with all the vector-like quark and lepton masses equal to the value shown in the $X$-axis, 
i.e. $M_{ \{ Q,U,D,L,E,N \} } = M_{VL}$,
with $Y_Q=1/6$ and $Y_L=-1/2$, and all the Yukawa couplings $\lambda=1$. 
All these points satisfy the $S$ and $T$ constraints at or better than $2\, \sigma$ level.
The color of the dots denote the lightest mass eigenvalue; 
the red, blue and green dots respectively stand for light, medium and heavy mass categories given in Table~\ref{MVLcats.TAB}.
These mass eigenvalue ranges are motivated by the direct collider limits discussed in 
Sections~\ref{VLQlimits.SEC}~and~\ref{VLLlimits.SEC}. 
\begin{table}
\begin{centering}
\caption{The three categories for the lightest mass eigenvalue of the vector-like quark and lepton.
``Light'' is with {\em either} the $M_q$ {\em or} the $M_l$ as shown, ``Medium'' is with {\em each} in the interval shown, 
and ``Heavy'' is with {\em both} above the values shown. 
\label{MVLcats.TAB}}
\begin{tabular}{|c|c|c|}
\hline 
 & $M_{q}$(GeV) & $M_{\ell}$ (GeV)\tabularnewline
\hline 
\hline 
Light & $\leq700$ & $\leq450$\tabularnewline
\hline 
Medium & $(700,1000)$ & $(450,750)$\tabularnewline
\hline 
Heavy & $>1000$ & $>750$\tabularnewline
\hline 
\end{tabular}
\par\end{centering}
\end{table}
In Fig.~\ref{mu-VSM1-MV-lam.FIG} we show the $\mu$ with all the vector-like quark and lepton masses set equal ($M_{ \{ Q,U,D,L,E,N \} } = M_{VL}$),
and all the $\lambda$ set equal ($\lambda_{ \{U,D,E,N \} } = \lambda_{VL}$), for $Y_Q = 1/6$ and $Y_L = -1/2$.
All these points satisfy the $S$ and $T$ constraints at or better than $2\, \sigma$ level.
The color of the dots denote the lightest mass eigenvalue; 
the red, blue and green dots respectively stand for light, medium and heavy mass categories given in Table~\ref{MVLcats.TAB}.  

\begin{figure}[!ht]
\begin{center}
\includegraphics[width=0.39\textwidth] {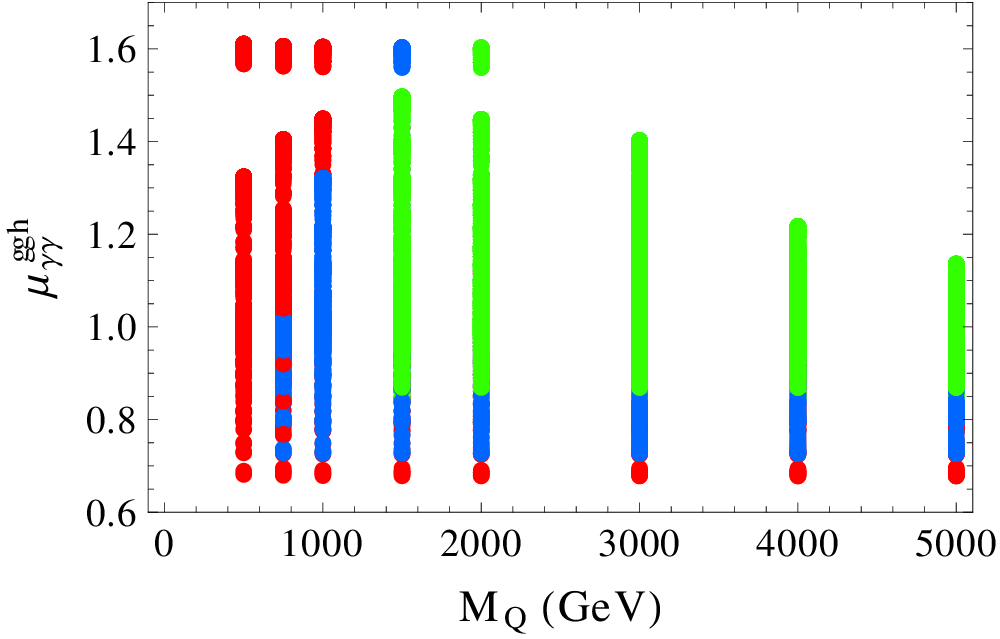}   
\vspace*{0.5cm}
\includegraphics[width=0.39\textwidth] {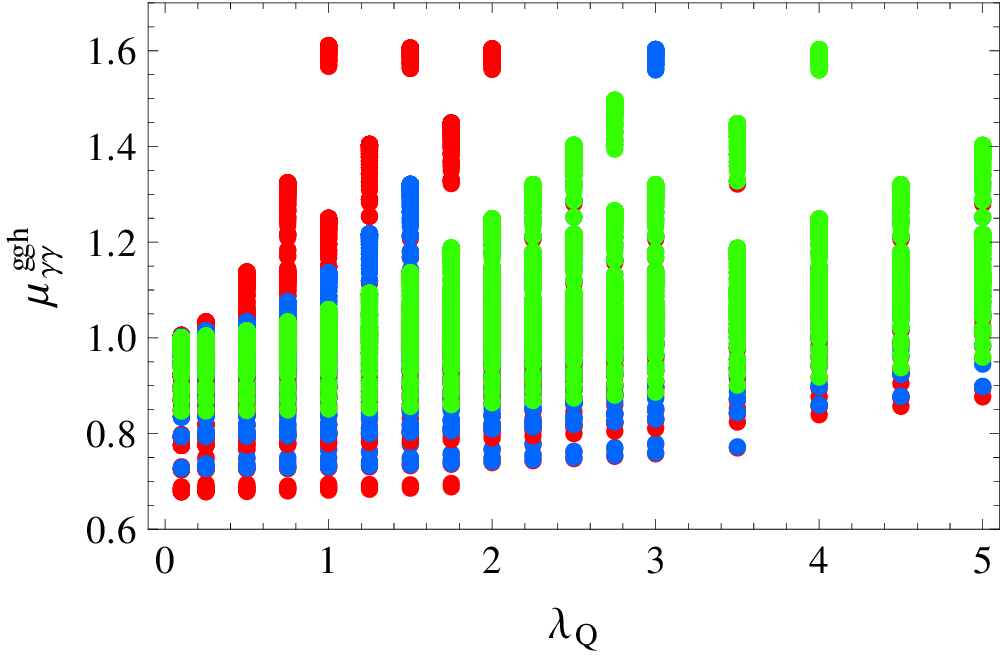}  
\includegraphics[width=0.39\textwidth] {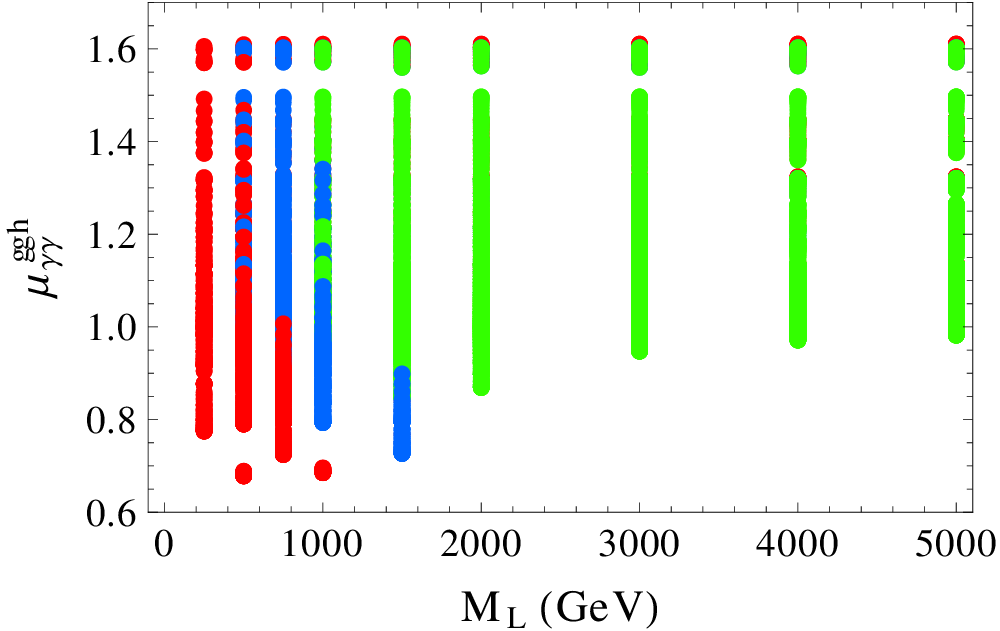}    
\includegraphics[width=0.39\textwidth] {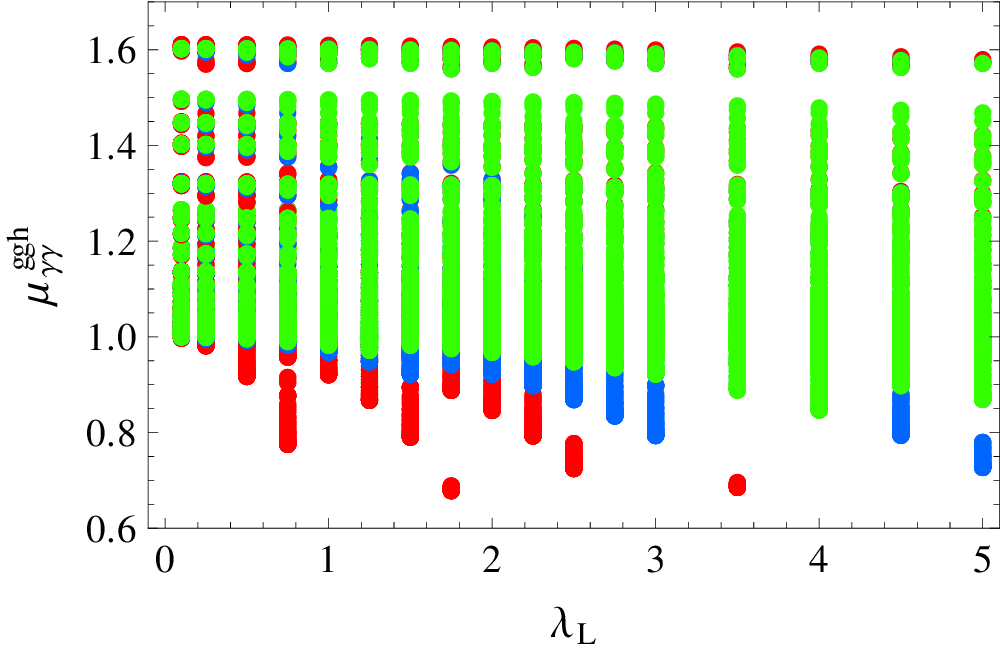}  
\caption{$\mu^{ggh}_{\gamma\gamma}$ in the VSM$_1$ model from a scan over all the vector-like quark and lepton masses 
in the range $(50,5000)~$GeV, and the Yukawa couplings in the range $(0.1,5)$, 
with the $Q,U,D,L,E,N$ vector-like fermion fields all present,  
with $M_Q=M_U=M_D$, $M_L=M_E=M_N$, $\lambda_U=\lambda_D\equiv \lambda_Q$, and $\lambda_E=\lambda_N\equiv \lambda_L$, 
for $Y_Q=1/6$ and $Y_L=-1/2$. 
All the points satisfy the $S$ and $T$ constraints at or better than $2\, \sigma$ level.
The color (or shade of gray, if viewing in gray-scale) of the dots denote the lightest mass eigenvalue; 
the red (dark gray), blue and green (light gray) dots respectively stand for light, medium and heavy mass categories given in 
Table~\ref{MVLcats.TAB}. 
\label{muAA-VSM1-lamQL-MQL-scan.FIG}}
\end{center}
\end{figure}

\begin{figure}[!ht]
\begin{center}
\includegraphics[width=0.39\textwidth] {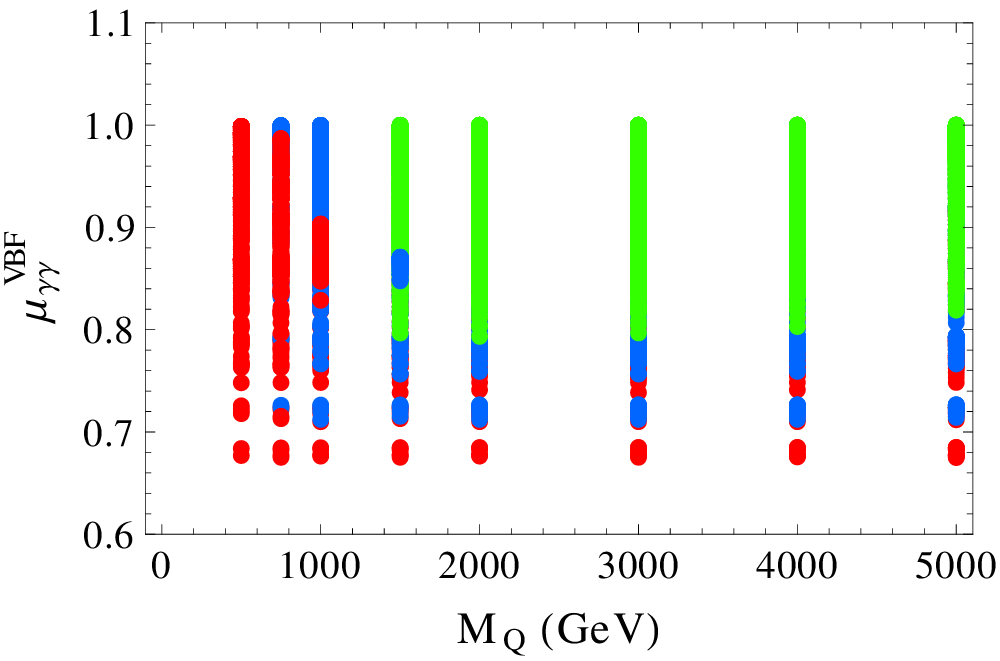}
\vspace*{0.5cm}
\includegraphics[width=0.39\textwidth] {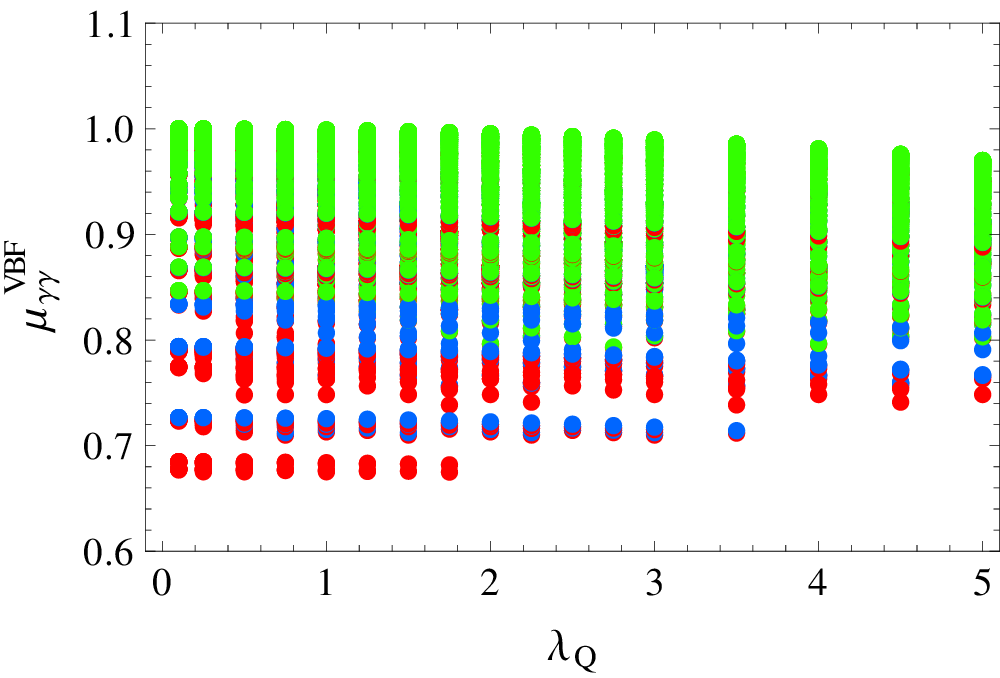}
\includegraphics[width=0.39\textwidth] {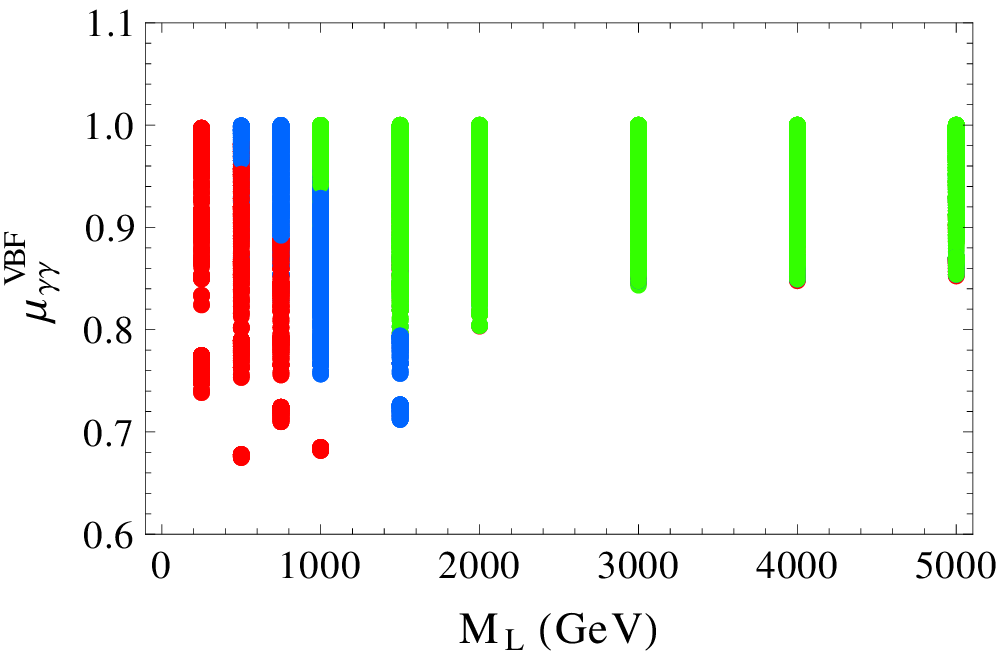}
\includegraphics[width=0.39\textwidth] {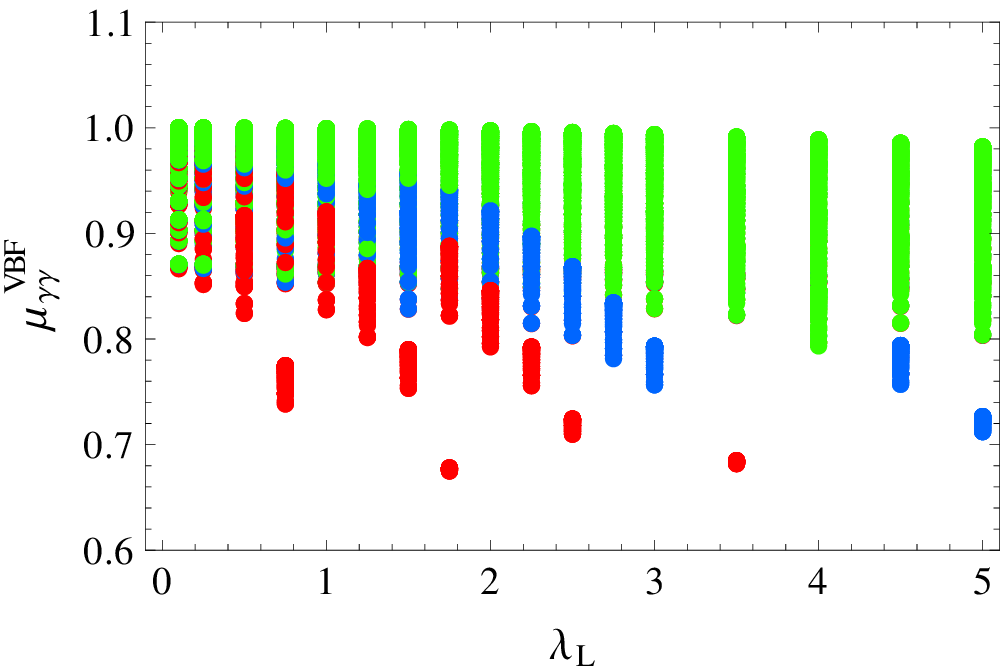}  
\caption{$\mu^{VBF}_{\gamma\gamma}$ in the VSM$_1$ model from a scan over all the vector-like quark and lepton masses 
in the range $(50,5000)~$GeV, and the Yukawa couplings in the range $(0.1,5)$, 
with the $Q,U,D,L,E,N$ vector-like fermion fields all present,  
with $M_Q=M_U=M_D$, $M_L=M_E=M_N$, $\lambda_U=\lambda_D\equiv \lambda_Q$, and $\lambda_E=\lambda_N\equiv \lambda_L$, 
for $Y_Q=1/6$ and $Y_L=-1/2$. 
All the points satisfy the $S$ and $T$ constraints at or better than $2\, \sigma$ level.
The color (or shade of gray, if viewing in gray-scale) of the dots denote the lightest mass eigenvalue; 
the red (dark gray), blue and green (light gray) dots respectively stand for light, medium and heavy mass categories given in 
Table~\ref{MVLcats.TAB}. 
\label{muAAvbf-VSM1-lamQL-MQL-scan.FIG}}
\end{center}
\end{figure}

\begin{figure}[!ht]
\begin{center}
\includegraphics[width=0.39\textwidth] {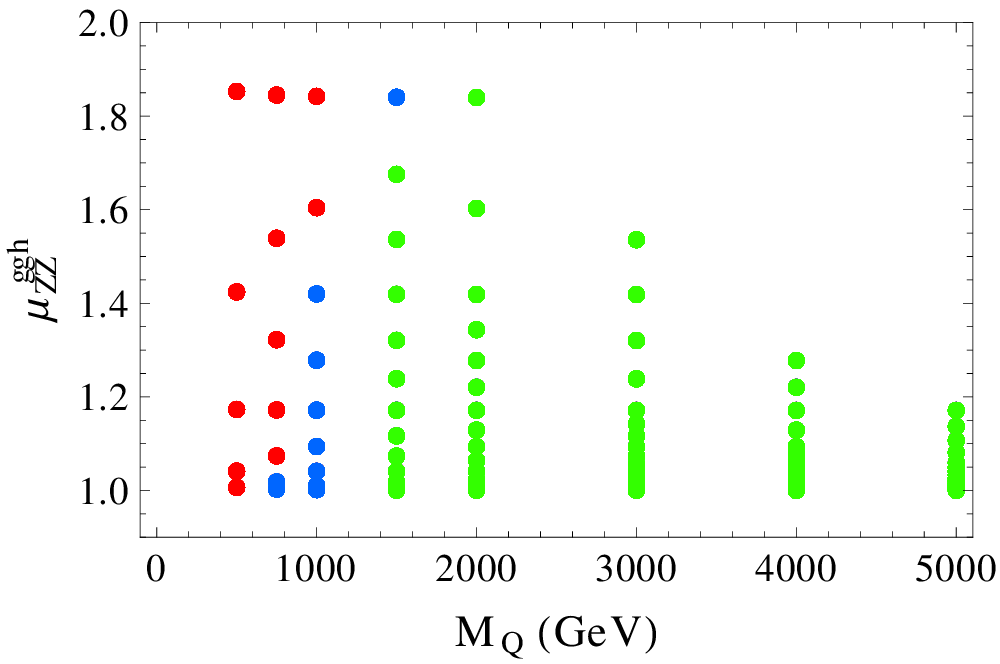}
\vspace*{0.5cm}
\includegraphics[width=0.39\textwidth] {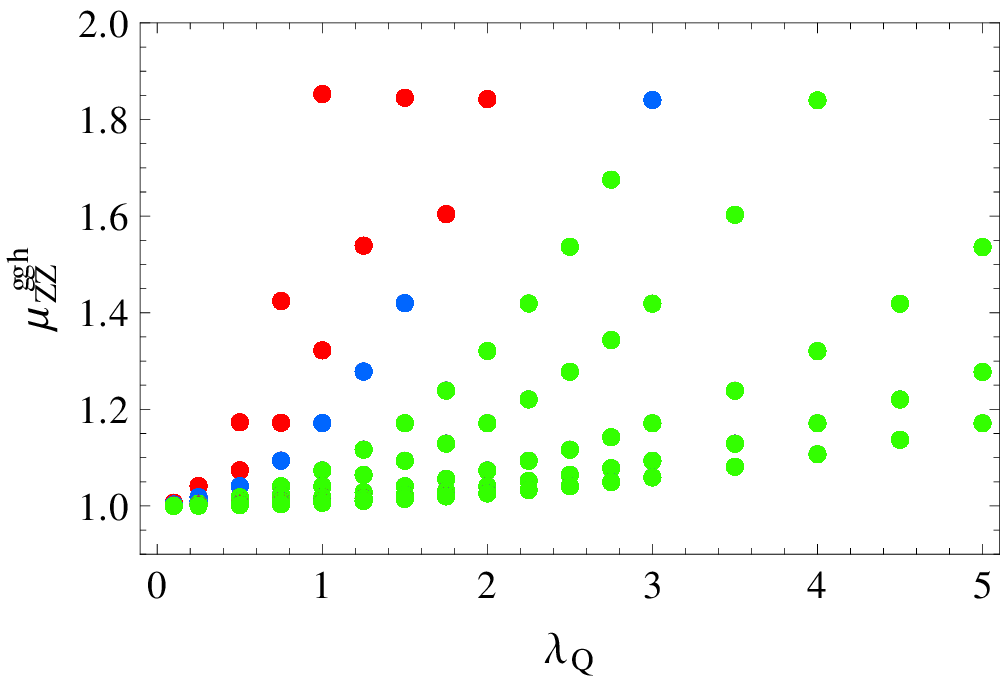}  
\includegraphics[width=0.39\textwidth] {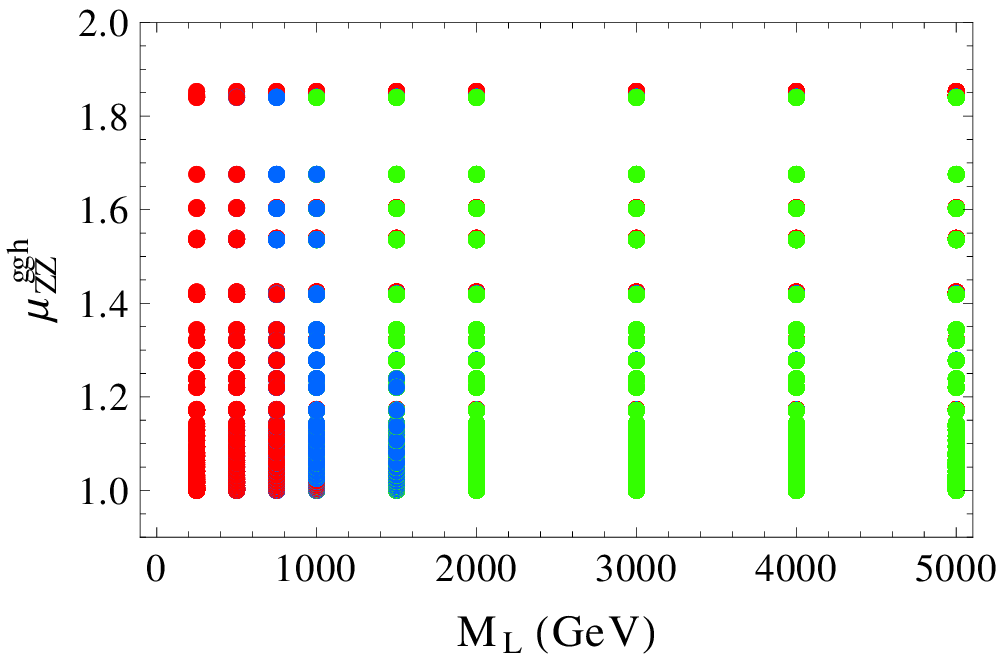}
\includegraphics[width=0.39\textwidth] {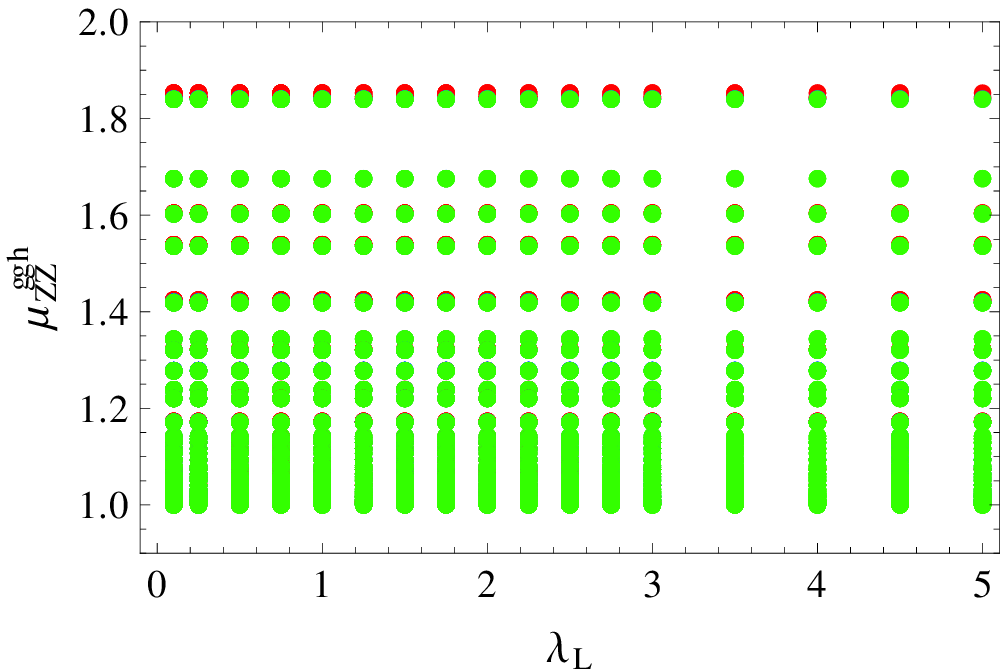}  
\caption{$\mu_{ZZ}$ in the VSM$_1$ model from a scan over all the vector-like quark and lepton masses 
in the range $(50,5000)~$GeV, and the Yukawa couplings in the range $(0.1,5)$,
with the $Q,U,D,L,E,N$ vector-like fermion fields all present,
with $M_Q=M_U=M_D$, $M_L=M_E=M_N$, $\lambda_U=\lambda_D\equiv \lambda_Q$, and $\lambda_E=\lambda_N\equiv \lambda_L$, 
for $Y_Q=1/6$ and $Y_L=-1/2$. 
All the points satisfy the $S$ and $T$ constraints at or better than $2\, \sigma$ level. 
The color (or shade of gray, if viewing in gray-scale) of the dots denote the lightest mass eigenvalue; 
the red (dark gray), blue and green (light gray) dots respectively stand for light, medium and heavy mass categories given in 
Table~\ref{MVLcats.TAB}. 
\label{muZZ-VSM1-lamQL-MQL-scan.FIG}}
\end{center}
\end{figure}

\begin{figure}[!ht]
\begin{center}
\includegraphics[width=0.39\textwidth] {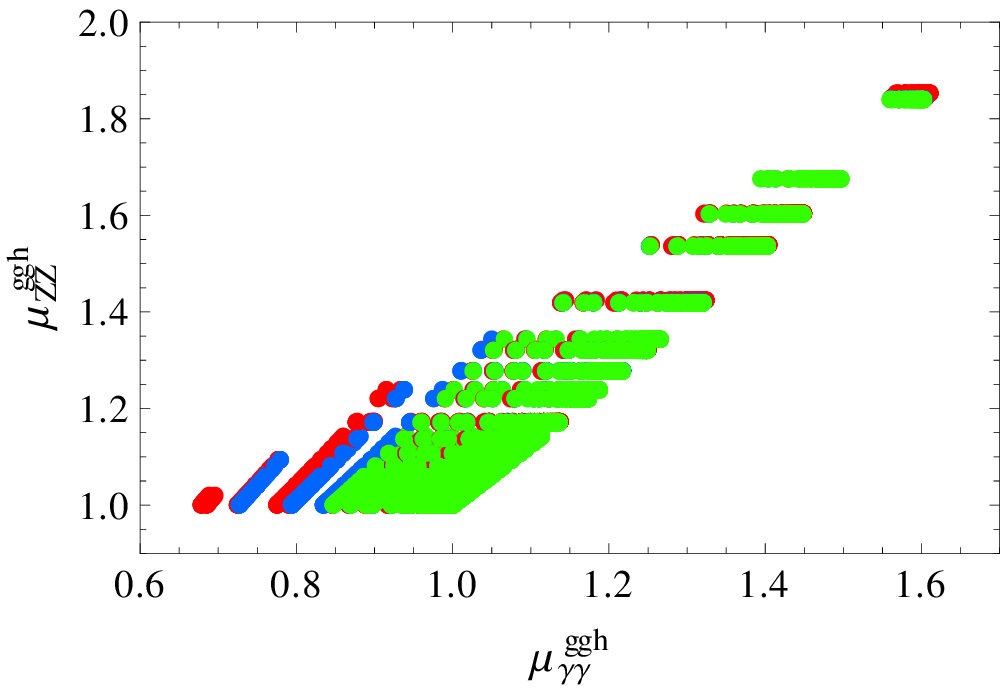}     
\includegraphics[width=0.39\textwidth] {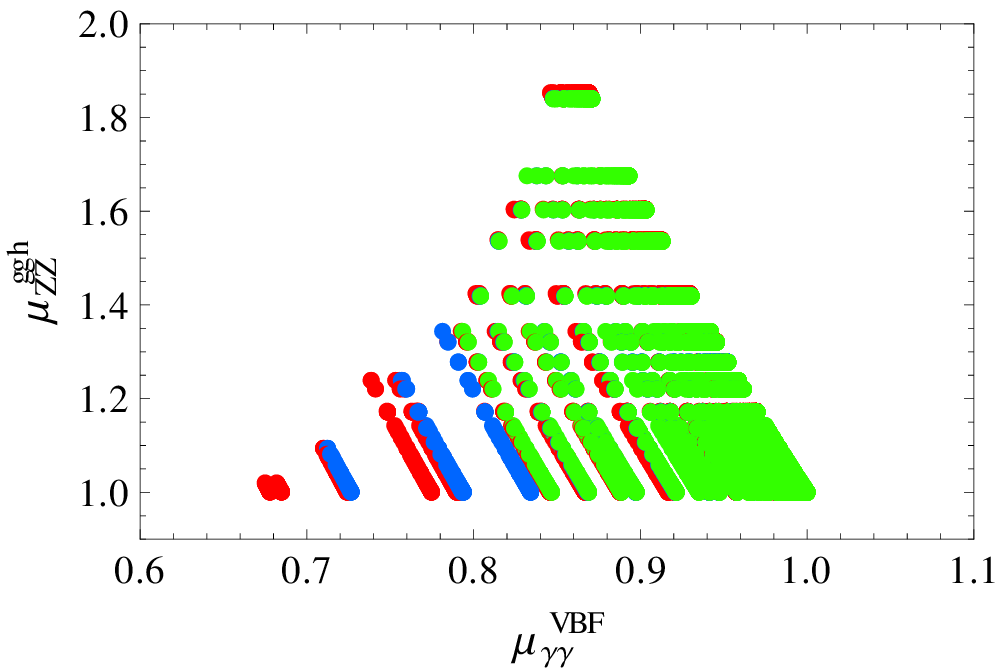}
\caption{$\mu^{ggh,VBF}_{\gamma\gamma}$--$\mu_{ZZ}$ correlation in the VSM$_1$ model from a scan over all the vector-like quark and lepton masses 
in the range $(50,5000)~$GeV, and the Yukawa couplings in the range $(0.1,5)$, 
with the $Q,U,D,L,E,N$ vector-like fermion fields all present,  
with $M_Q=M_U=M_D$, $M_L=M_E=M_N$, $\lambda_U=\lambda_D\equiv \lambda_Q$, and $\lambda_E=\lambda_N\equiv \lambda_L$, 
for $Y_Q=1/6$ and $Y_L=-1/2$. 
All the points satisfy the $S$ and $T$ constraints at or better than $2\, \sigma$ level.
The color (or shade of gray, if viewing in gray-scale) of the dots denote the lightest mass eigenvalue; 
the red (dark gray), blue and green (light gray) dots respectively stand for light, medium and heavy mass categories given in 
Table~\ref{MVLcats.TAB}. 
\label{muAAmuZZ-VSM1-lamQL-MQL-scan.FIG}}
\end{center}
\end{figure}

Next, we present some more results where we perform the scan in a more unconstrained fashion. 
We show 
in Fig.~\ref{muAA-VSM1-lamQL-MQL-scan.FIG} the signal strength $\mu^{ggh}_{\gamma\gamma}$,
in Fig.~\ref{muAAvbf-VSM1-lamQL-MQL-scan.FIG} the signal strength $\mu^{VBF}_{\gamma\gamma}$,
in Fig.~\ref{muZZ-VSM1-lamQL-MQL-scan.FIG} the signal strength $\mu_{ZZ}$
and in Fig.~\ref{muAAmuZZ-VSM1-lamQL-MQL-scan.FIG} the correlation between $\mu^{ggh,VBF}_{\gamma\gamma}$ and $\mu_{ZZ}$
in the VSM$_1$ model
by scanning over all the vector-like quark and lepton masses in the range $(50,5000)~$GeV, 
for $Y_Q=1/6$ and $Y_L=-1/2$, and the Yukawa couplings in the range $(0.1,5)$.
We set all the quark masses equal, i.e. $M_{\{Q,U,D\}} = M_Q$, and quark Yukawa couplings equal,
i.e. $\lambda_U = \lambda_D \equiv \lambda_Q$, 
and all the lepton masses equal, i.e. $M_{\{L,E,N\}} = M_L$ and all lepton Yukawa couplings equal,
i.e. $\lambda_E = \lambda_N \equiv \lambda_L$.   
All these points satisfy the $S$ and $T$ constraints at or better than $2\, \sigma$ level.
The color of the dots denote the lightest mass eigenvalue; 
the red, blue and green dots respectively stand for light, medium and heavy mass categories given in Table~\ref{MVLcats.TAB}. 
These results indicate the sizes of the deviations one can expect in the models that we have considered. 
In the next subsection, we ask to what degree such deviations are allowed by the present LHC Higgs data 
and the constraints on the model parameter space. 

\subsection{Fit to the LHC Higgs data}
\label{sec:higgs fit}

The ATLAS and CMS collaborations have extracted the effective $hgg$ and $h\gamma\gamma$ couplings from a combined fit to the 
various observed diboson Higgs decay modes. We give this result in Table~\ref{KaphgghAA.TAB}, where 
we quote the ATLAS result~\cite{Aad:2013wqa} directly as given, while we have translated
the CMS result~\cite{CMS-HiggsCombi} on the $95$\,\% C.L intervals of $\kappa_g = [0.63,1.05]$ and $\kappa_\gamma = [0.59,1.30]$ 
to the $1\,\sigma$ values shown. 
\begin{table}[!ht]
\caption{$\kappa_g$ and $\kappa_\gamma$ values from ATLAS~\cite{Aad:2013wqa} and CMS~\cite{CMS-HiggsCombi}. 
\label{KaphgghAA.TAB}}
\begin{centering}
\begin{tabular}{|c||c|c|}
\hline 
Coupling & ATLAS & CMS \tabularnewline
\hline 
\hline 
$\kappa_g$ & $1.04 \pm 0.14$ & $0.83\pm 0.11$
\tabularnewline
\hline 
$\kappa_\gamma$ & $1.2 \pm 0.15$ & $0.97\pm 0.18$
\tabularnewline
\hline 
\end{tabular}
\par\end{centering}
\end{table}
We perform a $\chi^2$ fit of the SM plus vector-like fermions model to these data.  
We treat the ATLAS and CMS channels shown in the Table as independent, implying four degrees of freedom ($dof$). 
We neglect correlations between $\kappa_g$ and $\kappa_\gamma$, which is a reasonable approximation. 

We compute the $\chi^2$ function 
\beq
\chi^2 = \sum_{i=1}^4 \left( \kappa^{\rm Exp}_i - \kappa^{\rm Th}_i \right)^2/\left(\sigma^{\rm Exp}_i\right)^2 \ ,
\eeq
where the $\kappa^{\rm Th}_i = \{ \kappa_g, \kappa_\gamma \}$ for the SM plus vector-like fermion models discussed in Sec.~\ref{VLFmodels.SEC},
and compared with the respective four $\kappa^{\rm Exp}_i$ ATLAS and CMS experimental values shown in Table~\ref{KaphgghAA.TAB}.
The $\kappa_i$ are given by 
\beq
\kappa_g = \sqrt{\frac{\Gamma_{gg}}{\Gamma^{SM}_{gg}}} \ \ ; \quad \kappa_\gamma = \sqrt{\frac{\Gamma_{\gamma\gamma}}{\Gamma^{SM}_{\gamma\gamma}}} \ .
\label{kapDef.EQ}
\eeq
As is standard (see for example, Ref.~\cite{LouisLyonsBook}), 
from the $\chi^2$ value for that model with vector-like fermions and the above LHC data (with $dof=4$), 
we compute ${\cal F}_{\chi^2}$ (which is twice the ``p-value''), the fraction of times a worse fit is obtained, 
which is the integral of the tail of the $\chi^2$ distribution from that $\chi^2$ value up to infinity. 
Roughly speaking, the regions of parameter space where ${\cal F}_{\chi^2} < 0.05$ can be taken to be excluded at about $2\,\sigma$ 
Gaussian equivalent, and regions where ${\cal F}_{\chi^2} < 0.01$ excluded at about $2.6\,\sigma$. 

For the SM alone, without the addition of vector-like fermions, we have $\kappa_{g,\gamma} = 1$ by definition, 
and from the data in Table~\ref{KaphgghAA.TAB} we obtain $\chi^2/dof = 1.07$
which yields ${\cal F}_{\chi^2} = 0.37$, an acceptable fit to the data. 
Next, we present $\chi^2$ and ${\cal F}_{\chi^2}$ for the SM plus vector-like fermions with the goal of identifying 
regions of vector-like fermion parameter space that have values of ${\cal F}_{\chi^2}$ bigger than about 0.05, which can be taken as the 
allowed regions for that model, given the present data.  

\begin{figure}[!ht]
\begin{center}
\includegraphics[width=0.39\textwidth]{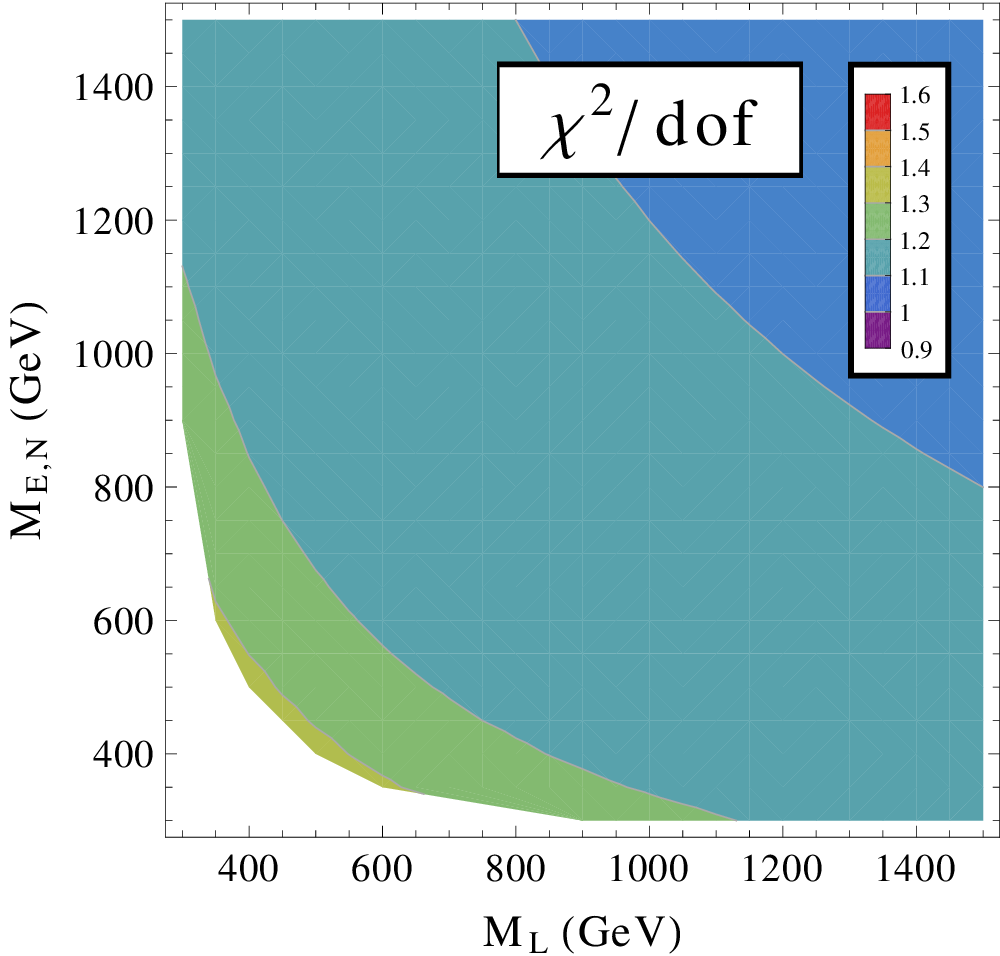}
\includegraphics[width=0.39\textwidth]{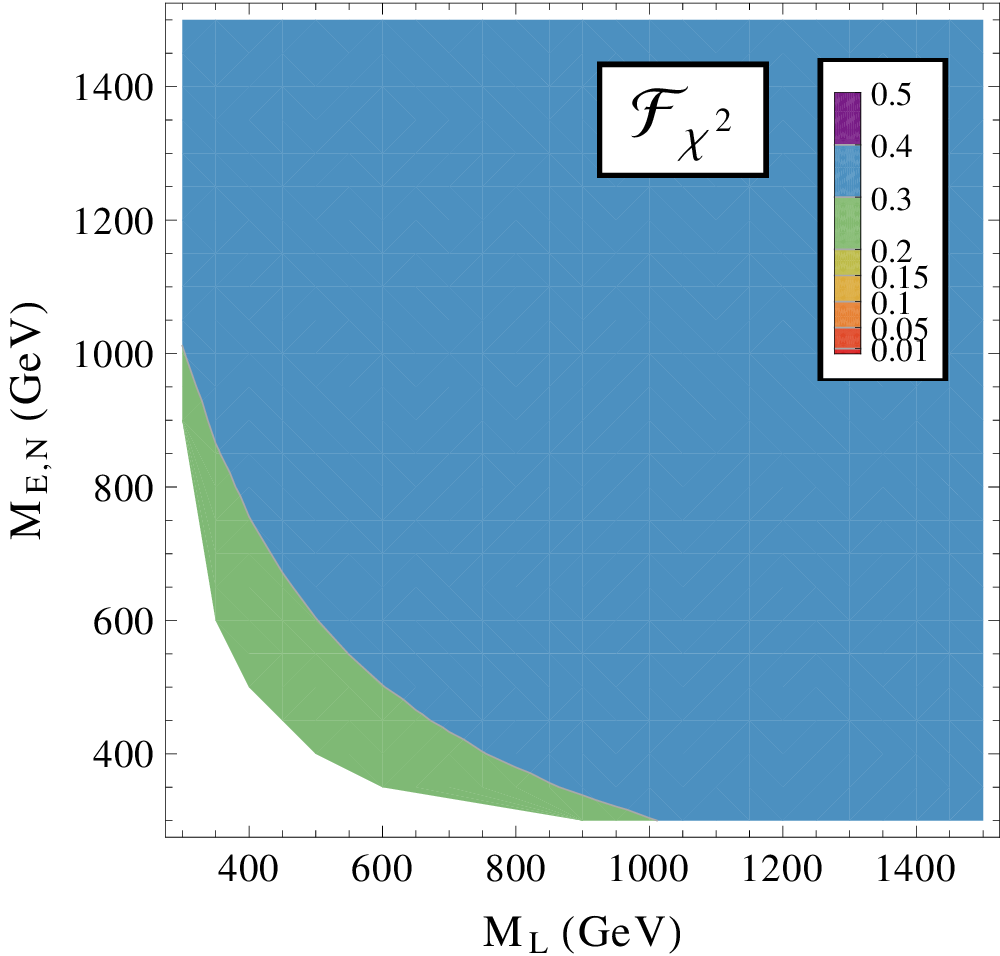}
\caption{For the  VL$_1$ model, for $Y_L = -1/2$, $\lambda_E = 1$, $\lambda_N = 1$, $M_E = M_N$. 
The colored regions are with all lepton mass eigenvalues $\geq 250~$GeV, 
and lie within the $2~\sigma$ ellipse in $S$ and $T$. 
In all the plots, $\chi^2$ values decrease and ${\cal F}$ values increase as we go toward larger masses, or lower $\lambda$. 
\label{VL1fitKap-M.FIG}}
\end{center}
\end{figure}

In Fig.~\ref{VL1fitKap-M.FIG} we show the $\chi^2/{\rm dof}$ and ${\cal F}_{\chi^2}$ for the VL$_1$ model 
with $Y_L = -1/2$, $\lambda_{L,N} = 1$, $M_E = M_N$, i.e., the vector-like singlet masses taken equal.
The parameter space shown in color is with all lepton mass eigenvalues $\geq 250~$GeV, 
and satisfy the $S$ and $T$ constraints at or better than $2~\sigma$.  

\begin{figure}[!ht]
\begin{center}
\includegraphics[width=0.39\textwidth]{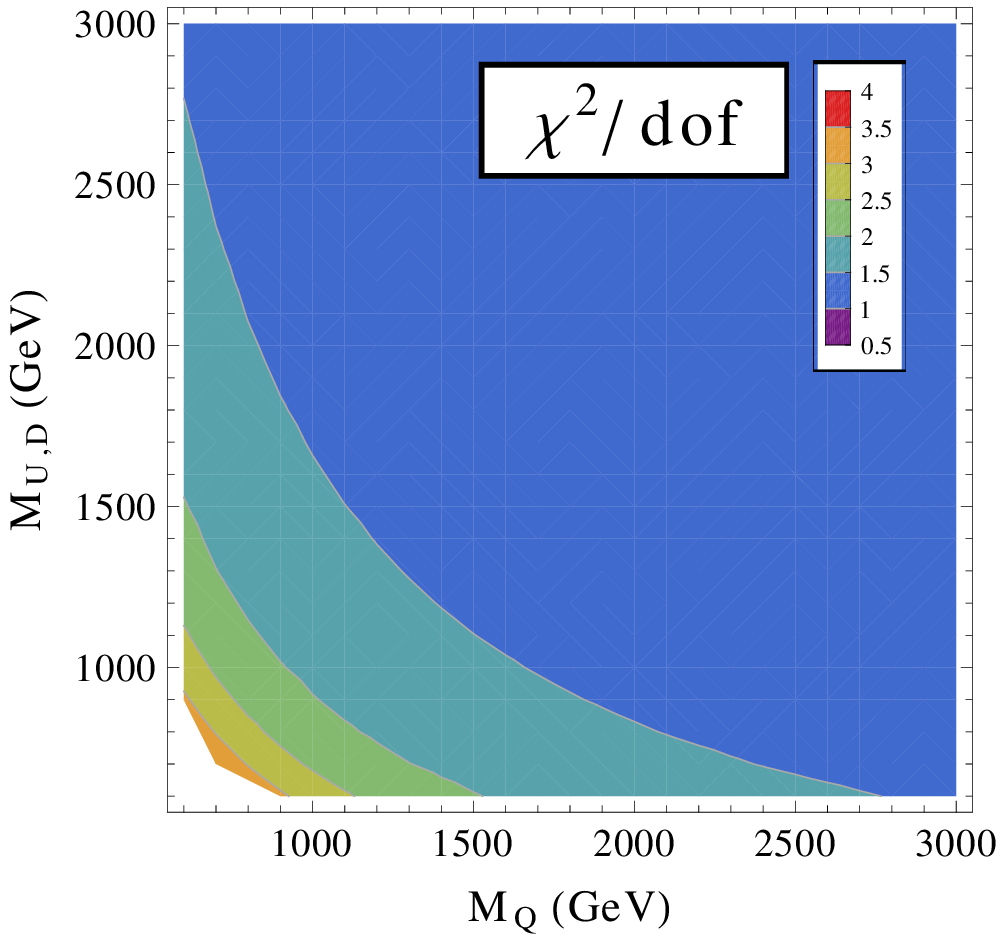}
\includegraphics[width=0.39\textwidth]{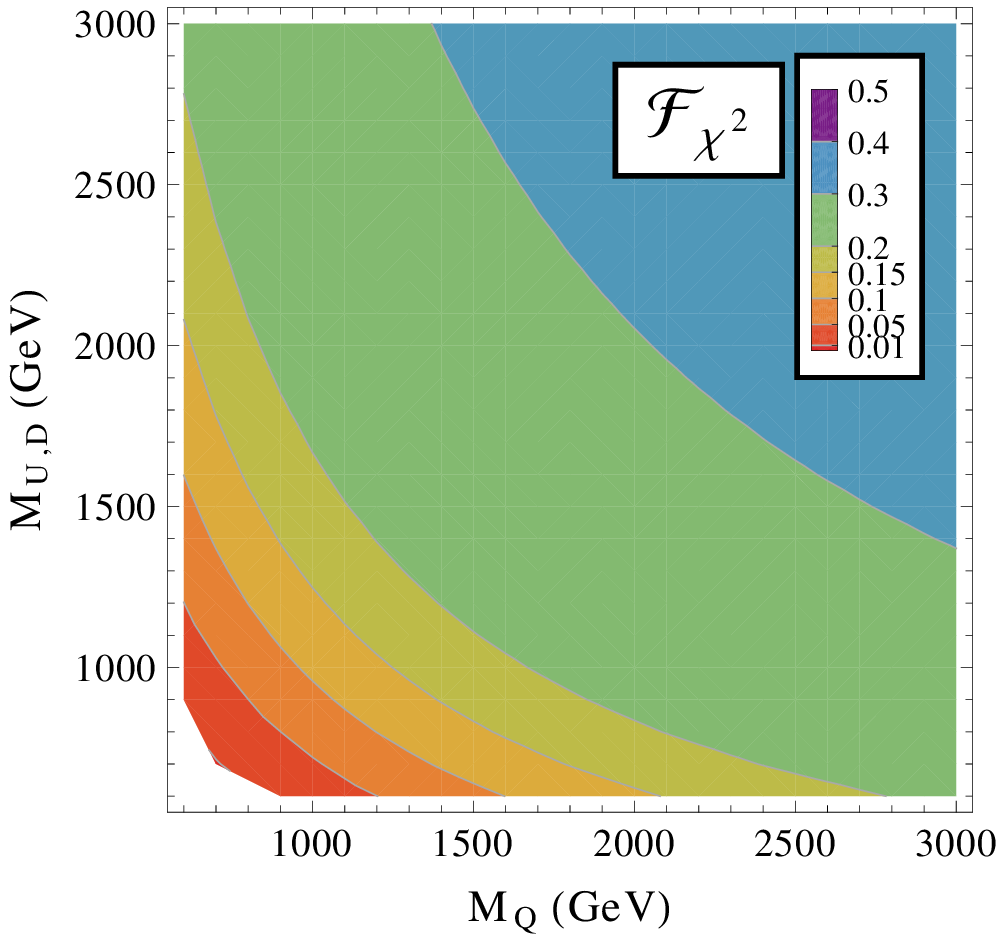}
\caption{For the  VQ$_1$ model, for $Y_Q = 1/6$, $\lambda_U = 1$, $\lambda_D = 1$, $M_U = M_D$. 
The colored regions are with all quark mass eigenvalues $\geq 500~$GeV, 
and satisfy the $S$ and $T$ constraints at or better than $2~\sigma$.  
In all the plots, $\chi^2$ values decrease and ${\cal F}$ values increase as we go toward larger masses, or lower $\lambda$. 
\label{VQ1fitKap-M.FIG}}
\end{center}
\end{figure}

In Fig.~\ref{VQ1fitKap-M.FIG} we show the $\chi^2/{\rm dof}$ and ${\cal F}_{\chi^2}$ for the VQ$_1$ model 
with $Y_Q = 1/6$, $\lambda_{U,D} = 1$, $M_U = M_D$, i.e., the vector-like singlet masses taken equal.
The parameter space shown in color is with all quark mass eigenvalues $\geq 500~$GeV, 
and satisfy the $S$ and $T$ constraints at or better than $2~\sigma$.

\begin{figure}[!ht]
\begin{center}
\includegraphics[width=0.39\textwidth]{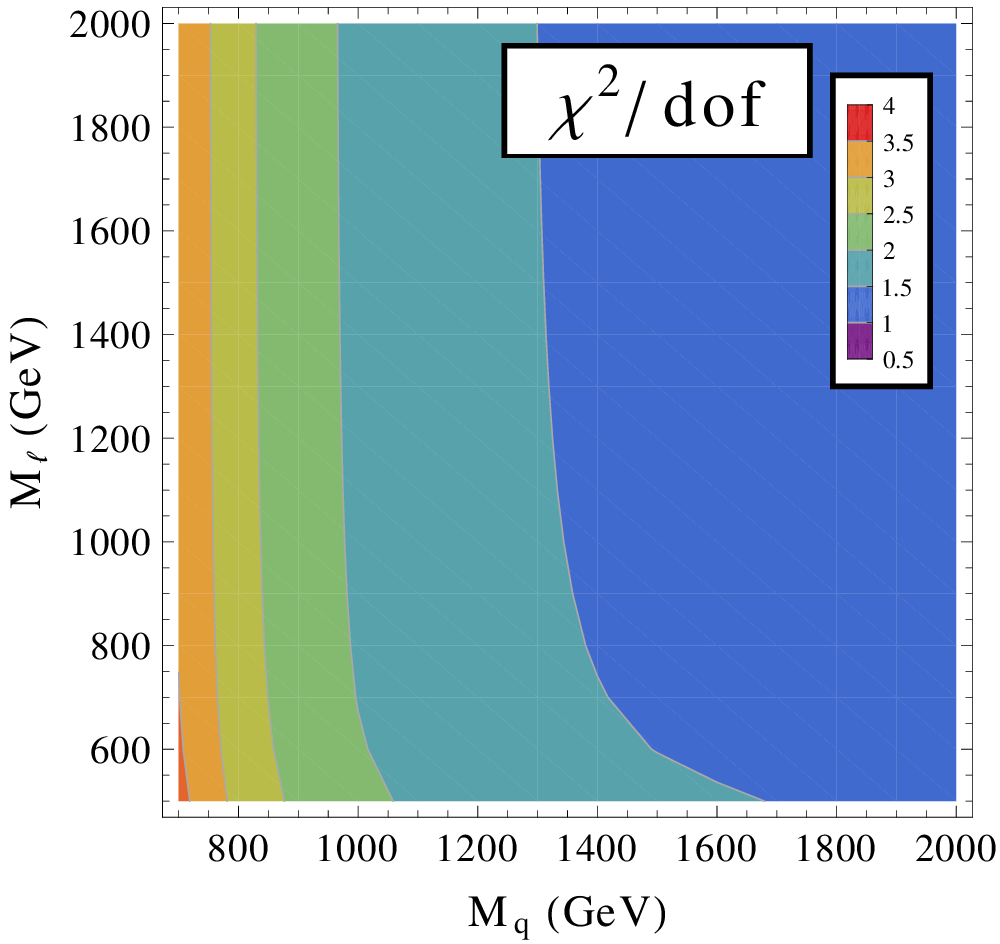}
\vspace*{0.5cm}
\includegraphics[width=0.39\textwidth]{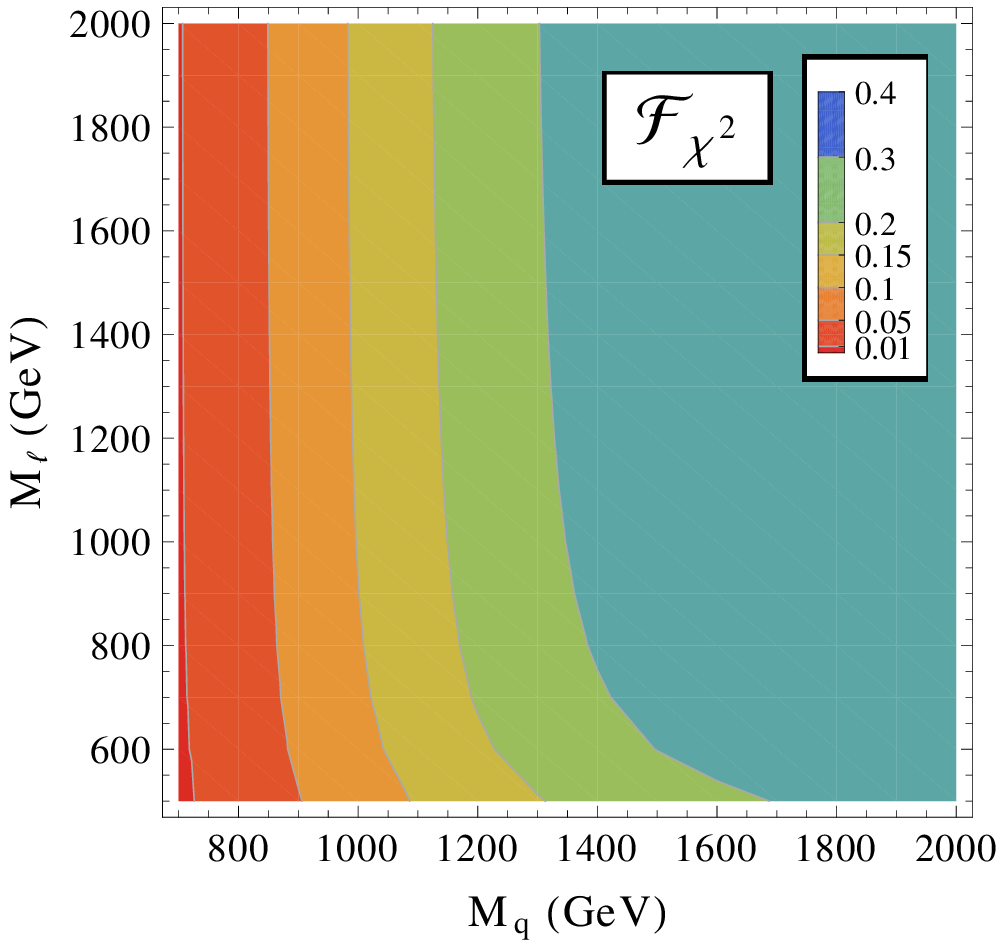}
\includegraphics[width=0.39\textwidth]{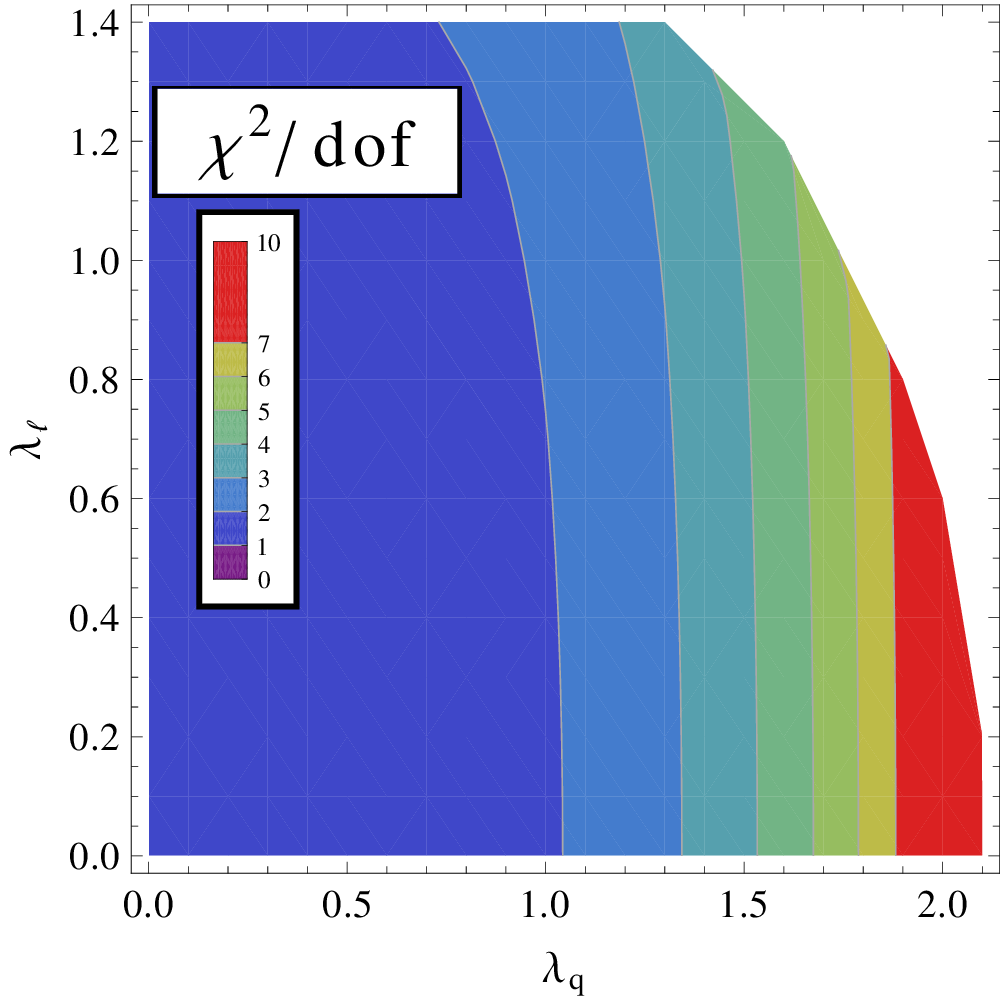}  
\includegraphics[width=0.39\textwidth]{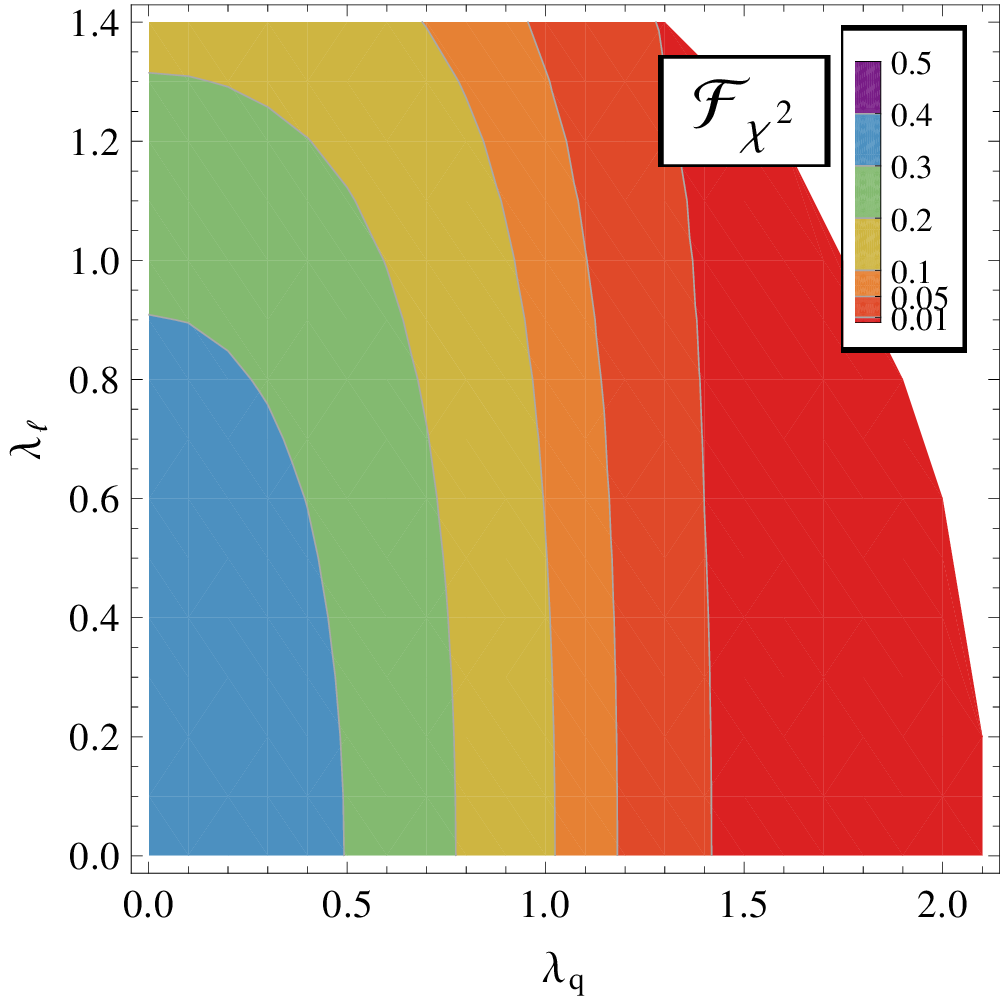}
\caption{For the VSM$_1$ model, for $Y_Q = 1/6$, $Y_L = -1/2$, $\lambda_q = 1$, $\lambda_\ell = 1$, $M_q = 1000$~GeV and $M_\ell = 500$~GeV. 
The colored regions are with all quark mass $\geq 500~$GeV, 
all lepton mass eigenvalues $\geq 250~$GeV, and satisfy the $S$ and $T$ constraints at or better than $2~\sigma$.
In all the plots, $\chi^2$ values decrease and ${\cal F}$ values increase as we go toward larger masses, or lower $\lambda$. 
\label{VSM1fitKap.FIG}}
\end{center}
\end{figure}

In Fig.~\ref{VSM1fitKap.FIG} we show the $\chi^2/{\rm dof}$ and ${\cal F}_{\chi^2}$ for the VSM$_1$ model
with $Y_Q = 1/6$, $Y_L = -1/2$, $\lambda_{U,D}\equiv \lambda_q = 1$, $\lambda_{E,N}\equiv \lambda_\ell = 1$, $M_{Q,U,D}\equiv M_q = 1000$~GeV, $M_{L,E,N} \equiv M_\ell = 500$~GeV, 
i.e. all the vector-like quark masses are taken equal, 
and all the vector-like lepton masses are taken equal (but not necessarily the same as the quark masses).  
The plots show the dependence on the two parameters that are varied with the other parameters fixed at the above mentioned values. 
The colored regions are with all quark mass eigenvalues $\geq 500~$GeV, 
all lepton mass eigenvalues $\geq 250~$GeV, and satisfy the $S$ and $T$ constraints at or better than $2~\sigma$.
For given masses, the $\chi^2$ approaches the SM value as $\lambda$ decreases; 
for example, for $\lambda=0.5$, $M_q = 1000$~GeV, $M_\ell = 500$~GeV 
we have $\chi^2/{\rm dof} = 1.26$, ${\cal F}_{\chi^2} = 0.28$.

\section{Conclusions}

In this paper we have surveyed numerous vector-like fermion extensions of the SM. Our purpose has been to illuminate the structure of the theories, detail the precision electroweak implications and constraints, investigate LHC direct limits, and investigate the implications of vector-like fermions for Higgs boson production and decay.  

The phenomenological implications of vector-like theories depend crucially on the details of the precise theory in question. Precision electroweak constraints depend on masses and couplings to vector bosons, flavor constraints depend on their flavor mixings with SM fermions, direct detection constraints depend on mass hierarchies and the size of the couplings and mixings to SM states that enable prompt or non-prompt decays into various model-specific final state ratios. In addition, the Higgs boson observables depend on overall mass scales and the ratios of the strongly interacting vector-like masses to electroweak interacting states.  Our aim was to demonstrate all of these features through examples and extensive computations.

One of the recent motivations for considering vector-like states added to the SM was to account for possible deviations in the $\mu_{\gamma\gamma}=\sigma(h)B(h\to \gamma\gamma)$ rate at the LHC.  As discussed in a Sec.~\ref{sec:higgs fit}, early indications from the LHC Higgs studies suggested an enhanced rate to two photons. Presently the central values of the ATLAS data are higher than the SM expected rates, and the central values of the CMS data are lower than the SM expected rates, making a combined total more consistent with the SM than originally thought. Nevertheless, it is one of the few observables that is sensitive to new physics in loops, and the uncertainties in the experimental measurements 
and QCD uncertainties~\cite{Baglio:2012et}
give plenty of room for large effects (tens of percent) from new physics. In our investigations we have shown that the addition of vector-like fermions with vector-like masses less than a few TeV, and with chiral couplings to the Higgs boson, leads to tens of percent shifts in Higgs production ($gg\to h\to XY$) and decay ($h\to\gamma\gamma$) observables. In time, measurements of Higgs observables and direct searches for vector-like states will go far to confirm or exclude this possibility near the weak scale.

\medskip
\noindent {\it Acknowledgements:}
We thank the CERN Theory group for hospitality where this work was initiated.  
We thank M.~Muhlleitner and M.~Spira for a discussion on the size of NNLO corrections to the $ggh$ amplitude.

\setcounter{section}{0}
\renewcommand\thesection{\Alph{section}}               
\renewcommand\thesubsection{\Alph{section}.\arabic{subsection}}
\renewcommand\thesubsubsection{\Alph{section}.\arabic{subsection}.\arabic{subsubsection}}

\renewcommand{\theequation}{\Alph{section}.\arabic{equation}}    
\renewcommand{\thetable}{\Alph{section}.\arabic{table}}          
\renewcommand{\thefigure}{\Alph{section}.\arabic{figure}}        

\section{Gauge boson 2-point functions with vector-like fermions}
\label{VV2pt-VL.SEC}
In a vector-like theory each vertex in Fig.~\ref{VV-mimj.FIG} is proportional to $P_L + P_R$, resulting in the total contribution 
$\Pi_{LL} + \Pi_{LR} + \Pi_{RL} + \Pi_{RR}$. 
A similar result holds for the derivative $\Pi^\prime$ also, where $\Pi^\prime = d\Pi/dq^2$. 
Since we have $\Pi_{LL} = \Pi_{RR}$ and $\Pi_{LR} = \Pi_{RL}$, the above contribution is $2 (\Pi_{LL} + \Pi_{LR})$. 
Using dimensional regularization and continuing to $d=4-\epsilon$ dimensions, we have (see for example, Ref.~\cite{PSBook})
\bea
\Pi^{(m_i m_j)}_{LL}(q^2)  &=&  -\frac{4}{(4\pi)^{d/2}} \int_0^1  dx \frac{\Gamma(2-d/2)}{\Delta^{2-d/2}} 
\left[ x(1-x) q^2 - \frac{1}{2} (x m_j^2 + (1-x) m_i^2) \right] \ ,  \label{PiLLq2.EQ} \nonumber \\ \\
\Pi^{(m_i m_j)}_{LR}(q^2) &=& -\frac{2}{(4\pi)^{d/2}} \int_0^1  dx \frac{\Gamma(2-d/2)}{\Delta^{2-d/2}} m_i m_j \ .
\label{PiLRq2.EQ}
\eea 
where $\Delta = x m_j^2 + (1-x) m_i^2 - x (1-x) q^2$, 
and we have dropped terms proportional to $q^\mu q^\nu$ assuming that gauge-bosons connect to massless fermion currents giving zero. 
We have
\beq
\lim_{d \rightarrow (4-\epsilon)} \frac{1}{(4\pi)^{d/2}} \frac{\Gamma(2-d/2)}{\Delta^{2-d/2}} = \frac{1}{(4\pi)^2} \left[ \frac{2}{\epsilon} - \gamma + \log(4\pi) - \log\Delta  \right] \ .
\label{GamOvDelta.EQ}
\eeq 
For notational ease, in the following we will define $\Pi \equiv \Pi_{LL} + \Pi_{LR}$, similarly for $\Pi^\prime$, 
and also, $\Pi^{\{m_i,m_j\}} = \Pi^{(m_i m_j)} + \Pi^{(m_j m_i)}$, $\Pi^{(m_i)} = \Pi^{(m_i m_i)}$.  
From Eqs.~(\ref{PiLLq2.EQ}),~(\ref{PiLRq2.EQ})~and~(\ref{GamOvDelta.EQ}), we obtain the following
\bea
\left. \Pi^{(m_i m_j)}(0) \right|_{\rm mass-dep} &=&  \frac{1}{32 \pi^2} \frac{1}{(m_j^2 - m_i^2)} \left\{ (m_j^2 - m_i^2) (m_i^2+m_j^2-4m_im_j) \right.  \nonumber \\
& &\left. - 2 (m_j^4 \log{m_j^2} - m_i^4\log{m_i^2}) + 4 m_i m_j (m_j^2 \log{m_j^2} - m_i^2 \log{m_i^2}) \right\}   \ , \nonumber \\
\left. \Pi^{\{m_i, m_j\}}(0) \right|_{\rm mass-dep} &=&  -\frac{1}{16 \pi^2} \frac{1}{(m_j^2 - m_i^2)} \left\{ 4 m_i m_j (m_j^2 - m_i^2) + 2 m_j^3 (m_j - 2 m_i) \log{m_j^2} \right. \nonumber \\
& &\left. + 2 m_i^3 (2m_j - m_i) \log{m_i^2} + m_i^4 - m_j^4 \right\} \ , \nonumber \\
\left. \Pi^{(m_i)}(0) \right|_{\rm mass-dep} &=& 0 \ ,
\label{PiVLMassDep.EQ}
\eea
where ``mass-dep'' denotes mass dependent finite parts excluding the $2/\epsilon - \gamma + 4\pi$ pieces.   
Again, from Eqs.~(\ref{PiLLq2.EQ}),~(\ref{PiLRq2.EQ})~and~(\ref{GamOvDelta.EQ}), we obtain
\bea
\Pi^{\prime {\ \{m_i,m_j\}}}_{LR}(0) &=& - \frac{m_i m_j \left[m_i^4 - m_j^4 - 4 m_i^2 m_j^2 \tanh^{-1}{\left((m_i^2-m_j^2)/(m_i^2 + m_j^2)  \right)} \right]}{8 \pi^2 (m_i^2-m_j^2)^3} \ , \nonumber \\
\Pi^{\prime {\ \{m_i,m_j\}}}_{LL}(0) &=& - \frac{1}{4 \pi^2} \left\{ \frac{1}{3} \left(-\frac12 + \frac2\epsilon -\gamma +\log{(4\pi)}\right) 
-\frac{1}{18 (m_i^2-m_j^2)^3} \left({\cal A} - {\cal B} \right)   \right\} \nonumber \\
{\cal A} &=& (m_i^2-m_j^2) \left[ -5m_i^4-5m_j^4+22m_i^2m_j^2+6(m_i^2-m_j^2)^2\log{m_im_j} \right. \nonumber \\
& & \left. + 18(m_i^4-m_j^4) {\rm cosech}^{-1}{\left(\frac{2m_im_j}{m_i^2-m_j^2}\right)} \right] \nonumber \\ 
{\cal B} &=& 12(m_i^2+m_j^2)(m_i^4+m_j^4-m_i^2m_j^2)\coth^{-1}\left(\frac{m_i^2+m_j^2}{m_i^2-m_j^2}\right) \ , \nonumber \\
\Pi^{\prime\ (m)}(0) &=& -\frac{2}{3(4\pi)^2} \left[ \frac2\epsilon-\gamma+\log{(4\pi) -\log{m^2}}  \right] \ , 
\label{PiPriVL.EQ}
\eea
which are valid for $(m_i \neq m_j)$, and we recall the definition $\Pi^\prime \equiv \Pi_{LL}^\prime + \Pi_{LR}^\prime$.
The $\log{m^2}$ terms are understood to be $\log{(m^2/\mu^2)}$ where $\mu$ is an arbitrary renormalization scale. 
For our numerical computations we set $\mu = M_Z$, 
but we have verified that our results remain the same even if a different value of $\mu$ is chosen.

\section{Explicit expressions for the mixing coefficients}
\label{MixCoeff.APP}

This appendix contains expressions for the various coefficients that we defined in order to simplify the formulae 
in Sec.~\ref{ColliderConstraints}.

\subsection*{Mixing through the Higgs boson}
We give here the explicit expressions for the mixing coefficients from section 5 between the new vector-like generation and the SM quarks through the Higgs boson. 

As explained in the text, $\alpha_{ij}$ are the new Yukawa couplings that arise from applying the matrices that diagonalise the mass matrix in Eq.~(\ref{MassLagrangian.EQ}) to the matrix of Yukawa couplings in the Lagrangian, given in Eq. (\ref{YukawaLagrangian.EQ}). In order to simplify the expressions derived for $\alpha_{ij}$, we define the following quantities
\begin{align}
\nonumber&a^u_1 = (V_L^u)^{12} = \frac{m_{ct}m_{tt} + m_{cc}m_{ct} + m_{cT}m_{tT}}{m_{tt}^2-m_{cc}^2},~~a^u_4 = (V_R^u)^{12} =  \frac{m_{ct}m_{tt} + m_{cc}m_{ct} + \mu_c \mu_t}{m_{tt}^2-m_{cc}^2} \\
\nonumber&a^u_2 = (V_L^u)^{13} = \frac{\mu_Tm_{cT} + \mu_cm_{cc} + \mu_tm_{ct}}{\mu_T^2-m_{cc}^2},~~a^u_5 = (V_R^u)^{13} = \frac{\mu_T \mu_c + m_{cT}m_{cc} + m_{tT}m_{ct}}{\mu_T^2-m_{cc}^2} \\
&a^u_3 = (V_L^u)^{23} = \frac{\mu_Tm_{tT} + \mu_cm_{ct} + \mu_tm_{tt}}{\mu_T^2-m_{tt}^2},~~a^u_6 = (V_R^u)^{23} = \frac{m_{ct}m_{cT} + \mu_t\mu_T+ m_{tt}m_{tT}}{\mu_T^2-m_{tt}^2}
\end{align}
Similar quantities can be defined for the down type expressions, with
\begin{align}
\nonumber&a^d_1 = (V_L^d)^{12} = \frac{m_{sb}m_{bb} + m_{ss}m_{sb} + m_{sB}m_{bB}}{m_{bb}^2-m_{ss}^2},~~a^d_4 = (V_R^d)^{12} =  \frac{m_{sb}m_{bb} + m_{ss}m_{sb} + \mu_s \mu_b}{m_{bb}^2-m_{ss}^2} \\
\nonumber&a^d_2 = (V_L^d)^{13} = \frac{\mu_Bm_{sB} + \mu_s m_{ss} + \mu_b m_{sb}}{\mu_B^2-m_{ss}^2},~~a^d_5 = (V_R^d)^{13} = \frac{\mu_B \mu_s + m_{sB}m_{ss} + m_{bB}m_{sb}}{\mu_B^2-m_{ss}^2} \\
&a^d_3 = (V_L^d)^{23} = \frac{\mu_B m_{bB} + \mu_s m_{sb} + \mu_b m_{bb}}{\mu_B^2-m_{bb}^2},~~a^d_6 = (V_R^d)^{23} = \frac{m_{sb}m_{sB} + \mu_b \mu_B+ m_{bb}m_{bB}}{\mu_B^2-m_{bb}^2} \ .
\end{align}
This results in the following coefficients $\alpha_{ij}$ for the up-type matrix 
\begin{align}
\nonumber&\alpha_{cc} = \lambda_{cc} + \lambda_{tc}(-a^u_1) + \left(-a^u_4\right) \left( \lambda_{ct}+ \lambda_{tt}(-a^u_1) \right) + \left(-a^u_5\right) \left( \lambda_{cT}+ \lambda_{tT}(-a^u_1)\right) \nonumber\\
\nonumber&\alpha_{ct} = \lambda_{ct} + \lambda_{tt}(-a^u_1) + a^u_4 \left( \lambda_{cc}+ \lambda_{tc}(-a^u_1) \right) + a^u_5 \left( \lambda_{cT}+ \lambda_{tT}(-a^u_1) \right) \nonumber\\
\nonumber&\alpha_{cT}  = \lambda_{cT} + \lambda_{tT}(-a^u_1) + a^u_5 \left( \lambda_{cc}+ \lambda_{tc}(-a^u_1) \right) + a^u_6 \left( \lambda_{ct}+ \lambda_{tt}(-a^u_1) \right) \nonumber\\
\nonumber&\alpha_{tc} = \lambda_{tc} +  \lambda_{cc}a^u_1+ \left(-a^u_4\right) \left( \lambda_{tt} + \lambda_{ct}a^u_1\right)+ \left(-a^u_5\right) \left(\lambda_{tT} + \lambda_{cT}a^u_1 \right) \nonumber\\
\nonumber&\alpha_{tt} = \lambda_{tt} +  \lambda_{ct}a^u_1+a^u_4 \left( \lambda_{tc} + \lambda_{cc} a^u_1\right)+ \left( -a^u_6\right) \left( \lambda_{tT} + \lambda_{cT}a^u_1 \right) \nonumber\\
\nonumber&\alpha_{tT} = \lambda_{tT} +  \lambda_{cT}a^u_1+ a^u_5 \left( \lambda_{tc} + \lambda_{cc}  a^u_1 \right)+ a^u_6 \left( \lambda_{tt} + \lambda_{ct}  a^u_1 \right) \nonumber\\
\nonumber&\alpha_{Tc} = \lambda_{tc} a^u_3 + \lambda_{cc} a^u_2+ \left( -a^u_4 \right)\left( \lambda_{tt}  a^u_3 + \lambda_{ct}a^u_2 \right)+\left( -a^u_5\right) \left( \lambda_{tT}  a^u_3 + \lambda_{cT}a^u_2 \right) \nonumber\\
\nonumber&\alpha_{Tt} = \lambda_{ct}a^u_2  + \lambda_{tt} a^u_3+a^u_4 \left( \lambda_{tc}  a^u_3 + \lambda_{cc}a^u_2 \right)+ \left( -a^u_6\right) \left( \lambda_{tT}  a^u_3 + \lambda_{cT}a^u_2 \right) \nonumber\\
&\alpha_{TT} = \lambda_{cT}a^u_2 + \lambda_{tT} a^u_3 + a^u_6  \left( \lambda_{tt}  a^u_3 + \lambda_{ct}a^u_2 \right)+a^u_5 \left( \lambda_{tc}  a^u_3 + \lambda_{cc}a^u_2\right)
\label{Alphas.EQ}
\end{align}
and similar coefficients for the down type, with $c \leftrightarrow s, ~~t \leftrightarrow b, ~~T \leftrightarrow B$.

The mass eigenvalues $M_i$ ($i= c,~t,~T$) are given by the following equations
\begin{align}
\nonumber & M_c = m_{cc} - a_1^u m_{tc} - a_2^u \mu_c - a_4^u \left( m_{ct} - a_1^u m_{tt} - a_2^u \mu_t\right) - a_5^u \left( m_{cT} - a_1^u m_{tT} - a_2^u \mu_T\right)
\\
\nonumber & M_t = m_{tt} + a_1^u m_{ct} - a_3^u \mu_t + a_4^u \left( m_{tc} + a_1^u m_{cc} - a_3^u \mu_c\right) - a_6^u \left( m_{tT} + a_1^u m_{cT} - a_3^u \mu_T\right)
\\
 & M_T = \mu_T + a_2^u m_{cT} + a_3^u m_{tT} + a_5^u \left( \mu_c + a_2^u m_{cc} + a_3^u m_{tc}\right) + a_6^u \left( \mu_t + a_2^u m_{ct} + a_3^u m_{tt}\right)
\label{EigenMasses.EQ}
\end{align}
\subsection*{Mixing through the Z boson}

The coefficients $\beta_{ij}$ defined in the text are given here explicitly. They govern the size of mixing between the vector-like and SM quarks via the Z boson discussed in section 5.

\begin{align}
&\nonumber\beta_{11} = \frac{1}{2}\left[1 -\left(\VLTc \right)^2\right]-\frac{2}{3}\sin^2\theta_W \\
&\nonumber\beta_{12} = -\frac{1}{2}\left( \VLTc\right)\left( \VLTt \right) \\
&\nonumber\beta_{13} = -\frac{1}{2} \left( \VLTc \right)  \\
&\nonumber\beta_{21} = -\frac{1}{2}\left( \VLTc \right)\left(\VLTt \right)  \\
&\nonumber\beta_{22} =  \frac{1}{2}\left[1 -\left(\VLTt \right)^2\right]-\frac{2}{3}\sin^2\theta_W  \\
&\nonumber\beta_{23} = -\frac{1}{2} \left( \VLTt \right)  \\
&\nonumber\beta_{31} = -\frac{1}{2} \left( \VLTc \right)  \\
&\nonumber\beta_{32} = -\frac{1}{2} \left( \VLTt \right)  \\
&\beta_{33} = -\frac{2}{3}\sin^2\theta_W
\end{align}

Similar expressions can be found for the down-type quark interactions with the Z, but are not listed here.


\end{document}